\documentclass[10pt]{iopart}
\usepackage{iopams}
\usepackage{t1enc}
\usepackage{setstack}
\usepackage{graphicx}
\usepackage[english]{babel}
\usepackage[T1]{fontenc}
\usepackage[latin1]{inputenc}
\usepackage{url}

\newcommand{\unaryminus}{\scalebox{0.75}[1.0]{\( - \)}}

\DeclareMathAlphabet{\mathantt}{OT1}{antt}{li}{it}
\DeclareMathAlphabet{\mathantt}{OT1}{pzc}{m}{it}

\graphicspath{{./figures/}}

\begin{document}
\title[Sparse source anomaly detection: accuracy and robustness]{Sparse source travel-time tomography of a laboratory target: accuracy and robustness of anomaly detection}
\author{S Pursiainen$^{1,2}$ and M Kaasalainen$^2$}
\address{$^1$Department of Mathematics and Systems Analysis, Aalto University, Helsinki, Finland}
\address{$^2$Tampere University of Technology, Department of Mathematics, PO Box 553, FI-33101 Tampere, Finland} 
\ead{sampsa.pursiainen@iki.fi}
\pagenumbering{arabic}

\begin{abstract}
This study concerned conebeam travel-time tomography. The focus was on a sparse distribution of signal sources that can be necessary in a challenging {\em in situ} environment such as in asteroid tomography. The goal was to approximate the minimum  number of source positions needed for robust detection of refractive anomalies, e.g., voids within an asteroid or a casting defects in concrete. Experimental ultrasonic  data were recorded utilizing as a target a 150 mm plastic cast cube  containing three stones with diameter between 22 and 41 mm. A signal frequency of 55 kHz (35 mm wavelength) was used.  Source counts from one to six were tested for different placements. Based on our statistical inversion approach and analysis of the results, three or four sources were found to lead to reliable inversion.  The source configurations  investigated were also ranked according to their performance. Our results can be used, for example, in the planning of planetary missions as well as in material testing.

\end{abstract}

\pacs{02.30.Zz, 42.30.Wb, 43.35.Yb, 96.30.Ys} \ams{65R32, 85A99, 85A99}

\section{Introduction}

This study concerned conebeam tomography in which a signal carried by a pulse wave was transmitted and received on opposite sides of a target domain and an unknown refractive index ${\mathtt n}$, the inverse of the signal velocity,  was to be recovered from travel time data. If the data are incomplete, the task in question is an ill-posed inverse problem \cite{kaipio2004}; i.e., the solution is non-unique and small errors in the data can cause large deviations in the solution. Moreover, the dependence of the data on the unknowns is non-linear. Our focus was on a scenario in which the distribution of sources needs to be sparse due to {\em in situ} limitations. Since the signal paths are non-linear and they depend on ${\mathtt n}$, the number of sources needed for reliable inverse results does not follow directly from problem parameters, e.g., spatial dimensionality. Furthermore, since general stability theorems comparable to those of surface reconstruction \cite{kaasalainen1992, kaasalainen2006,kaasalainen2011} are not available, we rely on numerical and experimental analysis continuing the work begun in \cite{pursiainen2013}. Guided by these aspects, the general objective of this study was to approximate the minimum number of source positions needed for robust detection of refractive anomalies.  Another main point was to conduct a laboratory experiment instead of simulations to avoid inverse crime due to model errors. This is paricularly important when assessing the robustness of methods and measurement setups.

The foremost application in mind was asteroid tomography in which the relative electric permittivity $\varepsilon_r$ satisfying ${\mathtt n} = 1/\sqrt{\varepsilon_r}$ is to be sought  based on radio frequency data gathered by an orbiter. Currently, there is a growing interest towards non-invasive imaging of asteroids to support future planetary research and extra-terrestrial mining activities. The first attempt for tomography of a planetary object has already been implemented in the CONSERT (comet nucleus sounding experiment by radiowave transmission) 
\cite{kofman2007,kofman2004,kofman1998,herique2011,herique2011b,herique2010,herique1999,landmann2010,nielsen2001,barriot1999} as  part of the Rosetta mission. CONSERT utilizes a single transmitter or transponder positioned on a comet surface. In this study, a similar scheme was explored covering source counts up to six positions. Optimizing the configuration with respect to the positioning and the number of sources is an important goal of mission planning in which high costs, long duration and very limited payload have to be taken into account. The robustness of the inversion method is also a central aspect since {\em a priori}  information of the unknown is likely to be scarce.  The presently available knowledge of the density and mineral content of asteroids indicates, for example, potential existence of internal voids \cite{belton2004}, motivating investigation of refractive anomalies, i.e.\ distinct local deviations in the refractive index. Moreover, the high permittivity of the asteroid minerals \cite{elshafie2013,virkki2014,herique2002,bottke2002} is likely to lead to strongly randomized or noisy reflections\footnote{\url{http://www.youtube.com/watch?v=gkCzjsEyX4w}} due to which the present inversion strategy utilizing the direct part of the signal can be advantageous \cite{pursiainen2013}. In addition to asteroid tomography, other potential applications of the present sparse source inversion approach include on-site material testing and inspection \cite{chai2010,chai2011,acciani2008,nawy2008,schickert2003,schickert2005,rens2000}, biomedical ultrasonography  \cite{ruiter2012, ranger2012, duric2007,nowicki2012,mudry2013}, as well as plenty of atmospheric, pedospheric, geological, and biological investigations utilizing travel time data  \cite{vecherin2008,jol2008,russell1988}, such as recovery of the root-zone structure of a tree \cite{attia_al_hagrey2007}.

The objectives above were studied in a laboratory experiment in which three 22--41 mm onyx stones were to be detected from the central part of a 150 mm synthetic reson cube based on travel time of a 55 kHz (35 mm wavelength) ultrasonic signal. The data was recorded on the surface of the cube using a conventional  PUNDIT (portable ultrasonic non-destructive digital indicating tester) device akin to concrete testing applications   \cite{bungey2006,raj2002,blitz1996}. Our experiment corresponded to the asteroid void localization task in terms of relative signal velocities. Six face-centered source positions were utilized covering all possible combinations and configurations of 1--6 positions. For each source, the direct (non-reflected) part of the signal was recorded in a regular 130-by-130 mm grid with 10 mm resolution on the opposite face, simulating the direct signal cone recordable at the orbit \cite{pursiainen2013}. 

We utilized a statistical inversion approach in which a subjective posterior probability distribution was maximized via the iterative alternating sequential (IAS) algorithm \cite{ohagan2004,pursiainen2013,  calvetti2009,calvetti2008,calvetti2007}. A conditionally Gaussian prior was used with variance either a fixed (f) or distributed according to the gamma (g) or inverse gamma (ig) hyperprior density. The first alternative results in a smoother ($L^2$-type) estimate whereas the latter two produce sharper and well-localized (e.g.\ $L^1$-type) solutions. The advantage of the current statistical framework is, especially, that it enables the comparison of (g), (ig) and (f), given the initial prior variance $\theta_0$, that is, a hyperparameter controlling the strength of the prior. Accuracy of the inverse results was measured in terms of  relative overlapping volume (ROV), i.e.\ the average overlap between the recovered and actual stones, and by evaluating the relative error in the value (REV) of the refractive index. 

The results were presented and  analyzed via frequentist descriptive statistics (box plots) \cite{mcgill1978}. With a suitable choice of $\theta_0$, already two sources led to successful anomaly detection. Inversion reliability was observed to increase along with the number of sources: three or four was found to be sufficient for  all tested values of $\theta_0$. The investigated source configurations were also ranked according to the source count and configuration. Future work will include further exploration of alternative inverse methods and target objects to further extend the knowledge of inversion accuracy regarding the present sparse source tomography scenario.  

\section{Materials and methods}

\begin{figure}[h!]\begin{center}\begin{footnotesize}
\begin{minipage}{4.2cm}\begin{center}
\includegraphics[width=2.36cm,draft=false]{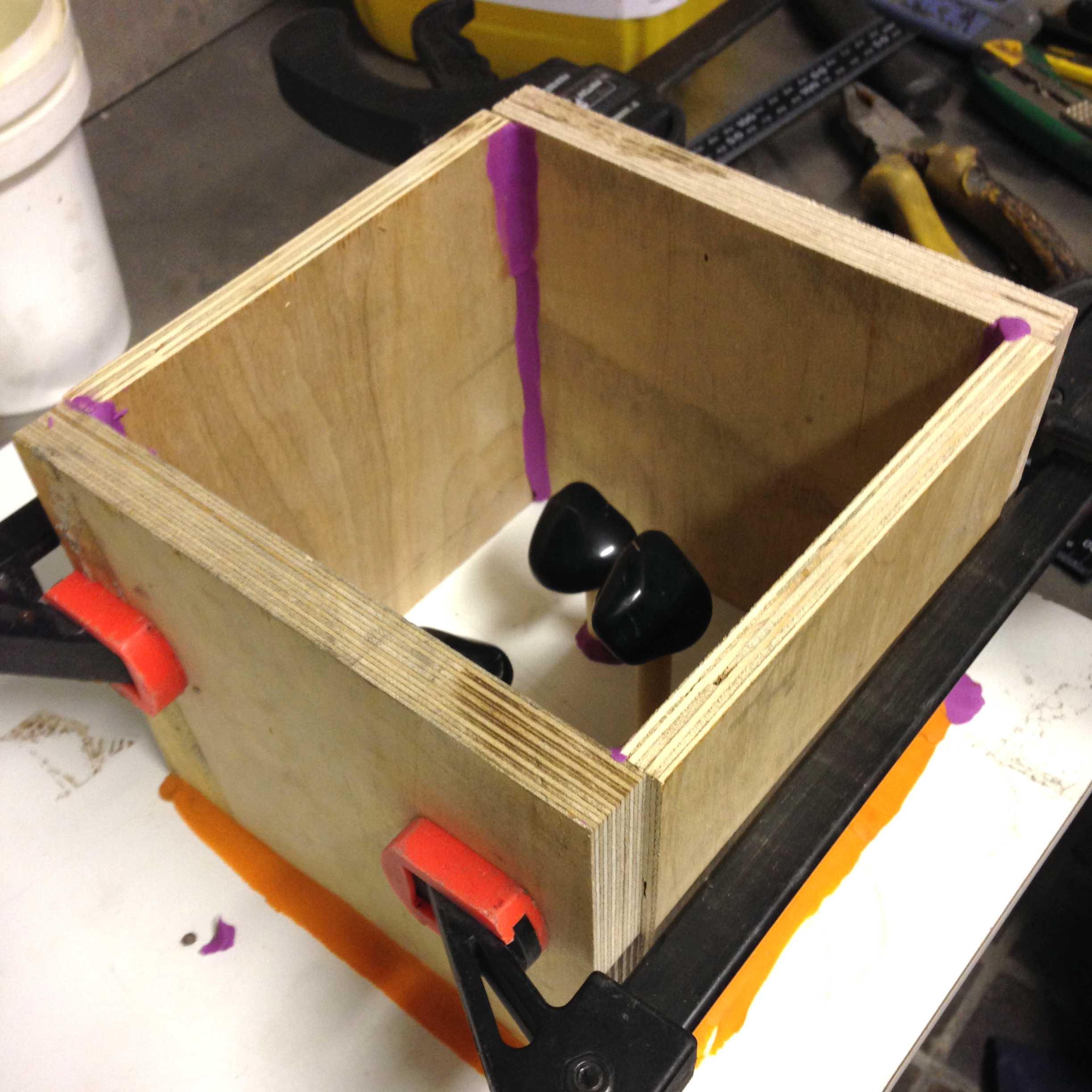} \\ Cube mould \end{center}
\end{minipage} 
\begin{minipage}{4.2cm}\begin{center}
\includegraphics[width=2.36cm,draft=false]{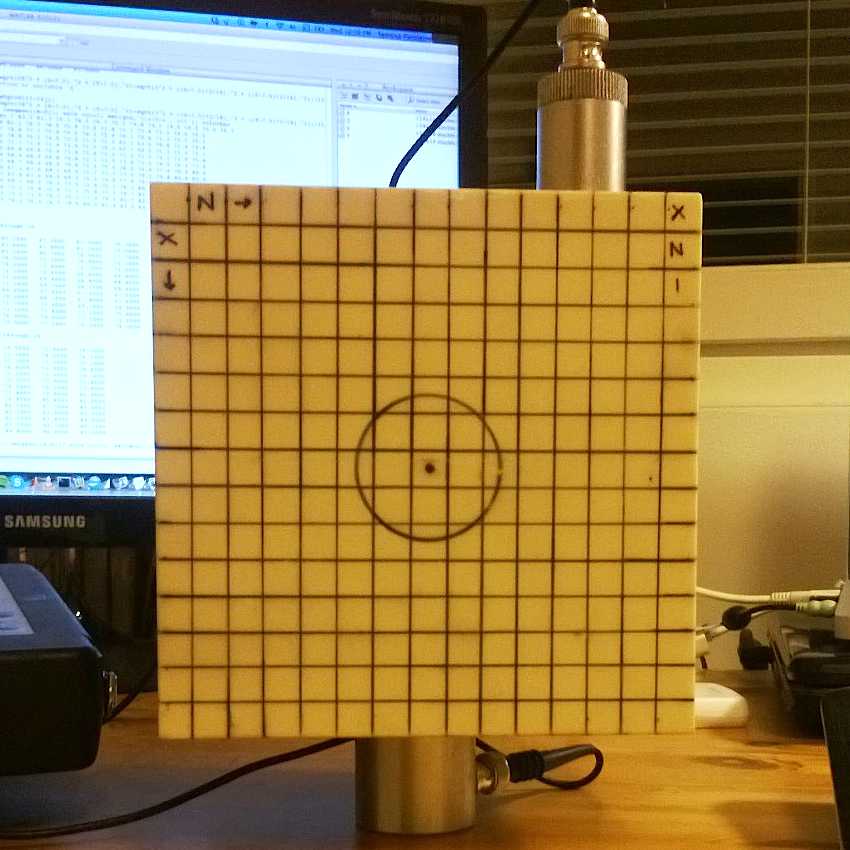}  \\ Transducers \end{center}
\end{minipage}
\begin{minipage}{4.2cm}\begin{center}
\includegraphics[width=2.36cm,draft=false]{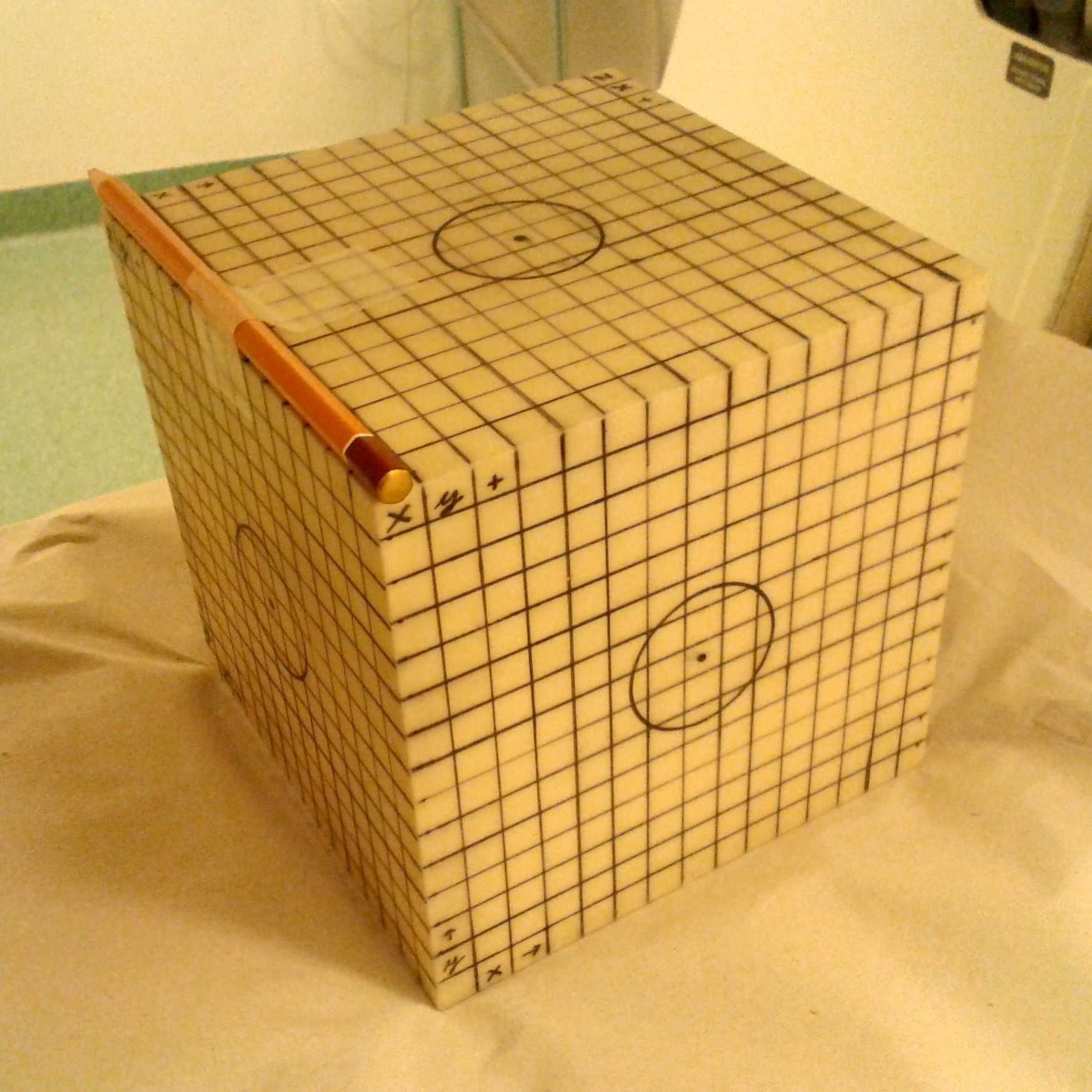}\\ CT scan  \end{center} \end{minipage}
\\ \vskip0.2cm
\begin{minipage}{4.2cm}\begin{center} 
\includegraphics[width=2.84cm,draft=false]{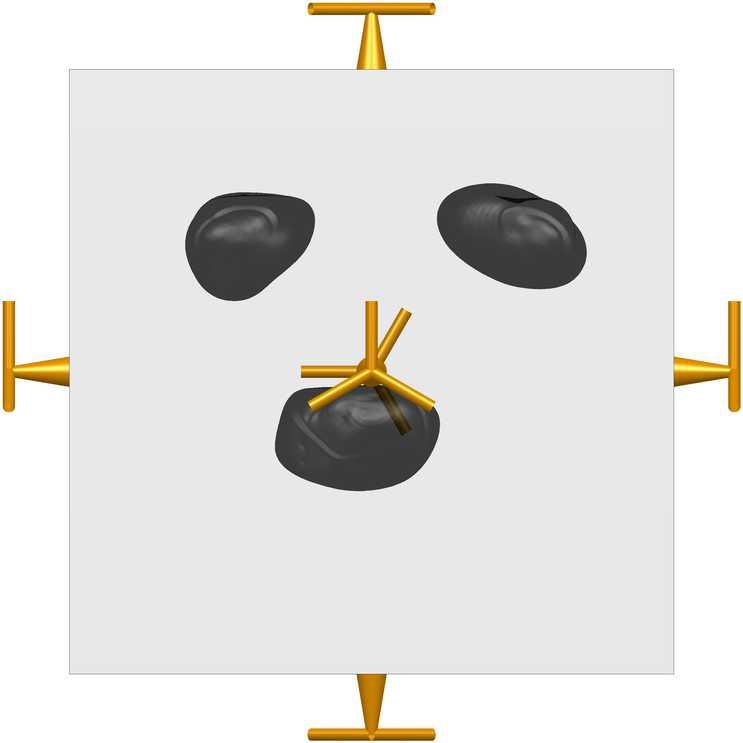} \\ $xy$-view\end{center}
\end{minipage}
\begin{minipage}{4.2cm}\begin{center} 
\includegraphics[width=2.84cm,draft=false]{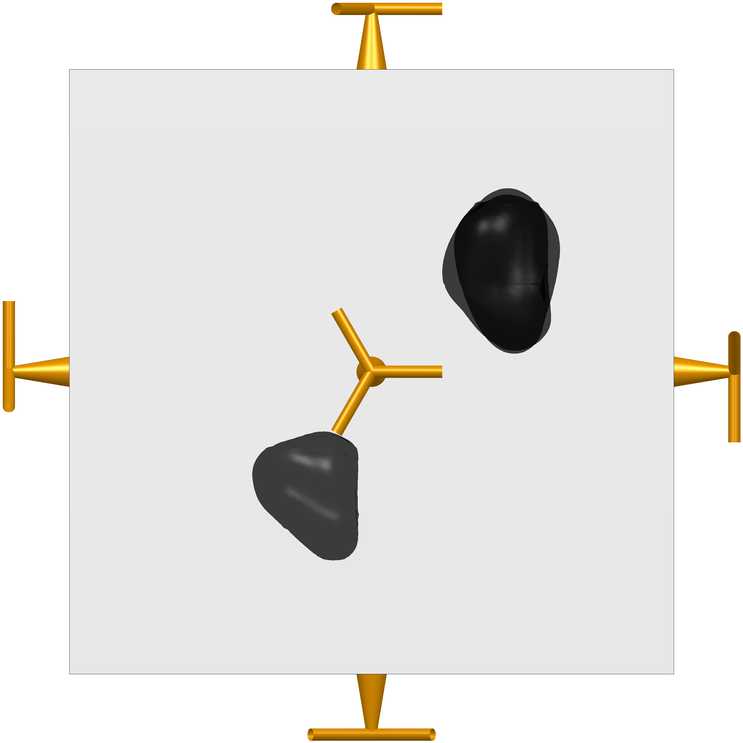} \\ $yz$-view \end{center}
\end{minipage} 
\begin{minipage}{4.2cm}\begin{center} 
\includegraphics[width=2.84cm,draft=false]{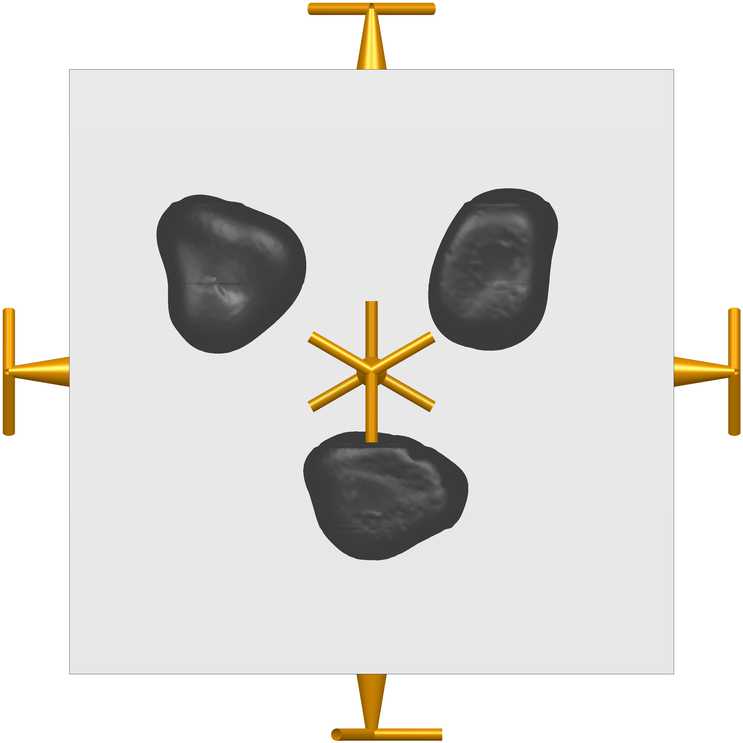} \\ zx-view \end{center}
\end{minipage} 
\end{footnotesize}
\end{center}
 \caption{Top row shows the cube mould with stones placed inside (left); ultrasonic measurement setup with a 25 mm and 37 mm transducer attached to the upward and downward face of the cube, respectively (center); and a picture of the CT scan (right). Bottom row illustrates a CT scan reconstruction of the stones from xy-, yz-, and zx-view (from left to right, respectively) with the six face-centered source positions indicated by the three branched antenna symbols.} \label{cube_fig}
\end{figure}

\begin{figure}[h!]\begin{center}\begin{footnotesize}
\begin{minipage}{5.5cm}\begin{center} \hskip0.5cm 
\includegraphics[width=4.2cm,draft=false]{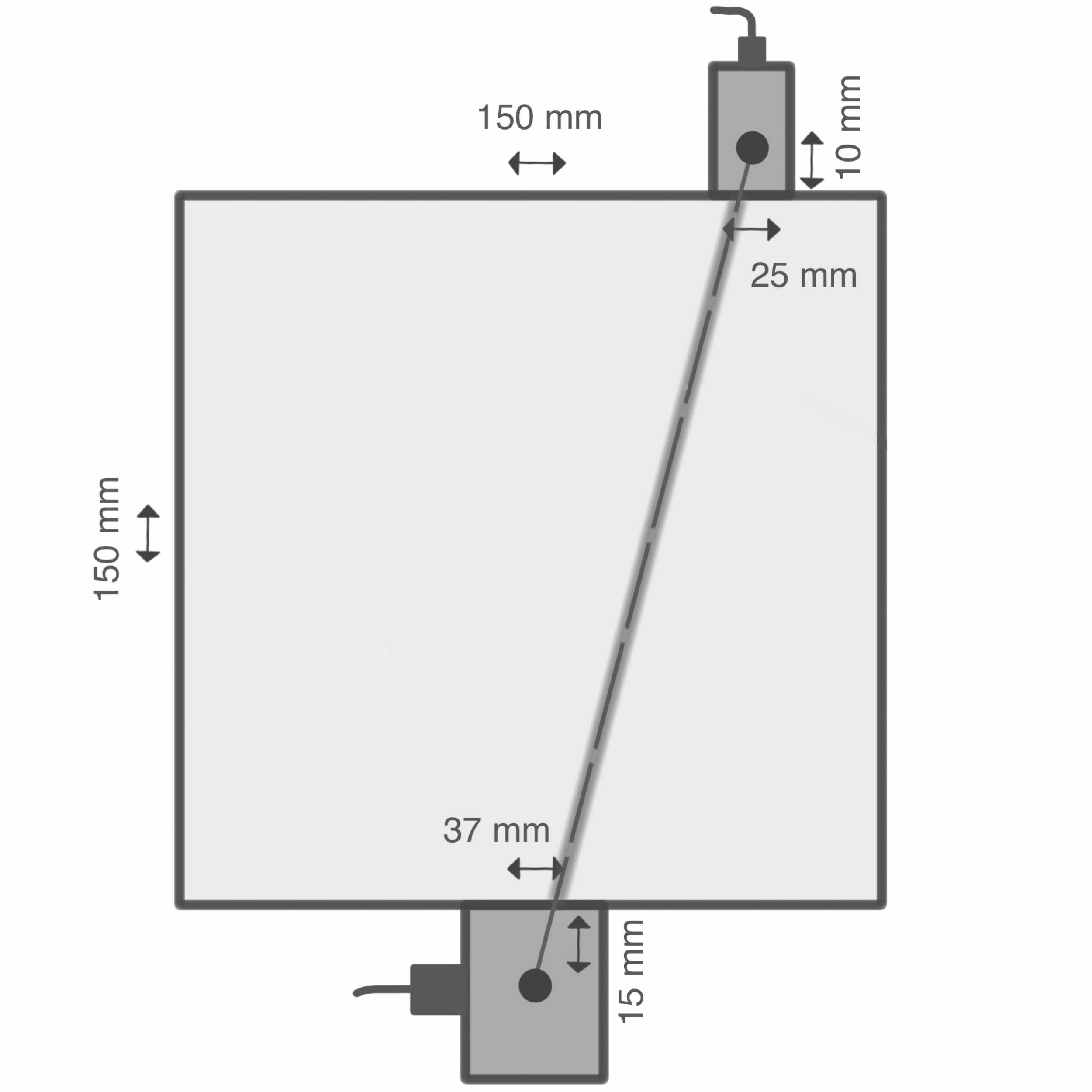} \end{center}
\end{minipage}
\end{footnotesize}
\end{center}
 \caption{A schematic picture of our linear forward model.  The signal path (bold)  between the transducer pair (dark grey) was predicted by a line segment (dashed) across the 150 mm  target cube (light grey)  intersecting the cylindrical axes of the 25 and 37 mm transducer at distances of 10 and 15 mm to the nearest face, respectively ($37/25 \approx 15/10$). } \label{forward_fig}
\end{figure}

\begin{figure}[h!]\begin{center}\begin{footnotesize}
\begin{minipage}{4.2cm}\begin{center} 
\includegraphics[width=2.7cm,draft=false]{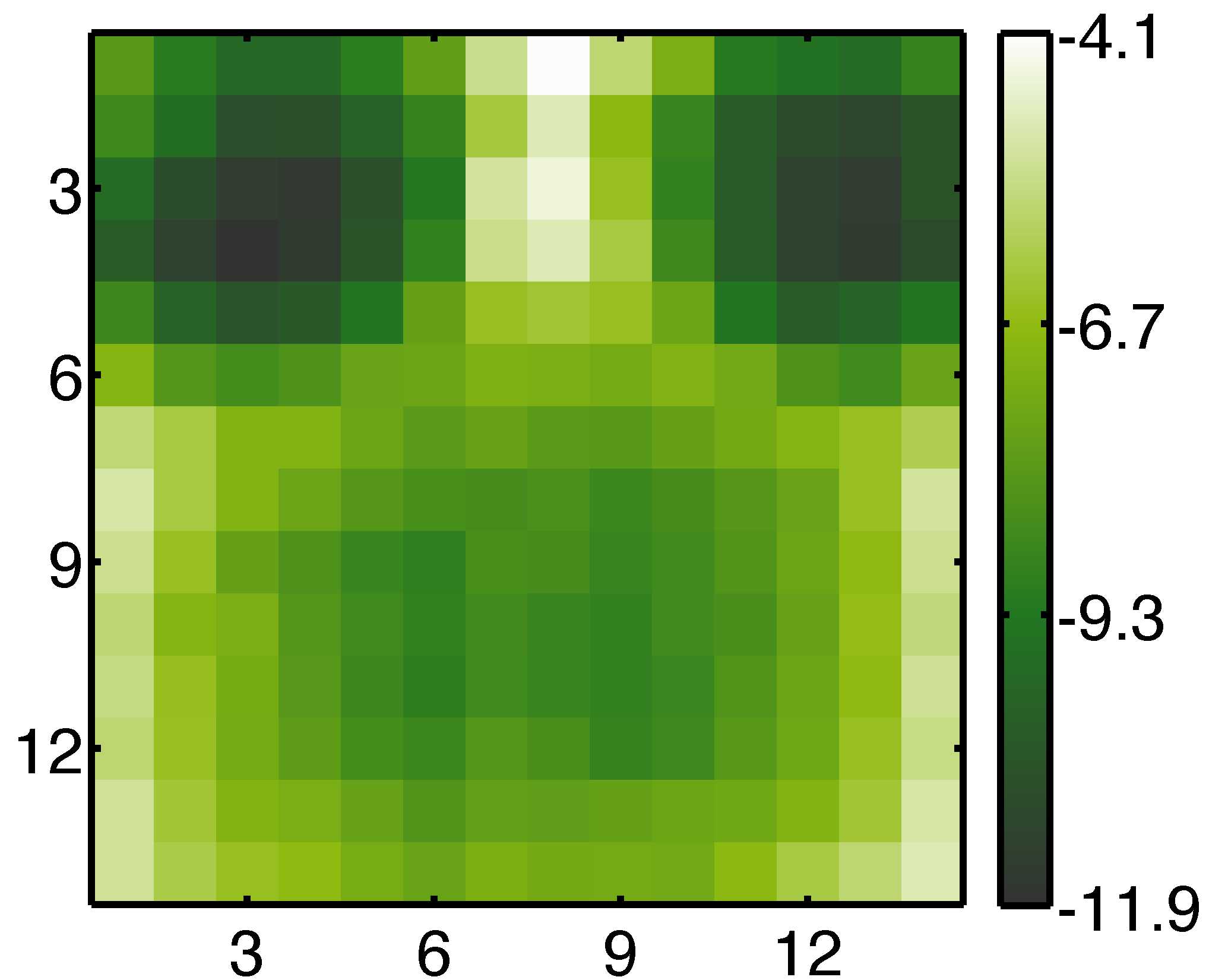} \\ $z\!=\!7.5$ cm, $xy$-view \end{center}
\end{minipage}
\begin{minipage}{4.2cm}\begin{center}
\includegraphics[width=2.7cm,draft=false]{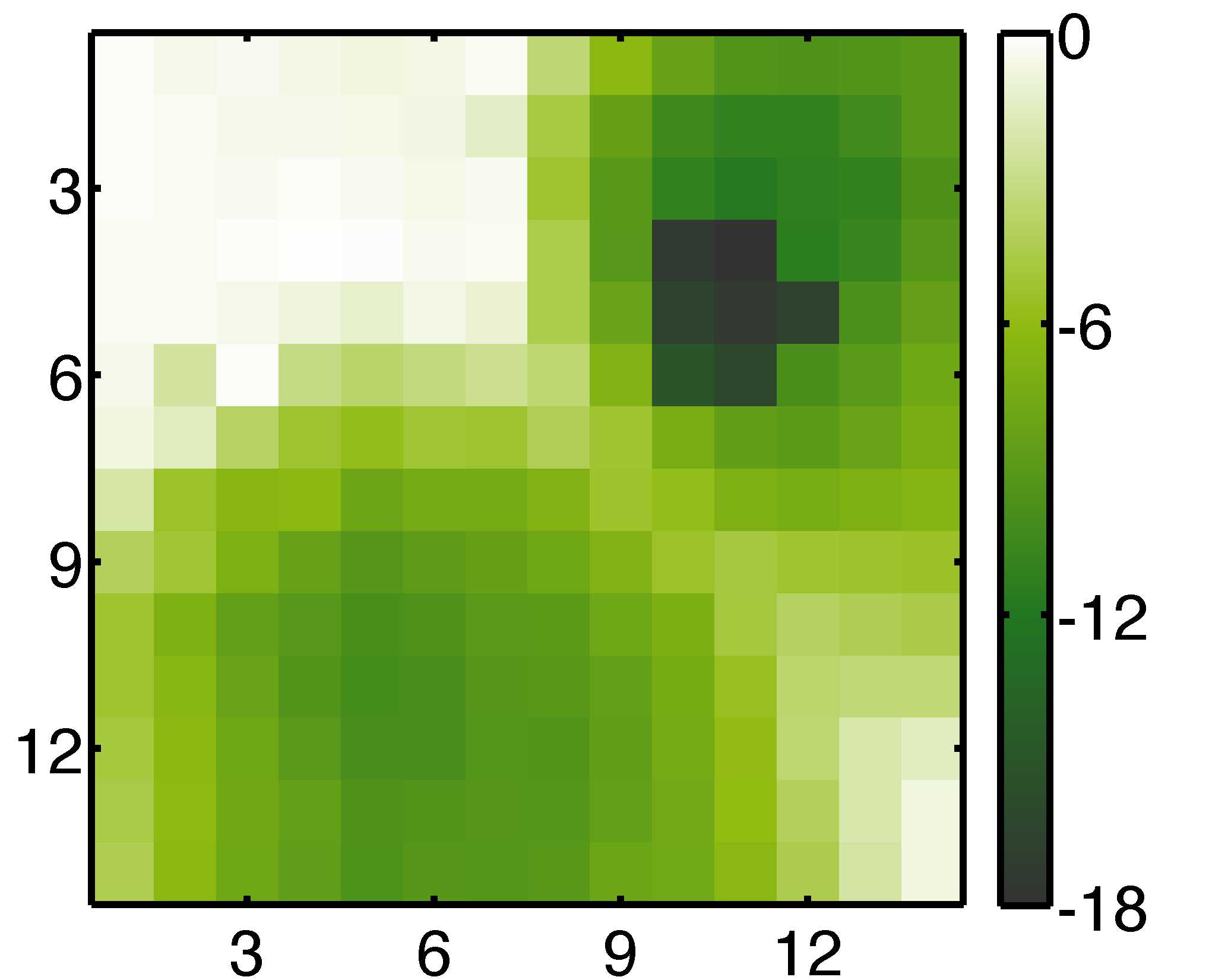} \\ $x\!=\!7.5$ cm,  $yz$-view \end{center}
\end{minipage} 
\begin{minipage}{4.2cm}\begin{center} 
\includegraphics[width=2.7cm,draft=false]{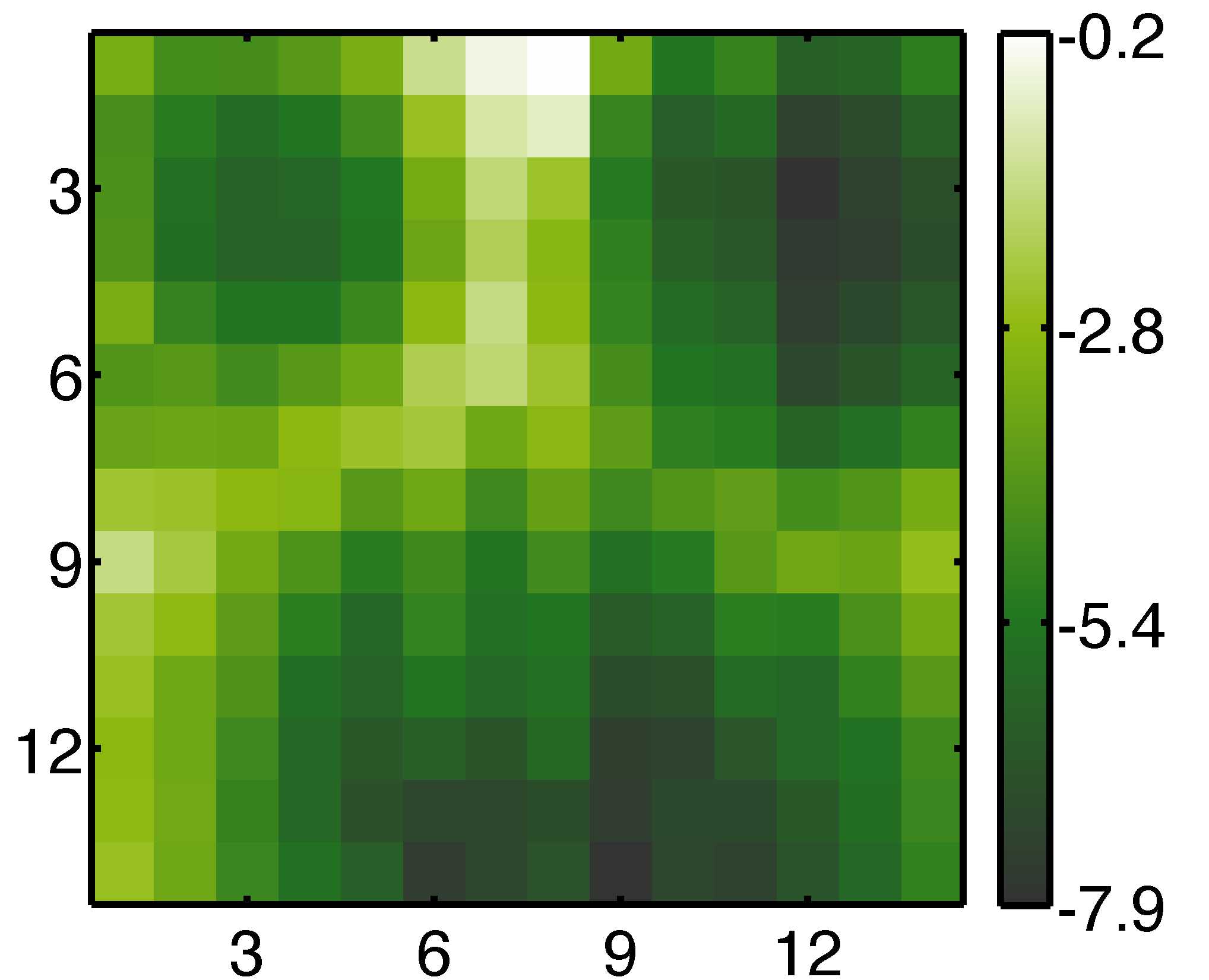}  \\ $y\! =\!7.5$ cm, $zx$-view\end{center}
\end{minipage} \\
\vskip0.2cm
\begin{minipage}{4.2cm}\begin{center} 
\includegraphics[width=2.7cm,draft=false]{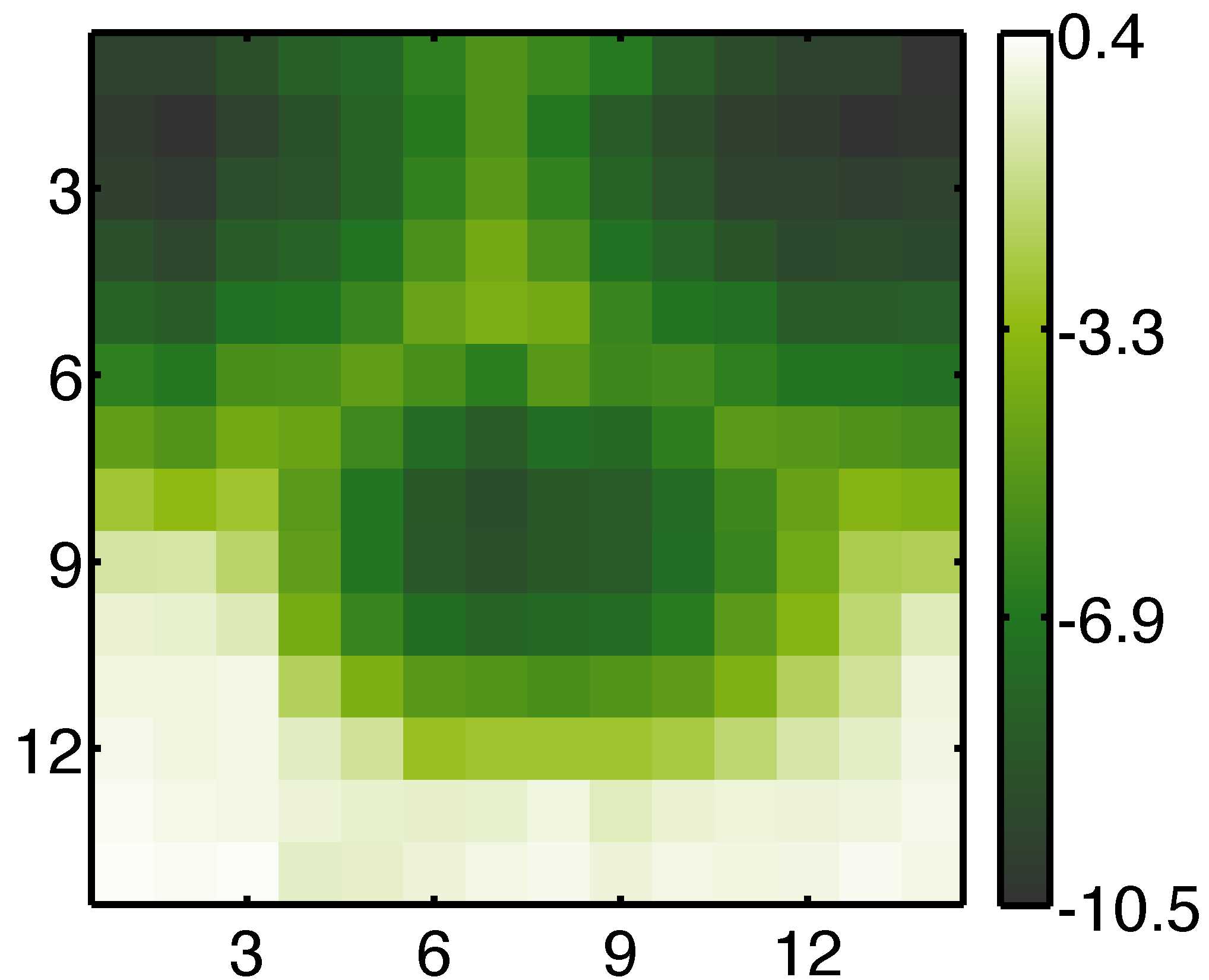}  \\ $z \! =\! \unaryminus7.5$ cm,  $\unaryminus xy$-view  \end{center}
\end{minipage}
\begin{minipage}{4.2cm}\begin{center} 
\includegraphics[width=2.7cm,draft=false]{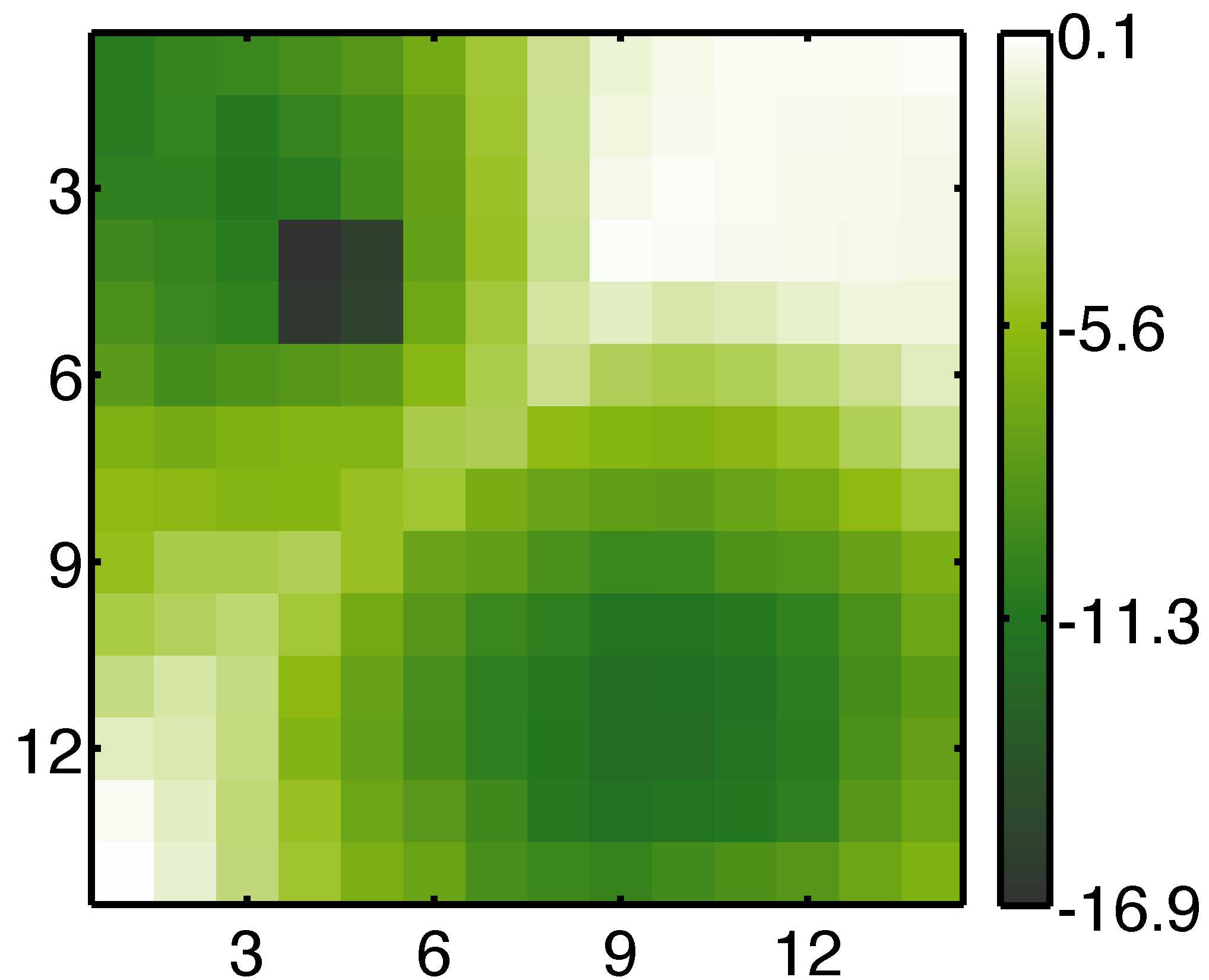} \\ $x\! =\! \unaryminus7.5$ cm, $\unaryminus yz$-view \end{center}
\end{minipage} 
\begin{minipage}{4.2cm}\begin{center} 
\includegraphics[width=2.7cm,draft=false]{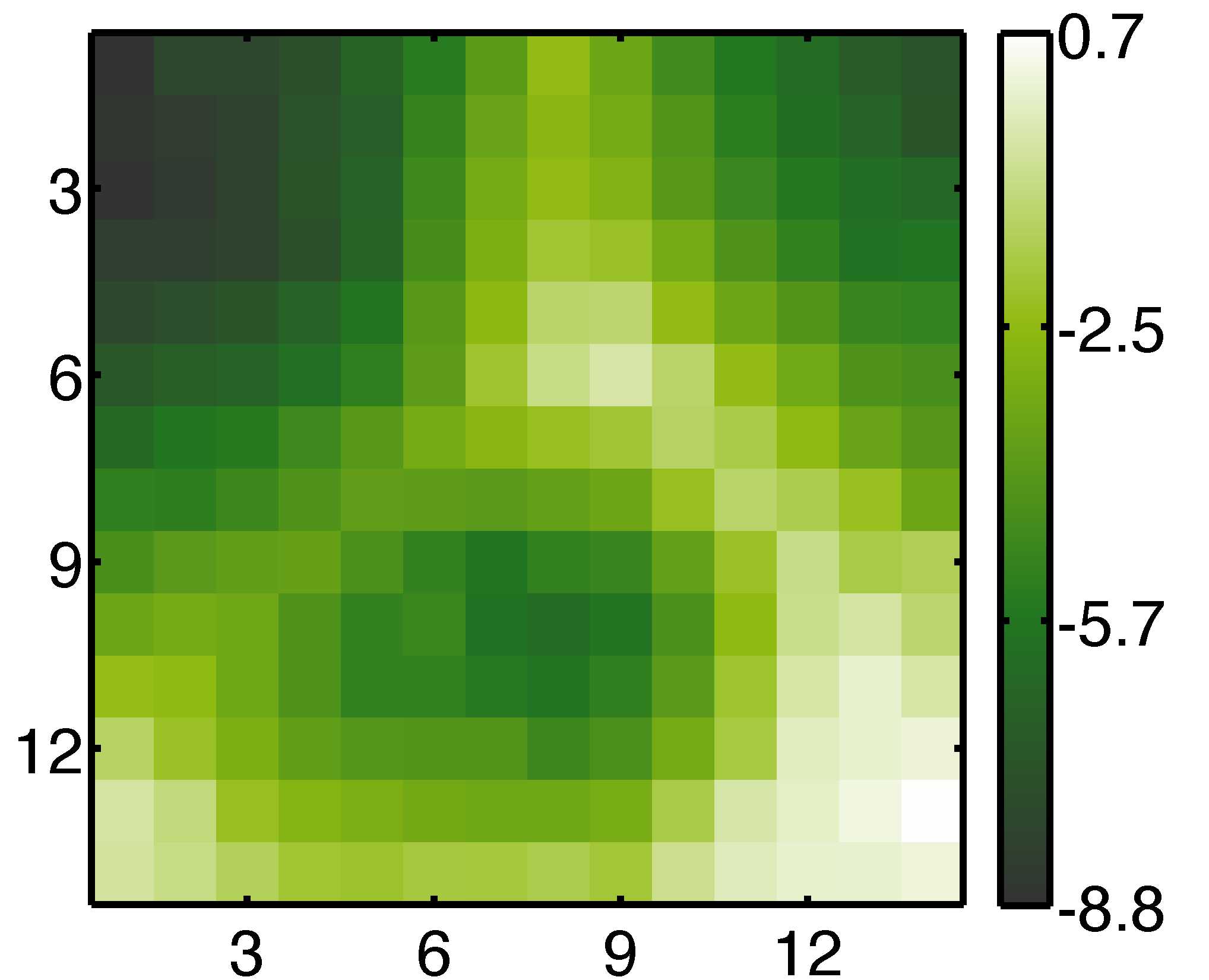} \\ $y \! = \!\unaryminus7.5$ cm, $\unaryminus zx$-view \end{center}
\end{minipage}
\end{footnotesize}
\end{center}
 \caption{Visualization of the ultrasonic travel time difference between the actual and the predicted signal. Each subfigure illustrates data over the face that is directed and oriented according to the given view. The coordinate values have been given with respect to an origin placed in the center of the cube. Note that the subfigures in the center column show part of the signal paths going through two stones which can be observed from  a large travel time difference.} \label{data_fig}
\end{figure}

\subsection{Target cube}

In this study, the target of measurements was a cast synthetic resin cube $\Omega$ with $\ell = 150$ mm side length. In the casting process, three polished natural onyx stones with diameters varying between 22 and 41 mm were first placed in the central part of a cube mould (Figure \ref{cube_fig}) with the support of pre-cast 10 mm thick rods, after which the rest of the mould was filled with molten synthetic resin. The faces of the cooled cube were then post-processed to achieve the targeted size. 

The longitudal sound velocity in the synthetic resin was observed to be approximately  1900 m/s and that within the stones 5200 m/s, i.e.\ the background velocity was 36.5 \% of the perturbed one. Considering the propagation of a radio signal in an asteroid \cite{elshafie2013,virkki2014,herique2002}, a similar percentual  change $p = 100 \varepsilon_r^{-1/2}$ in the velocity can occur, for instace, in basalt containing a vacuum void: e.g., desiccated Belleville basalt at 100 MHz frequency corresponds to $\varepsilon_r = 7.53$ \cite{elshafie2013} or $p = 36.4$ \%. Relying on carefully selected materials and thorough casting, the refractive index was estimated to be homogenous and isotropic inside the material boundaries for the applied signal wavelength. 

\subsection{Measurements}
\label{measurements}

Ultrasonic data were recorded with a conventional PUNDIT device measuring the travel time of a 55 kHz (wavelength $\lambda \approx 35$ mm) signal pulse between two cylindric transducers (Figure \ref{cube_fig}). To enable appropriate positioning, a regular 10 mm resolution grid was outlined on each cube face and a 40 mm diameter circle was drawn around the face centers. Grease lubricant was applied to all contact surfaces to maximize the signal energy transfer. To collect the data, the cube was first centered on top of a 37 mm diameter transducer.  Signal travel time was then recorded for the $14 \cdot 14 = 196$ internal grid nodes of the upward face with a smaller 25 mm transducer. This procedure was repeated two times per face averaging the measurements pointwise. The resulting data set included $6 \cdot 196 = 1176$ entries covering together six separate face-centric source positions, and oversampling the Nyquist criterion by the factor of 1.75 over each face \cite{nyquist2002,black1953}. The travel times within the set varied between 62.4 and 87.9 $\mu$s. Assuming that the measurement noise was normally distributed, the standard deviation was estimated to be around 0.3 $\mu$s. 

The internal structure of the cube was also analyzed via high resolution X-ray computed tomography (CT) scan \cite{mudry2013} to obtain a reference solution as well as to verify the quality of the casting. A structural illustration of the cube has been included in Figure \ref{cube_fig}. Original CT and ultrasonic data sets can be found included in the supportive material of this article at the IOP website.

\subsection{Forward model}

Our forward model predicted the travel time data vector ${\bf y} \in \mathbb{R}^N$ according to the formula \cite{born1999,hecht2002}
\begin{equation}
\label{forward_model_eq}
\quad {y}_i =  \int_{\mathcal{C}_i} {\mathtt n}_\delta \,  ds +  ({y}_{bg})_i + {g}_i, \quad \hbox{for} \quad i = 1,2,\ldots,N
\end{equation}
in which ${\mathtt n}_\delta$ denotes a perturbation of the refractive index; ${g}_i$ estimates the total noise due to different error sources; $({y}_{bg})_i = \int_{\mathcal{C}_i} {\mathtt n}_{bg} \,  ds $ is an entry of noiseless background data ${\bf {y}}_{bg} \in \mathbb{R}^N$ with ${\mathtt n}_{bg}=1/1935$ s/m; and $\mathcal{C}_i$ is a signal path. Given a pair of  transducer locations, $\mathcal{C}_i$ was predicted by
a line segment across the target cube (Figure \ref{forward_fig}) intersecting the cylindrical axes of the 25 and 37 mm transducers at distances of 10 and 15 mm to the nearest face, respectively. The resulting linear forward model approximated the effect of the finite transducer contact surface on the actual path length which, based on the measurements, was observed to be shorter than the distance between the contact surface center points. The distances 10 and 15 mm were chosen because they were observed to yield appropriate results: the difference between the actual and background travel time data ${\bf d} = {\bf {y}} - {\bf \overline{y}}_{bg}$ (Figure \ref{data_fig}) was close to zero for the paths not intersecting the stones. Also, the ratio between the values chosen was close to that of the cylindrical diameters, i.e.\ $37/25 \approx 15/10$. 

\subsection{Model discretization}

In the discretization of the forward model, ${\mathtt n}_\delta$ was assumed to be a piecewise constant function within a regular 3D lattice of $K$ voxels with center points 
${\bf r}_j$ (origin at the center of mass) and characteristic (indicator) functions $\chi_j$, $j = 1,2,\ldots,K$. The travel time data were approximated via 
$ 
{\bf d} = {\bf L} {\bf x}  + {\bf g}
$
in which ${\bf g}$ contained the total noise, ${\bf x} \in \mathbb{R}^K$ was the coordinate vector of ${\mathtt n}_\delta$ and ${\bf L} = {\bf A} {\bf W}$, with ${\bf A} \in \mathbb{ R}^{N \times K}$ and ${\bf W } \in \mathbb{R}^{K \times K}$. The entries of ${\bf A}$ were of the form $A_{ik}  = \int_{\mathcal{C}_i} \chi_k ds$ following from  the discretization of (\ref{forward_model_eq}), and ${\bf W}$ was a smoothing operator given by
\begin{equation} 
{ W}_{kj} \! = \! 
\frac{\exp{\Big(\unaryminus 
\frac{\| {\bf r}_j \unaryminus  {\bf r}_k \|_2^2 }{2 \varrho^2}\Big)}
}{\displaystyle\sum_{\|  {\bf r}_j \unaryminus  {\bf r}_i     \|_1 \leq 3 \varrho} \kern-3ex \textstyle \exp{ \Big(  \unaryminus \frac{ \|{\bf r}_j \unaryminus  {\bf r}_i   \|_2^2 }{2 \varrho^2}} \Big)},  \quad \! \!  \hbox{if} \! \! \quad \|  {\bf r}_j \unaryminus  {\bf r}_k    \|_1 \! \leq \! 3 \varrho, \, \, \|{\bf r}_k\|_1 \! \leq \! {\textstyle \frac{\ell}{2} \unaryminus  3 \varrho}, 
\end{equation} and otherwise $W_{kj}=0$ \cite{pursiainen2013}. The mask size $M = 3 \varrho$ defining the degree of smoothing was chosen to obtain appropriately regular reconstructions for statistical analysis of the results, e.g., to avoid ghost inclusions or fragmenting due to the strongly localizing prior model \cite{pursiainen2013}. Note that our smoothing technique reorganized ${\bf A}$ without limiting the number of degrees of freedom (or resolution) in the essential (center) part of the target. Consequently, smoothing can be interpreted as adapting the forward model to the applied wavelength, which was necessary due to oversampling (Section \ref{measurements}).

\subsection{Inverse model}
\label{section:inverse_model}

Our inverse approach was based on maximizing a posterior probability density \cite{ohagan2004} given by the product $p({\bf x}, {\bf z} \! \mid \! {\bf d}) \propto p({\bf d} \! \mid \! {\bf x}) p({\bf x} \! \mid \! {\bf z}) p({\bf z})$ of a conditionally zero-mean Gaussian prior $p({\bf x} \! \mid \! {\bf z})$, hyperprior $p({\bf z})$ of a latent variance hyperparameter  ${\bf z}$ and likelihood $p({\bf d} \! \mid \! {\bf x})$ following from a zero-mean Gaussian density $p({\bf g})$ assumed for the total noise $ {{\bf g}} = {\bf d} - {\bf L} {\bf x}$. In our model, the covariance matrices  of $p({\bf x} \! \mid \! {\bf z})$ and $p({\bf g})$ were of the form ${\bf D}_{{\bf z}} = \hbox{diag}(z_1,z_2,\ldots,z_K)$ and $\frac{1}{2} \sigma^{-2} {\bf I}$, respectively. A gamma (g; $\zeta=0$) and an inverse gamma (ig; $\zeta=1$) hyperprior yield a posterior of the form {\setlength\arraycolsep{2 pt}
\begin{equation}
\label{prior_2}
\fl p({\bf x}, {\bf z} \! \mid \! 
{\bf d}) \propto {\rm exp} \Bigg(\!\! - \! \frac{\| {\bf d} \unaryminus  {\bf
L} {\bf x}\|^2}{2\sigma^2} \! - \! \frac{\| {\bf D}^{\unaryminus \frac{1}{2}}_{\bf z} {\bf x} \|^2}{2}\! - \! \sum_{k=1}^K  \Big( \frac{z_k}{\theta_0}\Big)^{\! 1 \unaryminus 2\zeta} \!
+ \! \Big( \! ( \! \unaryminus 1 \!)^\zeta \beta \unaryminus \frac{3}{2} \Big) \sum_{k=1}^K \log z_k \! \! \Bigg)
\end{equation}
Here, the shape and scale parameter $\beta$ and $\theta_0$ control the sensitivity of the joint prior $p({\bf x},{\bf z}) = p({\bf x} \! \mid \! {\bf z}) p({\bf z})$ to detect a perturbation.  A fixed prior variance (f), i.e.\ a Dirac delta hyperprior $p({\bf z}) = \delta_{\theta_0}({\bf z})$, again yields a Gaussian marginal posterior given by 
\begin{equation} \label{prior_2} \fl p({\bf x} \! \mid \! 
{\bf d}) \propto {\rm exp} \Bigg(\!\! - \! \frac{\| {\bf d} \unaryminus  {\bf
L} {\bf x}\|^2}{2\sigma^2} \! - \! \frac{\| {\bf D}^{\unaryminus \frac{1}{2}}_{\bf \theta_0} {\bf x} \|^2}{2}\! \Bigg)   \quad \hbox{with} \quad {\bf D}_{\bf \theta_0} = \theta_0 {\bf I}. \end{equation}
Since gamma and inverse gamma hyperpriors give high propabilities for well-localized anomalous distributions of the prior variance ${\bf z}$, they make the posterior favor sharper and more localized distributions of ${\bf x}$ compared to the delta hyperprior which forces the variance $\theta_0$ to be constant.

\label{section_ias_algorithm}

The maximizer of the posterior was estimated via the following iterative alternating sequential (IAS)  {\em maximum a posteriori} (MAP) algorithm \cite{pursiainen2013,  calvetti2009,calvetti2008,calvetti2007}:  
\begin{description} 
\item[(1)] Choose $m \in \mathbb{N}$. Set ${\bf z}^{(0)} = (\theta_0, \theta_0, \ldots, \theta_0)$ and $i=1$; \item[(2)] Find the maximizer ${\bf x}^{(i)}$ of $p({\bf x} \! \mid \! {\bf d},{\bf z}^{(i-1)})$; \item[(3)] Find the maximizer ${\bf z}^{(i)}$ of $p({\bf z} \! \mid \! {\bf d},{\bf x}^{(i)})$; \item[(4)] If $i <m$, set $i = i+1$ and go back to 2. \end{description}
The IAS method provides a fast way to produce a MAP estimate, since  $m$ can typically be a relatively low number, e.g.\ $m=20$, and since  ${\bf x}^{(i)}$ and ${\bf z}^{(i)}$ can be obtained by optimizing a quadratic function \cite{calvetti2009} in a straightforward manner. When $\beta=1.5$, the  gamma and inverse gamma hyperprior can be shown to result in $L^1$- and minimum support type estimates (g) and (ig) of $\label{xg} {\bf x}^{(g)} = {\rm argmin} \{ \| {\bf d} - {\bf L} {\bf x} \|_2^2 + \gamma \| {\bf x} \|_1\}$, $\gamma = \sigma^2 \sqrt{2/\theta_0}$ and  $\label{xig} {\bf x}^{(ig)} = {\rm argmin} \{ \| {\bf d} - {\bf L} {\bf x}   \|_2^2 + \gamma \sum_{k = 1}^{K} |x_k|^2/(|x_k|^2 + 2 \theta_0)\}$, $\gamma = 12 \sigma^2$ \cite{calvetti2009}. A fixed prior variance leads to an $L^2$-type estimate (f) of $\label{xf} {\bf x}^{(f)} = {\rm argmin} \{ \| {\bf d} - {\bf L} {\bf x}   \|_2^2 + \gamma \| {\bf x }\|^2 \}$, $\gamma = \sigma^2 / \theta_0$.

\subsection{Choice of $\beta$ and $\theta_0$}

Given a space of candidate solutions $\mathcal{V}$, the subspaces of detectable and indetectable perturbations can be approximated, respectively, by $\mathcal{S}^+_\epsilon = \{ {\bf x} \in \mathcal{V} \, | \, \| {\bf L}{\bf x} \| \geq \epsilon \}$ and $\mathcal{S}^-_\epsilon = \{ {\bf x} \in \mathcal{V}  \, | \, \|  {\bf L}{\bf x} \| < \epsilon \}$ with $\epsilon$ chosen based on the total noise level of the inversion procedure \cite{pursiainen2008,liu1995}. 
Due to a non-empty $\mathcal{S}^-_\epsilon $, the solution of the inverse problem is subjective: it is heavily dependent on the prior model. The better the extent of the data, the smaller the $\mathcal{S}^-_\epsilon$, and the more robust the inversion. 
In this study, these aspects were explored for all possible configurations of face-centered sources, estimates of the types (f), (g) and  (ig), and initial prior variances (scaling parameter values) $\theta_0 = 10^k, k=-1,0,2,3$ with the shape parameter $\beta$ set to the lowest possible value $\beta = 1.5$. The interval of $\theta_0$ was chosen based on the authors' experience of the limits for under and over sensitivity of the prior to localize anomalies. 

\subsection{Source configurations}

The investigated configurations {\bf 1}, {\bf 2a}, {\bf 2b}, {\bf 3a}, {\bf 3b}, {\bf 4a}, {\bf 4b}, {\bf 5} and {\bf 6} of face-centered sources have been visualized in Figure \ref{source_configurations}. These were divided into two categories $\hbox{I}=\{\hbox{{\bf 1}, {\bf 2a}, {\bf 3a}, {\bf 4a}, {\bf 6}}\}$ and $\hbox{II}=\{\hbox{{\bf 2b}, {\bf 3b}, {\bf 4b}, {\bf 5}}\}$ of  primary  and secondary interest, respectively. This division was made to test our prior expectance of the performance of the configurations (which was corroborated by the results as shown below). The former category includes the single source ({\bf 1}) as well as the multiple directional alternative of two and three sources ({\bf 2a} and {\bf 3a}), since enchancing the directionality of the configuration can be seen as the principal way to increase the coverage of the data. We also deemed multiple symmetries interesting  ({\bf 4a} and {\bf 6}). Other settings were categorized as of secondary interest. The details of the configurations have been listed in Table \ref{table_sources}.

\begin{table}
\caption{Details of the investigated source configurations. }\label{table_sources}
\begin{indented}
\item[]
\begin{tabular}{@{}llllllllll}
\\ \br 
Number of sources & 1 & \centre{2}{2} &\centre{2}{3} &\centre{2}{4} & 5& 6 \\
Number of configurations & 6 & \centre{2}{15} & \centre{2}{20} & \centre{2}{15} & 6 & 1 \\
\mr
Configuration type & {\bf 1} & {\bf 2a} & {\bf 2b} & {\bf 3a} & {\bf 3b} & {\bf 4a} & {\bf 4b} & {\bf 5} & {\bf 6} \\
Number of members & 6 & 12 & 3 & 8& 12 & 3& 12& 6 & 1 \\
Category & I & I & II & I & II & I & II & II & I \\
\br
\end{tabular}
\end{indented}
\end{table}

\begin{figure}[h!]\begin{center}\begin{footnotesize}
\begin{minipage}{3.8cm}\begin{center}
\includegraphics[width=1.5cm,draft=false]{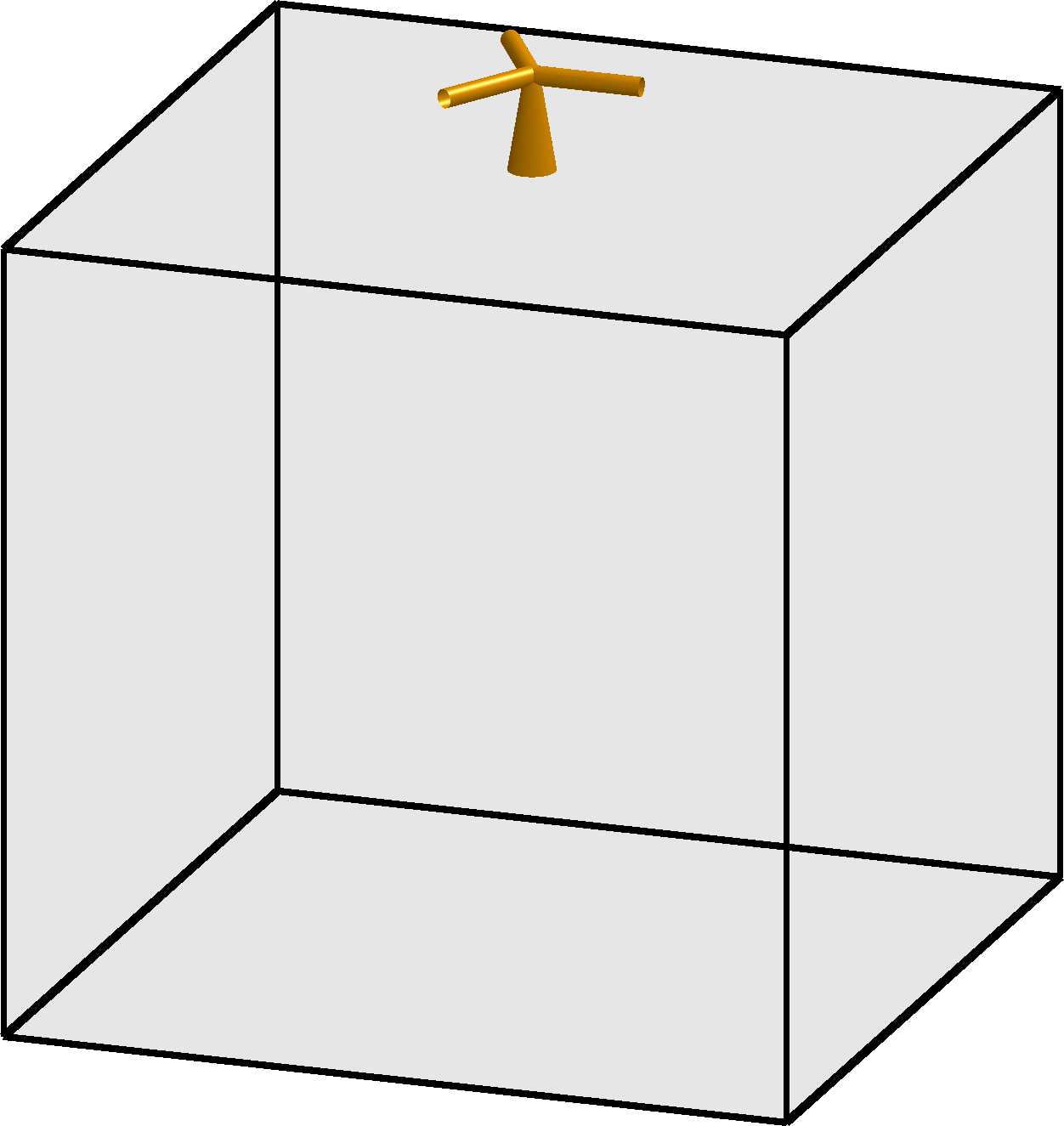}  \\ {\bf 1}
\end{center}
\end{minipage}
\begin{minipage}{3.8cm}\begin{center}
\includegraphics[width=1.5cm,draft=false]{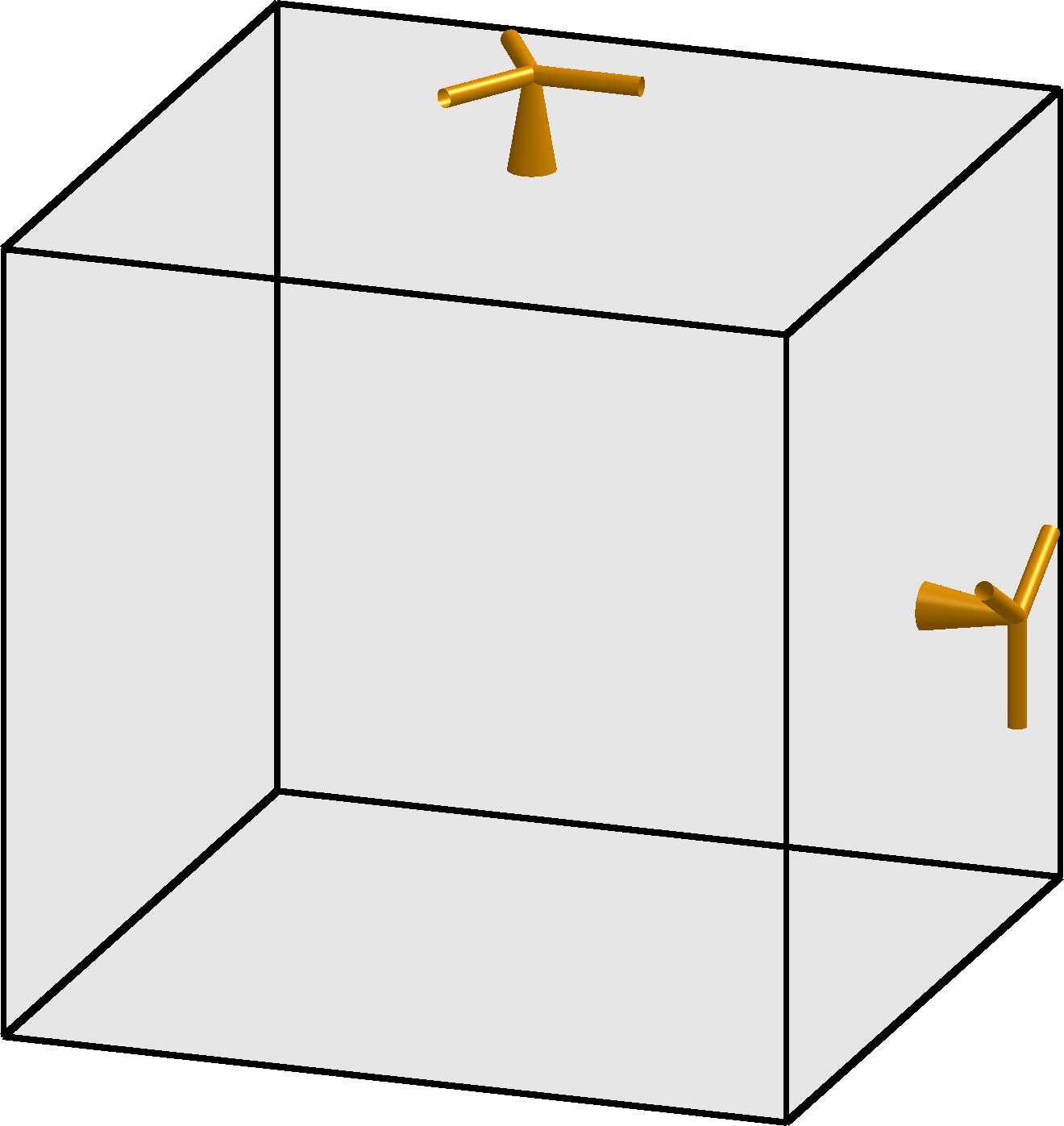} \\ {\bf 2a} \end{center}
\end{minipage} 
\begin{minipage}{3.8cm}\begin{center}
\includegraphics[width=1.5cm,draft=false]{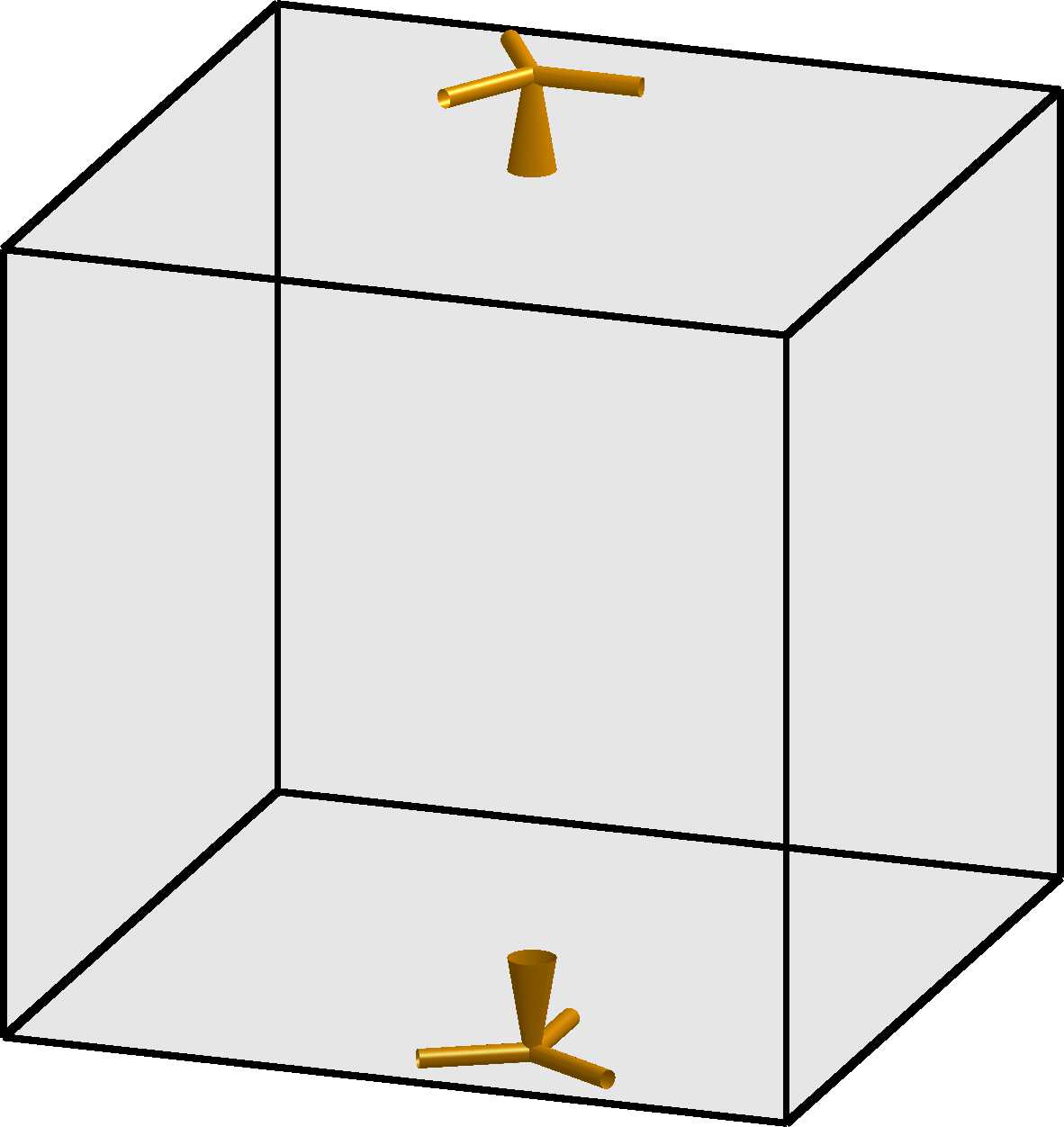}  \\ {\bf 2b} \end{center}
\end{minipage}\\ \vskip0.1cm
\begin{minipage}{3.8cm}\begin{center}
\includegraphics[width=1.5cm,draft=false]{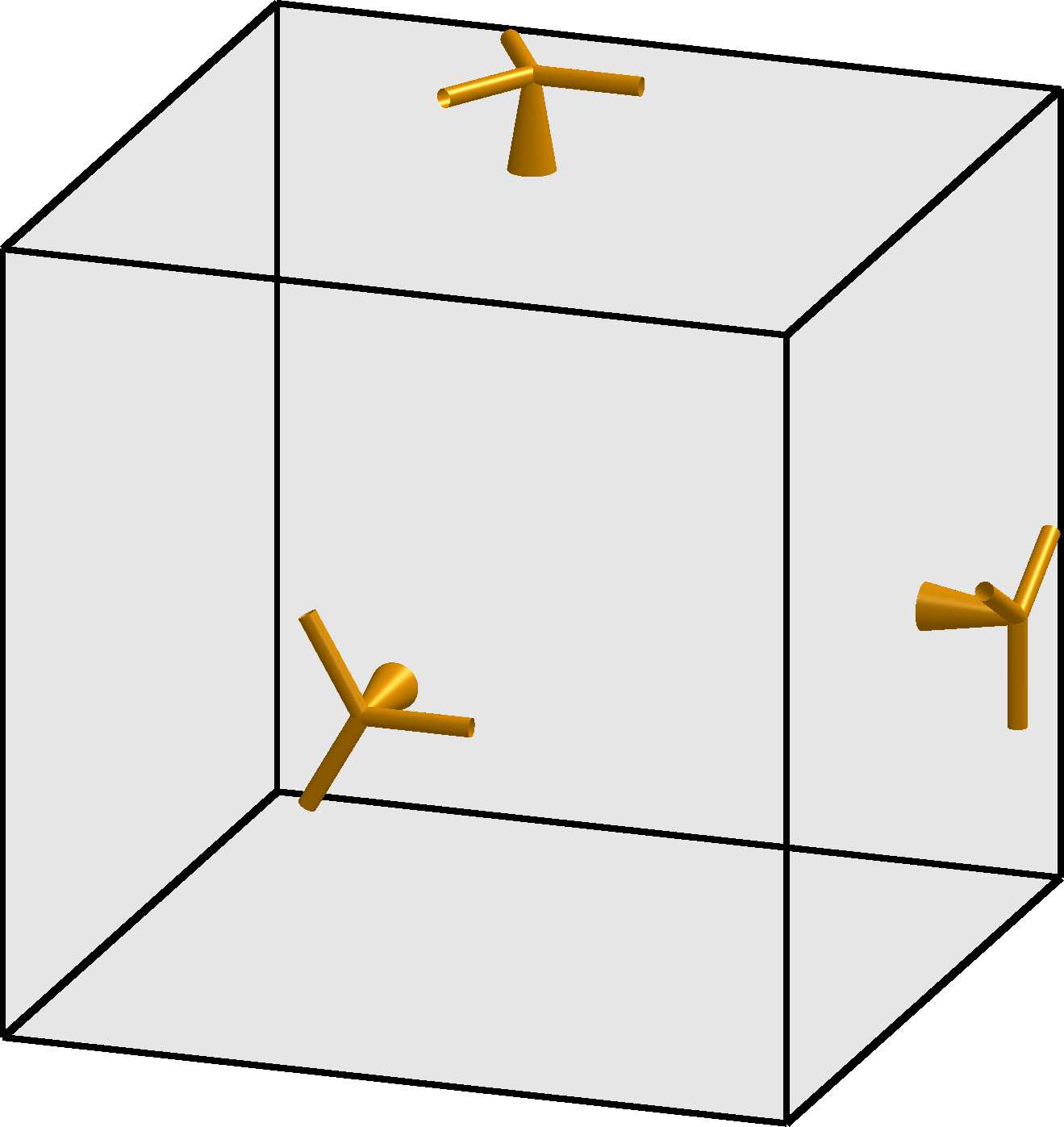}  \\ {\bf 3a} \end{center}
\end{minipage}
\begin{minipage}{3.8cm}\begin{center}
\includegraphics[width=1.5cm,draft=false]{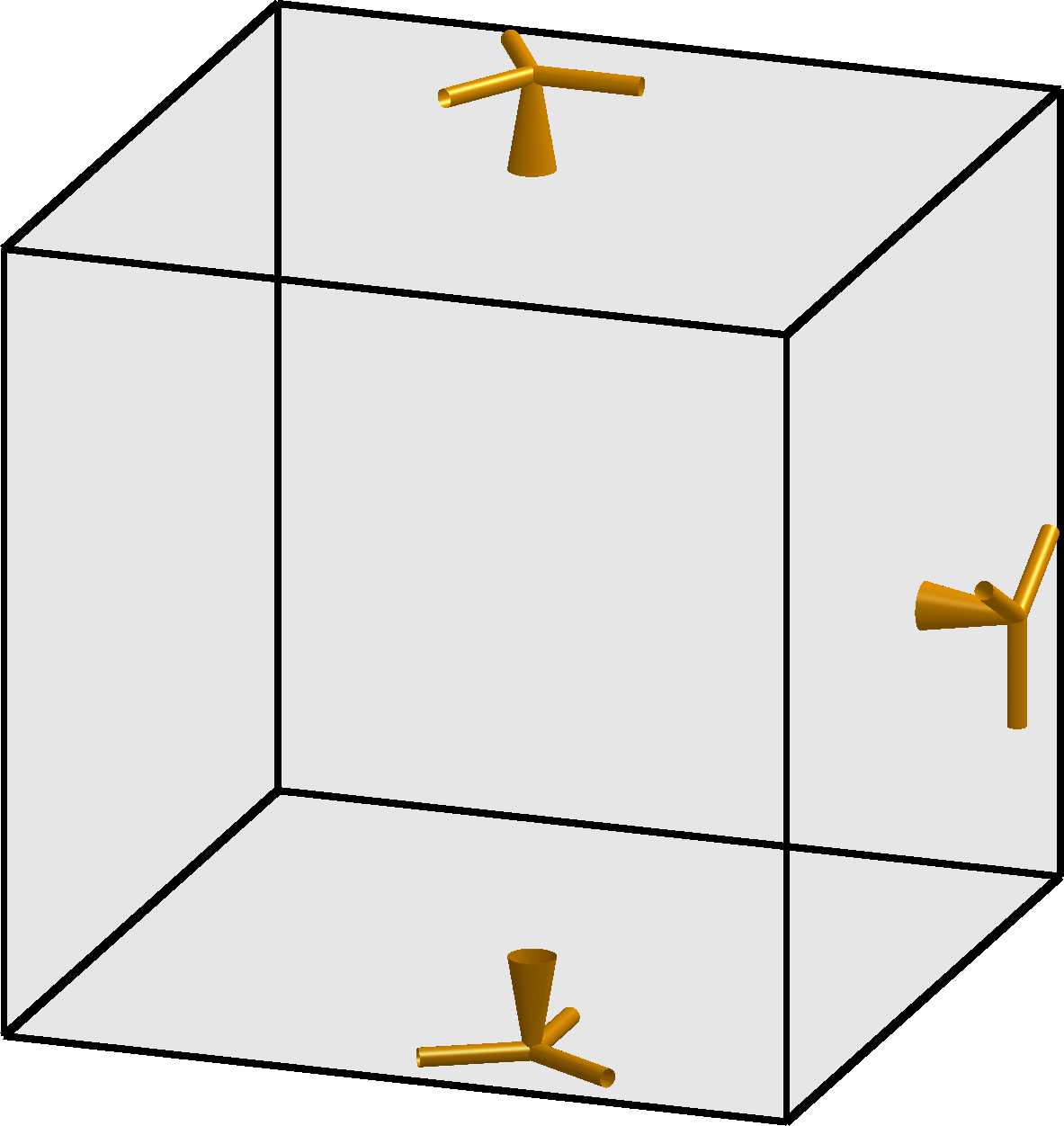} \\ {\bf 3b} \end{center}
\end{minipage} 
\begin{minipage}{3.8cm}\begin{center}
\includegraphics[width=1.5cm,draft=false]{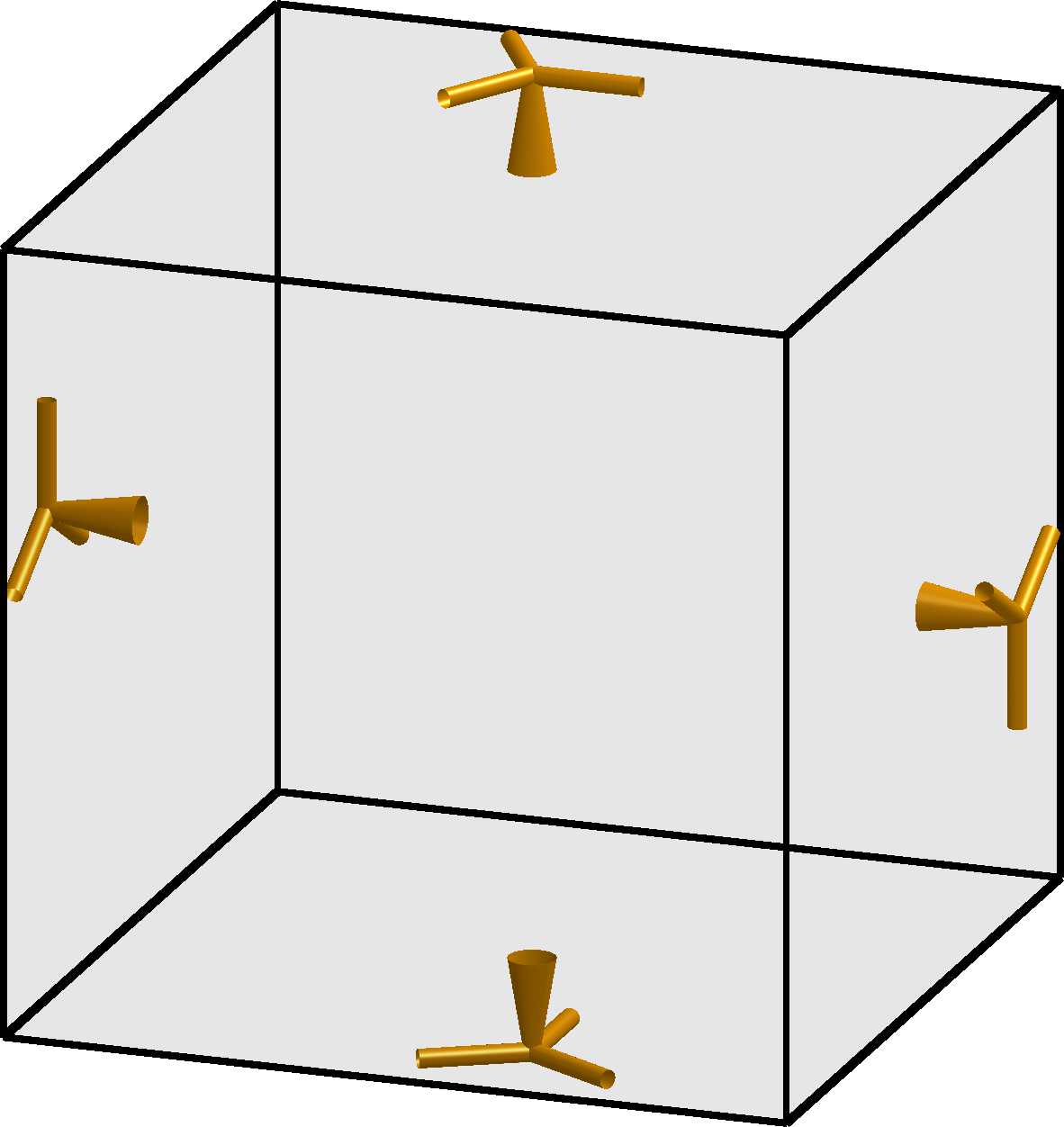}  \\ {\bf 4a} \end{center}
\end{minipage}\\ \vskip0.1cm
\begin{minipage}{3.8cm}\begin{center}
\includegraphics[width=1.5cm,draft=false]{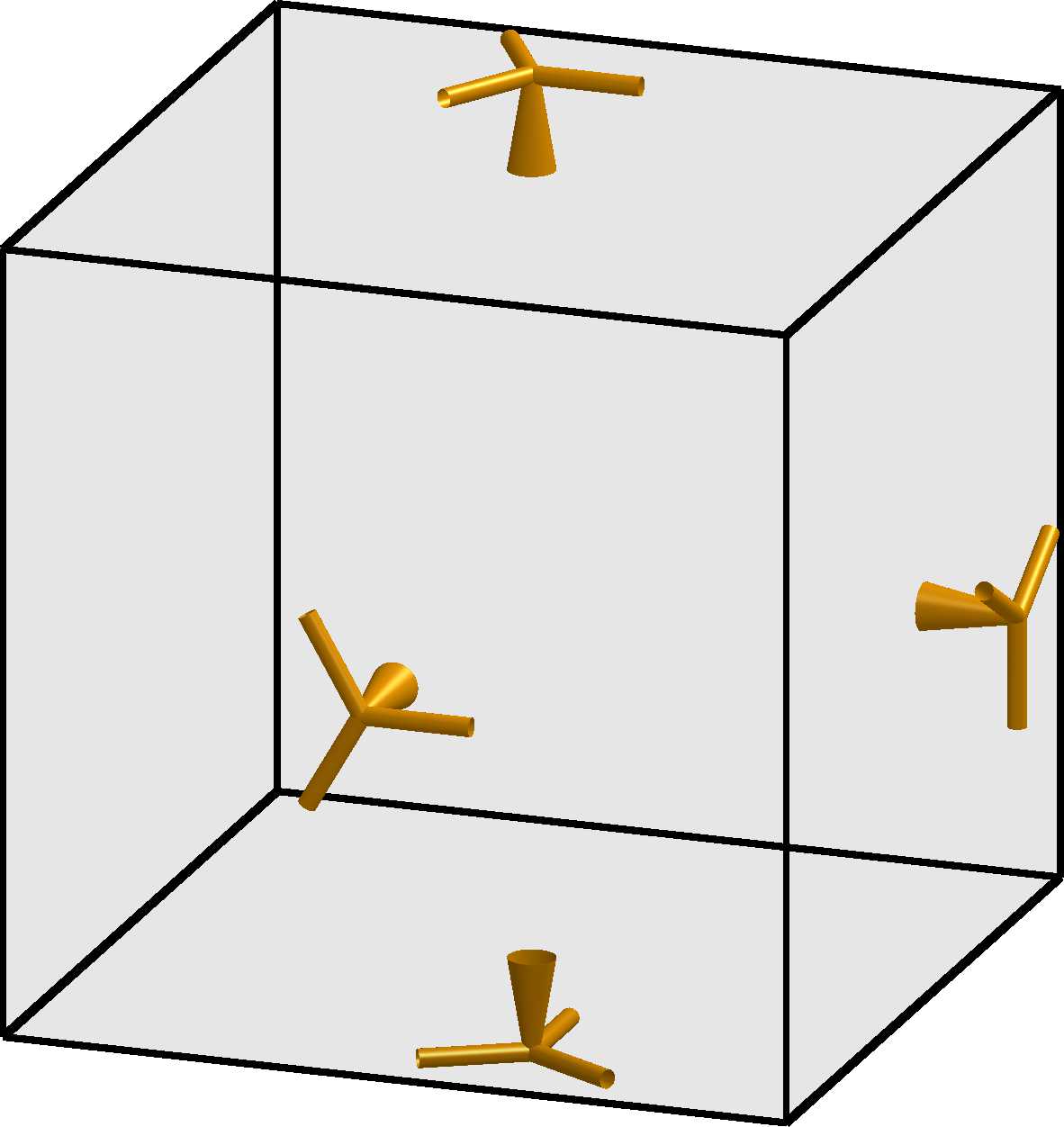} \\ {\bf 4b} \end{center}
\end{minipage}
\begin{minipage}{3.8cm}\begin{center}
\includegraphics[width=1.5cm,draft=false]{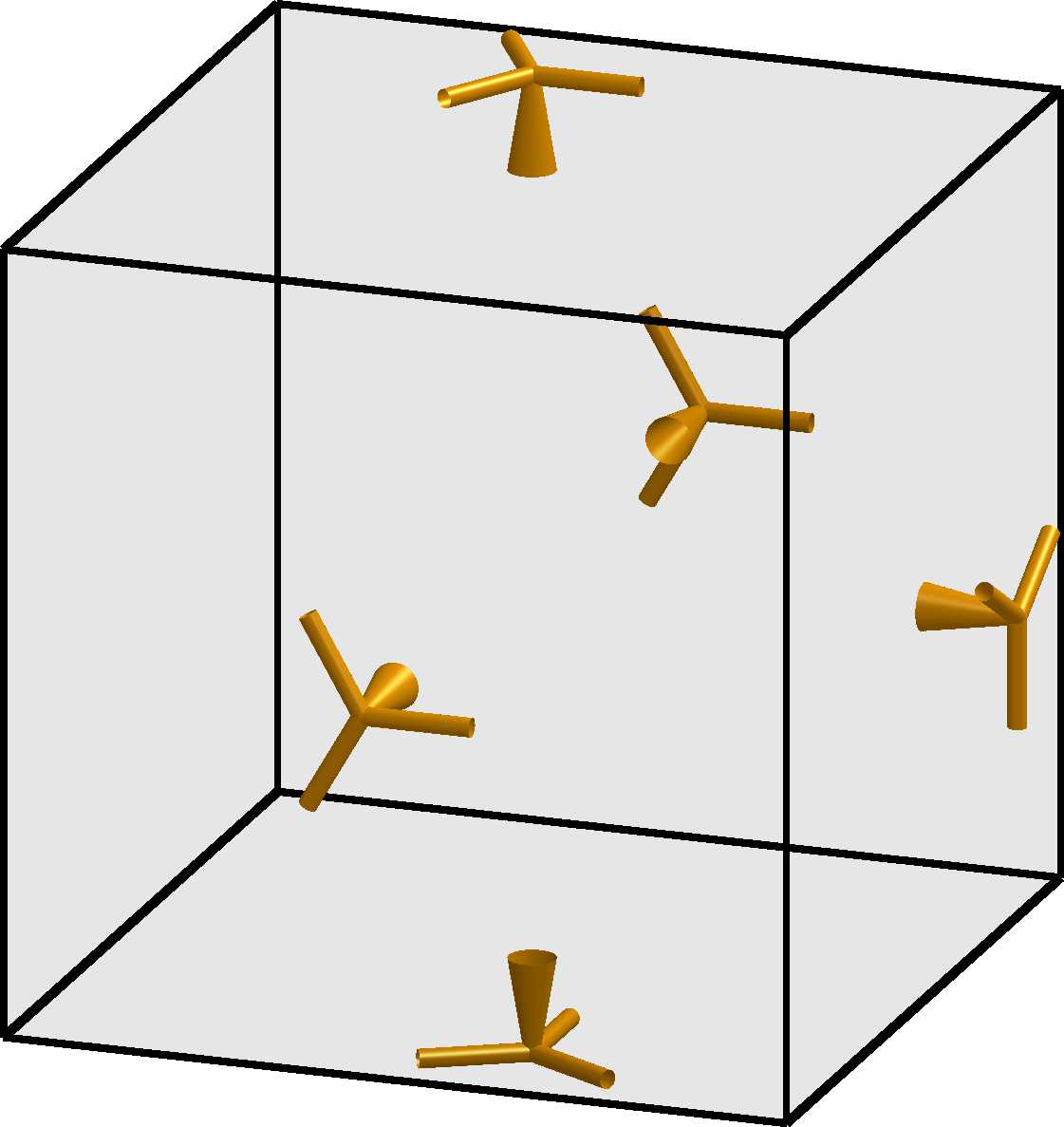}  \\ {\bf  5}  \end{center}
\end{minipage}
\begin{minipage}{3.8cm}\begin{center}
\includegraphics[width=1.5cm,draft=false]{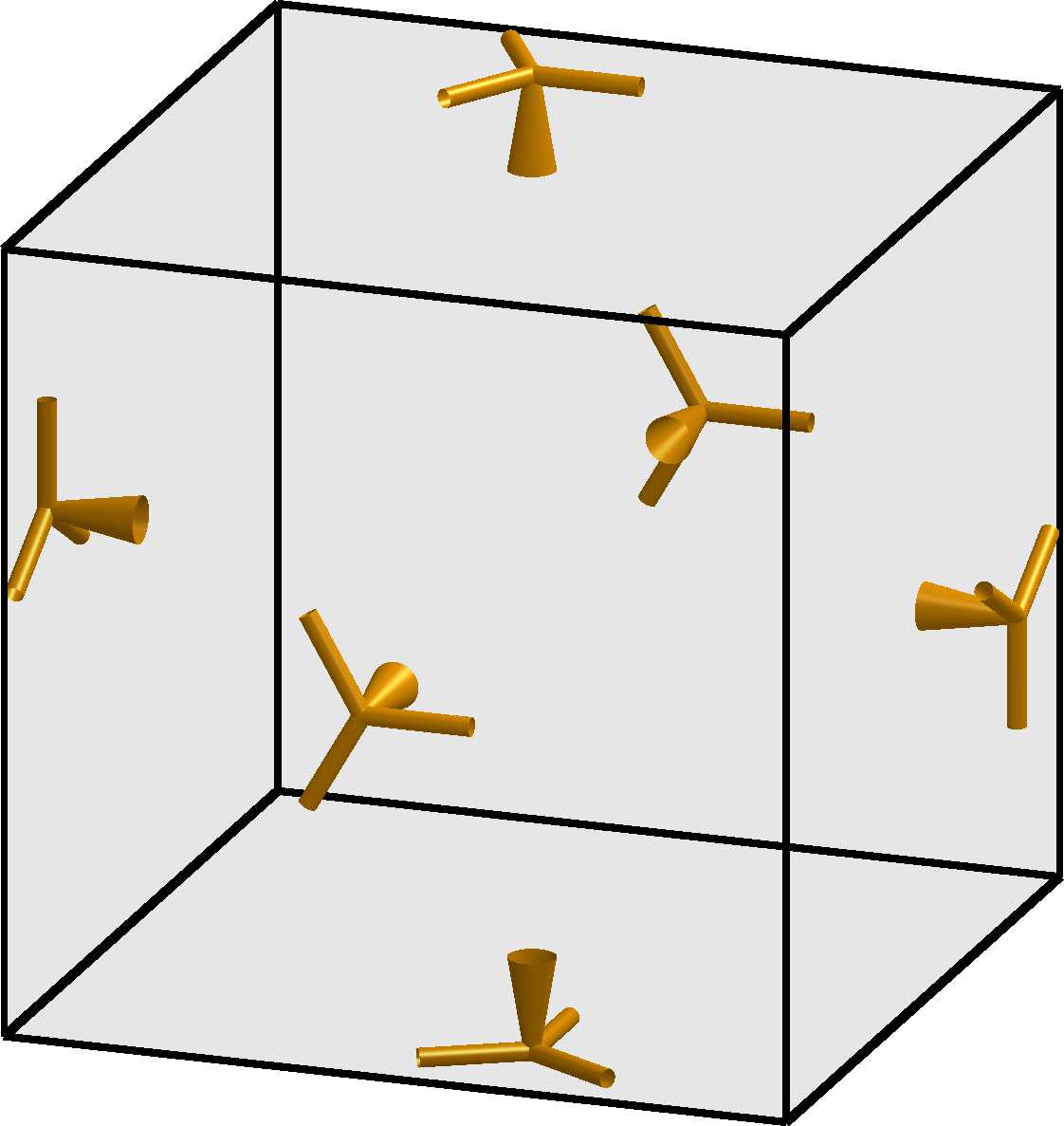}  \\ {\bf  6} \end{center}
\end{minipage}
\end{footnotesize}
\end{center}
 \caption{The source configurations investigated in this study with the antenna symbols indicating the source positions.  } \label{source_configurations}
\end{figure}

\subsection{Accuracy of inversion}

The accuracy of the reconstructed ${\mathtt n}_\delta$ was examined via the relative overlapping volume (ROV) and the relative error in value (REV), defined by
 \begin{equation} \fl
\hbox{ROV}  =  \frac{100 \tau}{3} \sum_{i=1}^3 \frac{\int_{\mathcal{R}\cap {\Omega_i} } \, d V}{\int_{{\Omega_i}} \, d V} \qquad \!\!\! \hbox{and} \qquad \! \!\! \hbox{REV} =   100 \tau \frac{ \sum_{i=1}^3 \int_{\mathcal{R} \cap \Omega_i} ({\mathtt n}_{bg} +  {\mathtt n}_\delta) \, d V}{{\mathtt n}_s \sum_{i=1}^3 \int_{\mathcal{R} \cap \Omega_i} \, d V} - 100
\end{equation}
in which $\Omega_i$ denotes the i-th stone to be detected, ${\mathtt n}_s$ is the actual refractive index of a stone, $\mathcal{R}$ is a set in which ${\mathtt n}_\delta$ is smaller than a limit such that  $\int_{\mathcal{R}} \, d V = \sum_{i=1}^3 \int_{\Omega_i} \, d V$, and $\tau$ is a hard threshold with respect to minimum overlap criterion $\nu$, that is, \begin{equation} \hbox{$\tau=1$, if} \, \, \min_i \Big\{  \frac{\int_{\mathcal{R}\cap {\Omega_i} } \, d V}{\int_{{\Omega_i}} \, d V} \Big\} \geq \nu^3, \, \, \hbox{and otherwise $\tau=0$}.
\end{equation} ROV describes the percentage of average stone-wise overlap between the actual stones and a reconstructed set $\mathcal{R}$ with the volume equal to that of the stones. REV measures the percentual error between ${\mathtt n}_s$ and the integral mean of ${\mathtt n}_\delta + {\mathtt n}_{bg}$  over the overlapping part of $\mathcal{R}$. If ${\mathtt n}_\delta$ is properly recovered, then $\tau=1$, meaning that the diameter of  $\mathcal{R} \cap \Omega_i$ must be at least around $\nu$ compared to that of $\Omega_i$ for $i=1,2,3$. Otherwise, if the overlapping part $\mathcal{R} \cap {\Omega_i}$ covers less than $\nu^3$ percent of the stone $\Omega_i$ for some $i$, then $\tau=0$, $\hbox{ROV}=0$ and $\hbox{REV}=-100$, indicating that one or more of the stones were mislocalized or badly recovered. A perfect reconstruction would lead to ROV=100 and REV=0. 
The statistics of the results regarding ROV and REV were analyzed via box plots visualizing the median, maximum, minimum, and (linearly interpolated) quartiles 1 \& 3. 

\subsection{Computational details}

A regular 18-by-18-by-18 lattice of voxels was utilized in the forward simulation together with a mask size $M=5$ ($\varrho \approx \lambda/2$), that was, based on our experience, the smallest value to prevent fragmenting. As a result, the localization problem was restricted to the innermost 10-by-10-by-10 part of the grid. The standard deviation (STD) of the total noise ${\bf g}$ was assigned the value $\sigma=1$ $\mu$s  corresponding to the estimated STD of the measurement noise (0.3 $\mu$s) plus possible forward modeling and simulation inaccuracies (this is a typical case of modelling errors dominating over random ones). Minimum overlap criterion was given the value $\nu=2/5$, meaning 40 \% minimum correspondence in diameter, that is around 6.4 \% in volume. Also alternative noise and overlap parameter values $\sigma = 2$ $\mu$s, $\nu=1/5$ and $\nu=3/5$ were tested. A total number of 315 estimates were computed. Each MAP estimate (reconstruction) was computed via twenty rounds of the IAS algorithm.   

\section{Results}

\begin{table}[t]
\caption{Minimum, median and maximum of ROV and REV for each configuration type according to the reconstruction type with the greatest (best) median value. }\label{table_rov}
\begin{indented}
\item[]
\begin{tabular}{@{}lllllllll}
\\ \br
& \centre{4}{ROV} & \centre{4}{REV} \\
& \crule{4} & \crule{4} \\
Config. & Type & Min & Median  & Max & Type & Min & Median & Max \\
\mr
 {\bf 6}  &  (g)  &  52    & 59   &  60    &  (ig)  & -15   &-11   & -8   \\
{\bf 5}  &   (g)  &  37    & 60   &  64    &  (ig)  & -16   &-12   & -5   \\
 {\bf 4a} &  (g)  &  39    & 56   &  67    &  (ig)  & -27   &-15   &  2    \\
 {\bf 4b} &  (g)  &  32    & 58   &  69    &  (ig)  & -100  & -11  &   3   \\
{\bf 3a} &   (ig) &    0   &  56  &   72   &   (ig) & -100  & -11  &   3   \\
 {\bf 3b} &  (g)  &  26    & 50   &  63    &  (ig)  & -100  & -15  &  16   \\
{\bf 2a} &   (g)  &   0    & 49   &  68    &  (ig)  & -100  & -18  &  26   \\
{\bf 2b} &   (g)  &   0    & 31   &  48    &  (g)   & -100  & -43  &  -4   \\
{\bf 1}  &   (g)  &   0    &  0   &  49    &  (ig)  & -100  &-100  &  37   \\
 \br
\end{tabular}
\end{indented}
\end{table}
\begin{table}[t]
\caption{Ranking of the minimum, median and spread (distance between the minimum and maximum) of the samples given in Table \ref{table_rov}. The cases in which $\hbox{ROV}=0$ or $\hbox{REV}=-100$ were given the lowest ranking (9).  The sum of the ranked positions provides an overall ranking of the source configurations in ascending order.  }\label{table_ranking}
\begin{indented}
\item[]
\begin{tabular}{@{}lllllllll}
\\ \br
 & \centre{4}{ROV} & \centre{4}{REV}\\
& \crule{4} & \crule{4} \\
Config.\   & Min & Median  & Spread  & Sum  & Min & Median & Spread & Sum\\
\mr
{\bf 6}  &   1   &  2    &  1    &  4   &   1   &  1  &   2  &   4  \\
{\bf 5}  &    3   &  1    &  3    &  7   &  2    & 4   &  1   &  7 \\
 {\bf 4a} &   2   &  4    &  2    &  8   &  3    & 5   &  3   & 11  \\
 {\bf 4b} &   4   &  3    &  4    & 11   &  9    & 2   &  9   & 20  \\
{\bf 3a} &    9   &  5    &  9    & 23   &  9    & 3   &  9   & 21  \\
 {\bf 3b} &   5   &  6    &  5    & 16   &  9    & 6   &  9   & 24  \\
{\bf 2a} &    9   &  7    &  9    & 25   &  9    & 7   &  9   & 25  \\
{\bf 2b} &    9   &  8    &  9    & 26   &  9    & 8   &  9   & 26  \\
{\bf 1}  &    9   &  9    &  9    & 27   &  9    & 9   &  9   & 27    \\
\br
\end{tabular}
\end{indented}
\end{table}
\begin{figure}[h!]\begin{center}\begin{footnotesize}
\begin{minipage}{6cm}\begin{center}
\includegraphics[width=4.2cm,draft=false]{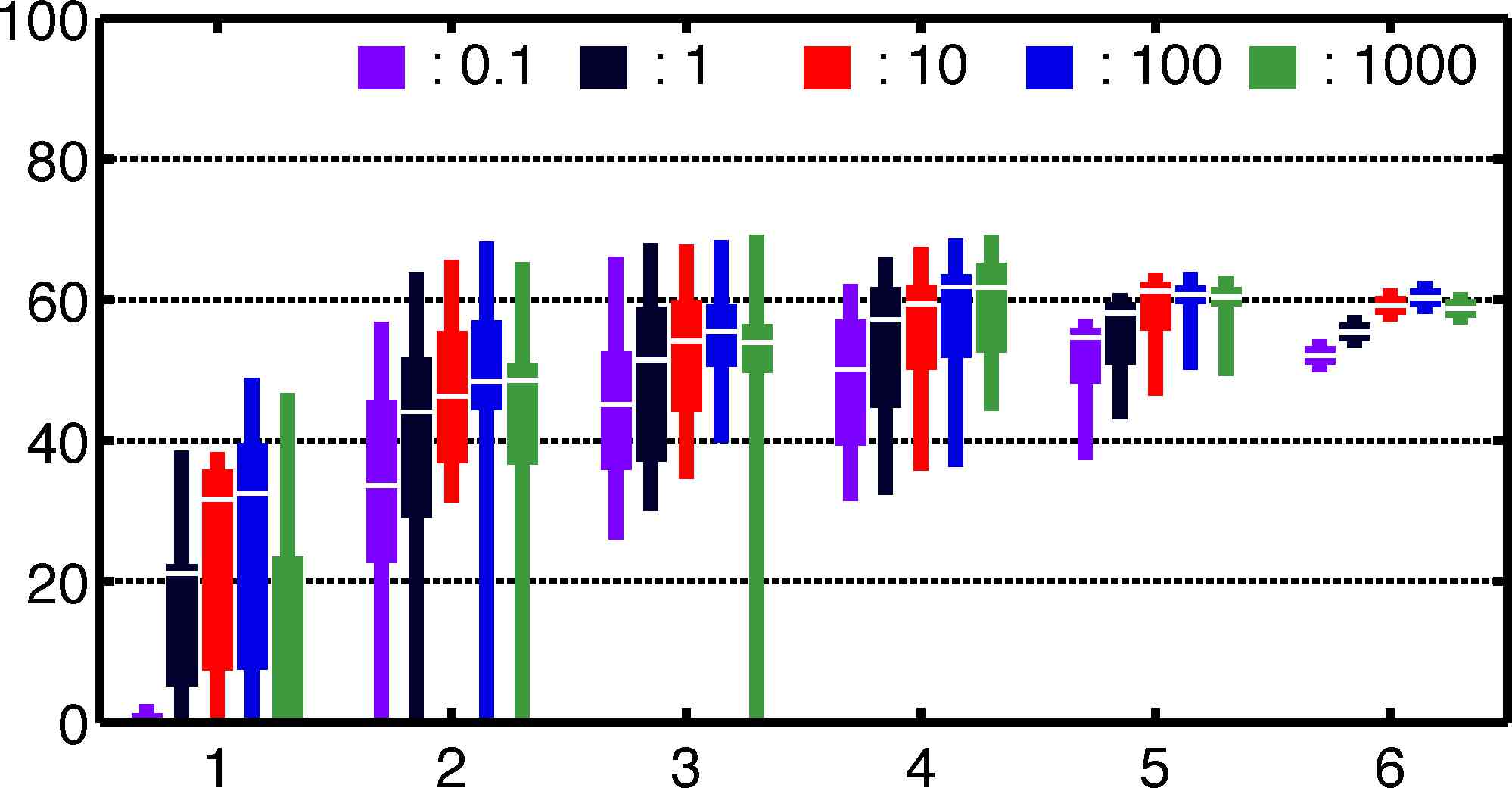} \\ ROV, (g) \end{center}
\end{minipage}
\begin{minipage}{6cm}\begin{center}
\includegraphics[width=4.2cm,draft=false]{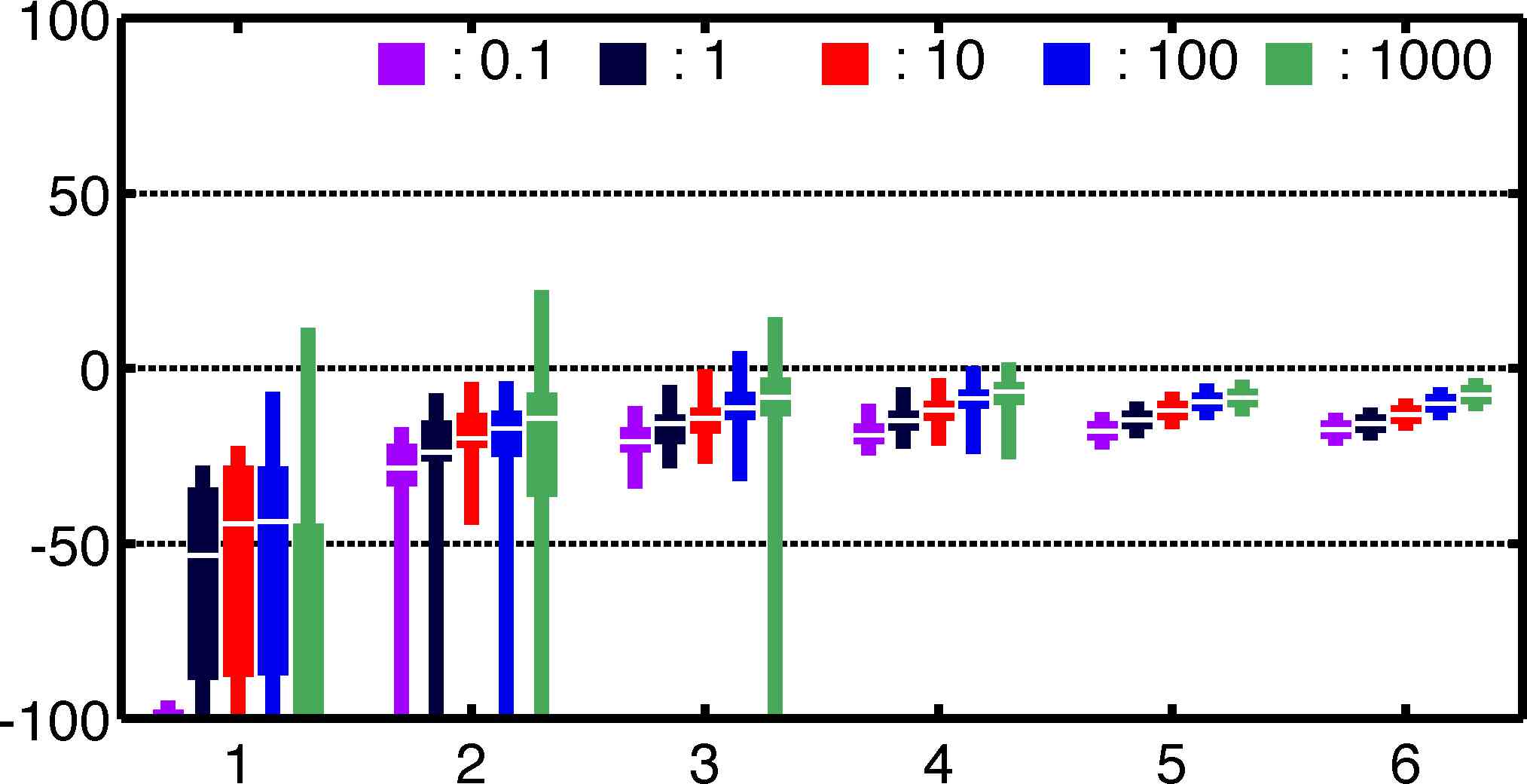} \\ REV, (g) \end{center}
\end{minipage} \\ \vskip0.1cm
\begin{minipage}{6cm}\begin{center}
\includegraphics[width=4.2cm,draft=false]{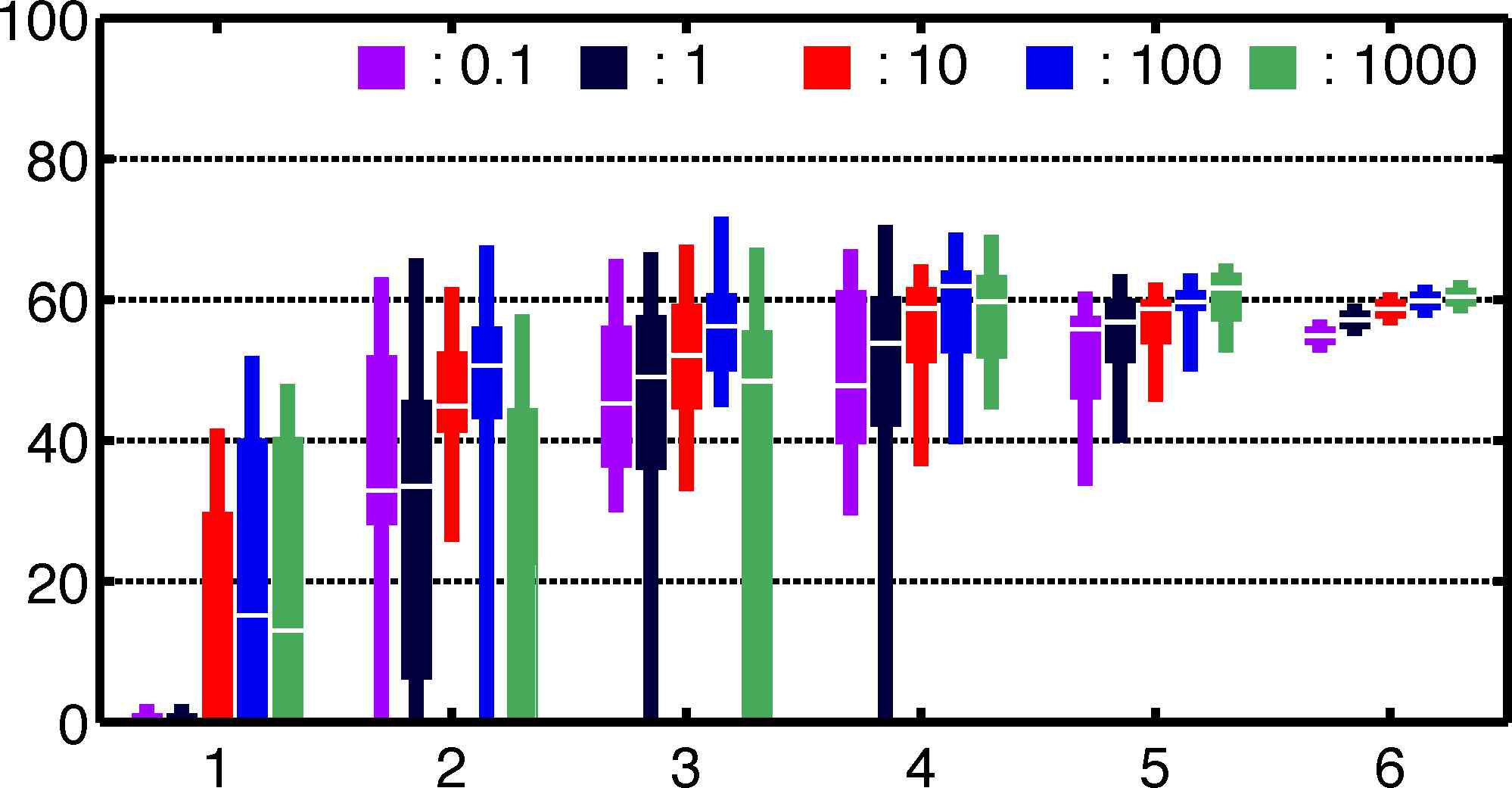} \\ ROV, (ig) \end{center}
\end{minipage}
\begin{minipage}{6cm}\begin{center}
\includegraphics[width=4.2cm,draft=false]{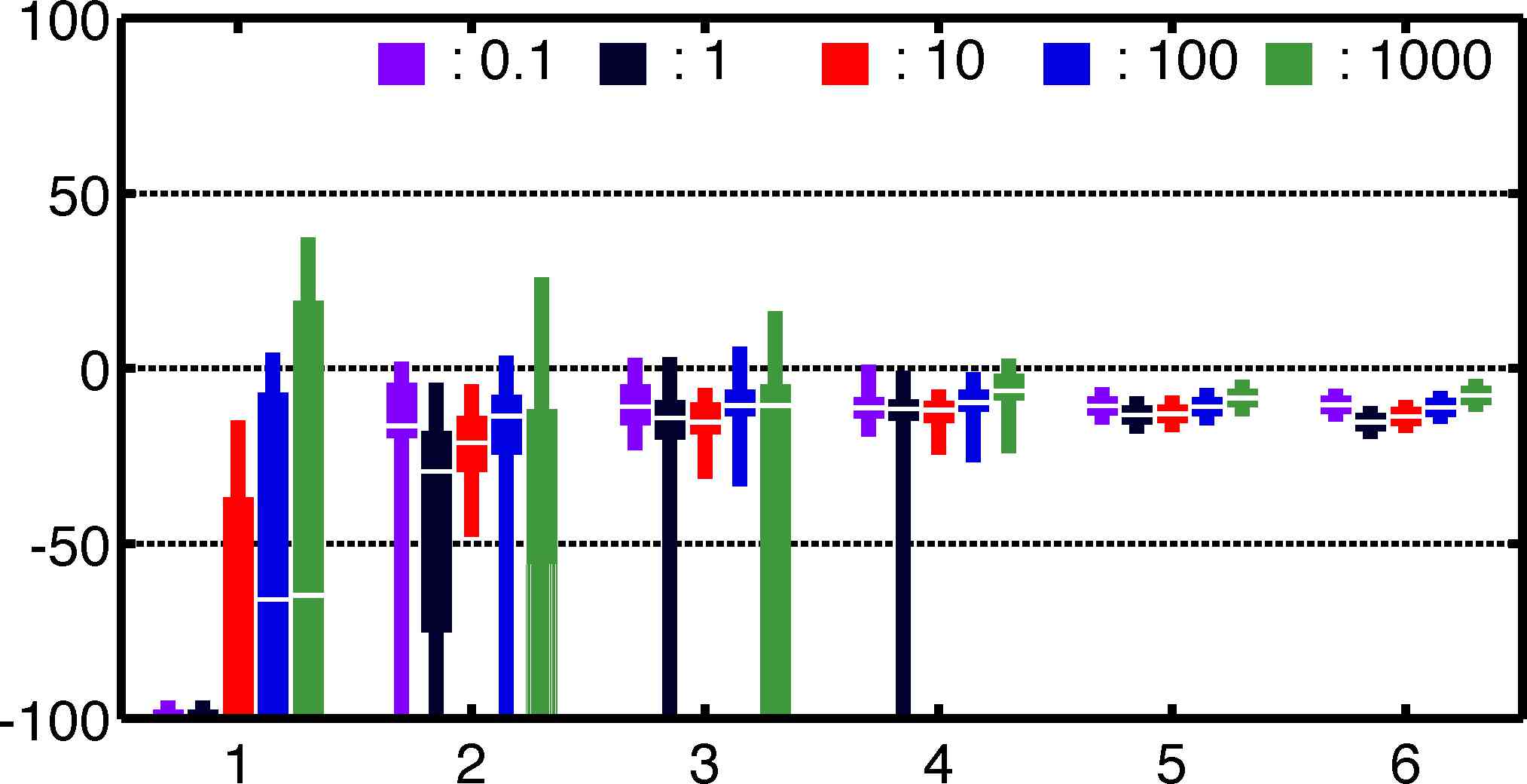} \\ REV, (ig) \end{center}
\end{minipage}\\ \vskip0.1cm
\begin{minipage}{6cm}\begin{center}
\includegraphics[width=4.2cm,draft=false]{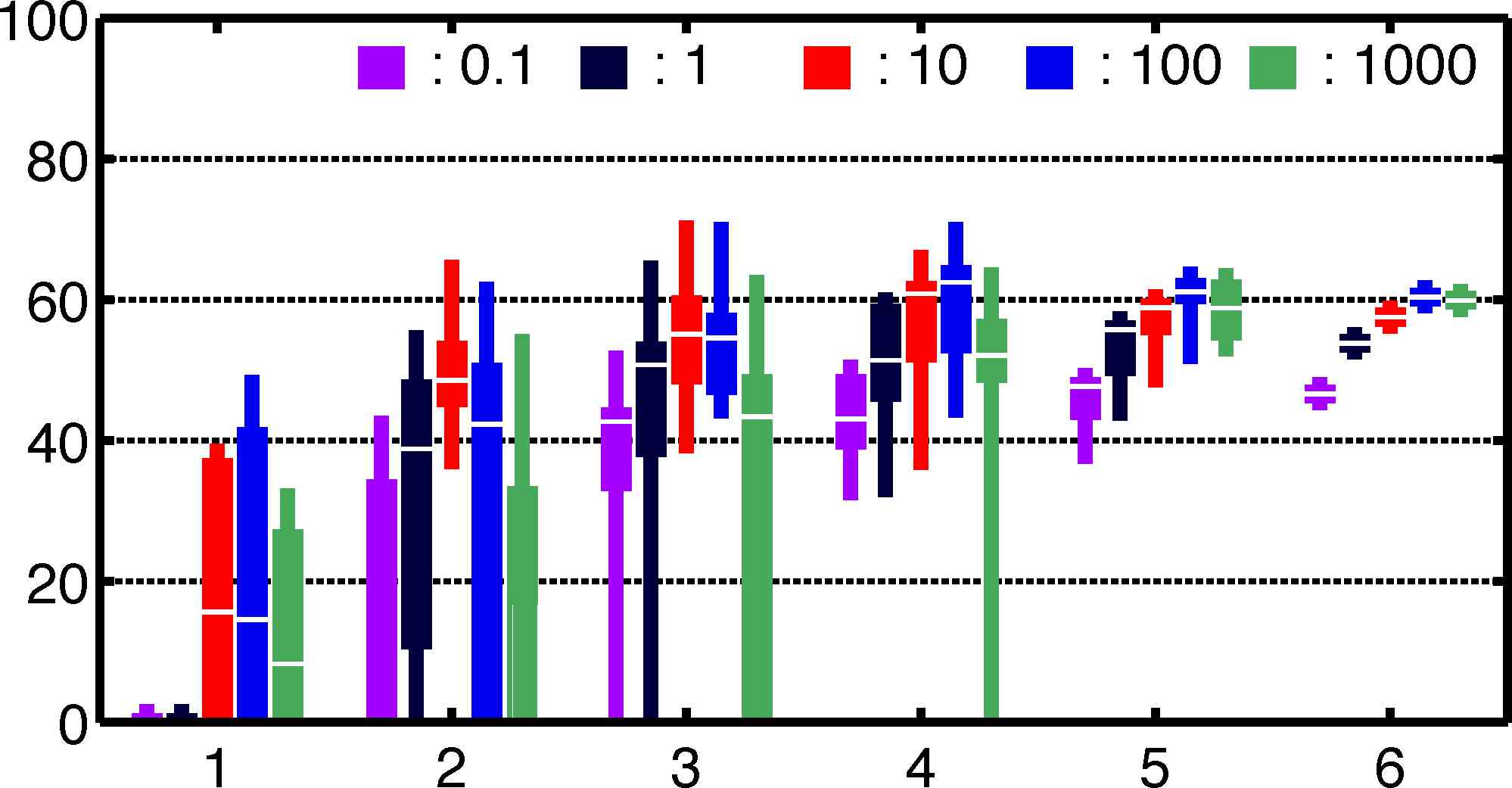} \\ ROV, (f) \end{center}
\end{minipage}
\begin{minipage}{6cm}\begin{center}
\includegraphics[width=4.2cm,draft=false]{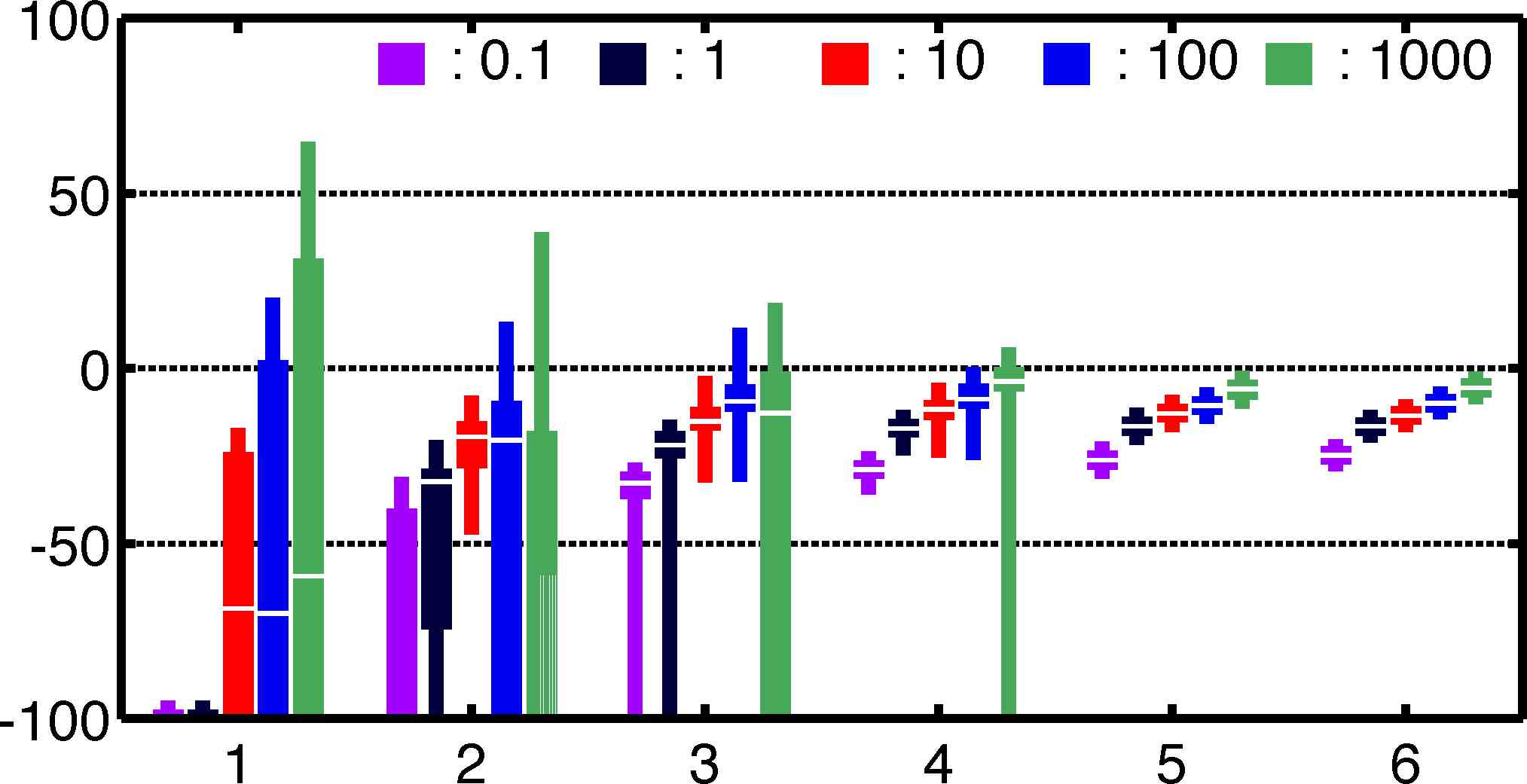} \\ REV, (f) \end{center}
\end{minipage}
\end{footnotesize}
\end{center}
 \caption{Box plots of ROV (left) and REV (right) with the rows  from top to bottom corresponding, respectively, to the reconstruction types (g), (ig) and (f). Each subfigure covers source counts 1--6 and initial prior variances $\theta_0 = 10^k, k=-1,0,1,2,3$. In each bar, the narrow part shows the maximum and minimum, the thicker part corresponds to the quartiles 1 \& 3, and the white line gives the median.} \label{comparison_1}
\end{figure}
\begin{figure}[h!]\begin{center}\begin{footnotesize}
\begin{minipage}{6cm}\begin{center}
\includegraphics[width=4.2cm,draft=false]{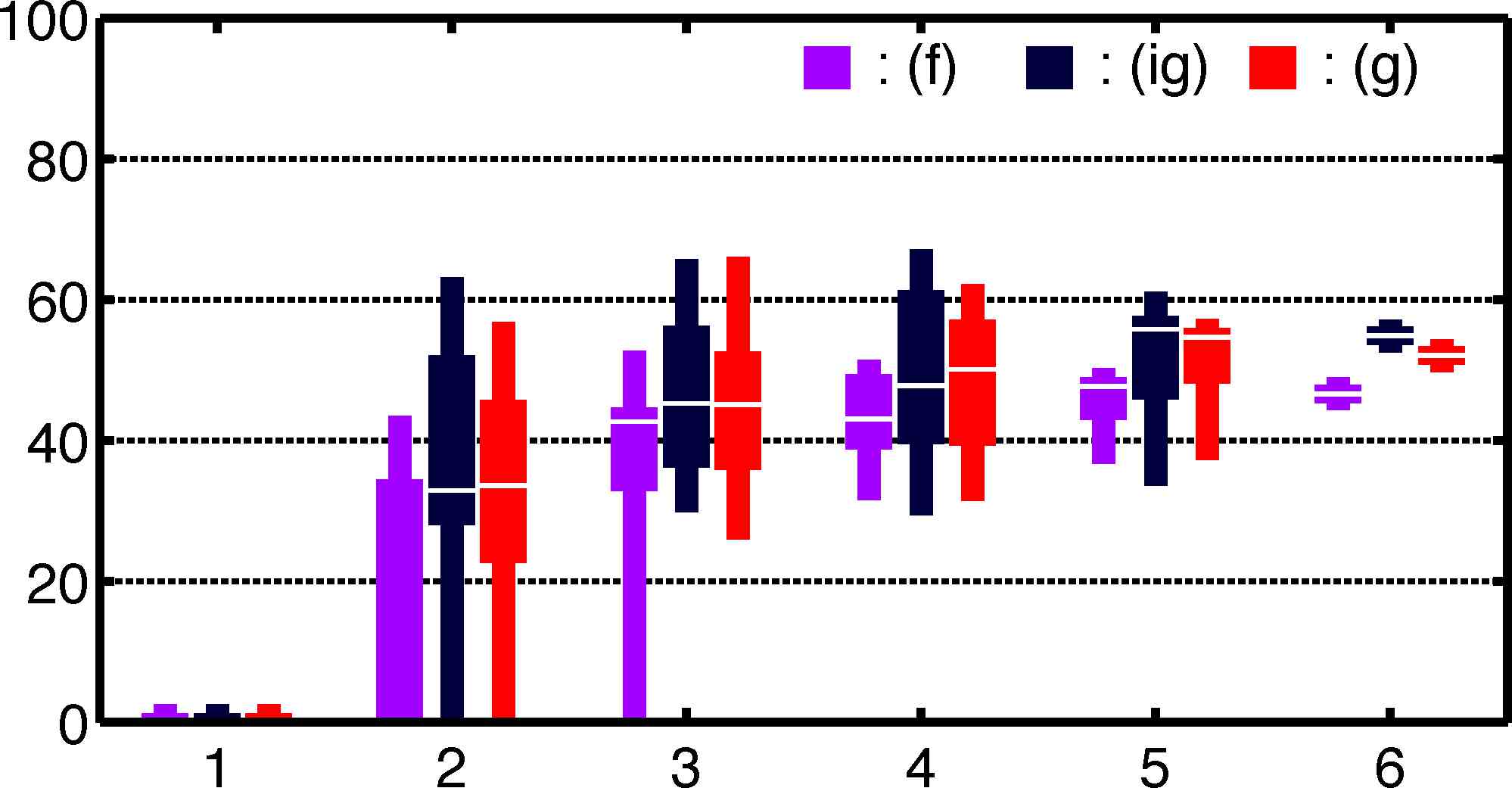} \\ ROV, $\theta_0 = 0.1$  \end{center}
\end{minipage}
\begin{minipage}{6cm}\begin{center}
\includegraphics[width=4.2cm,draft=false]{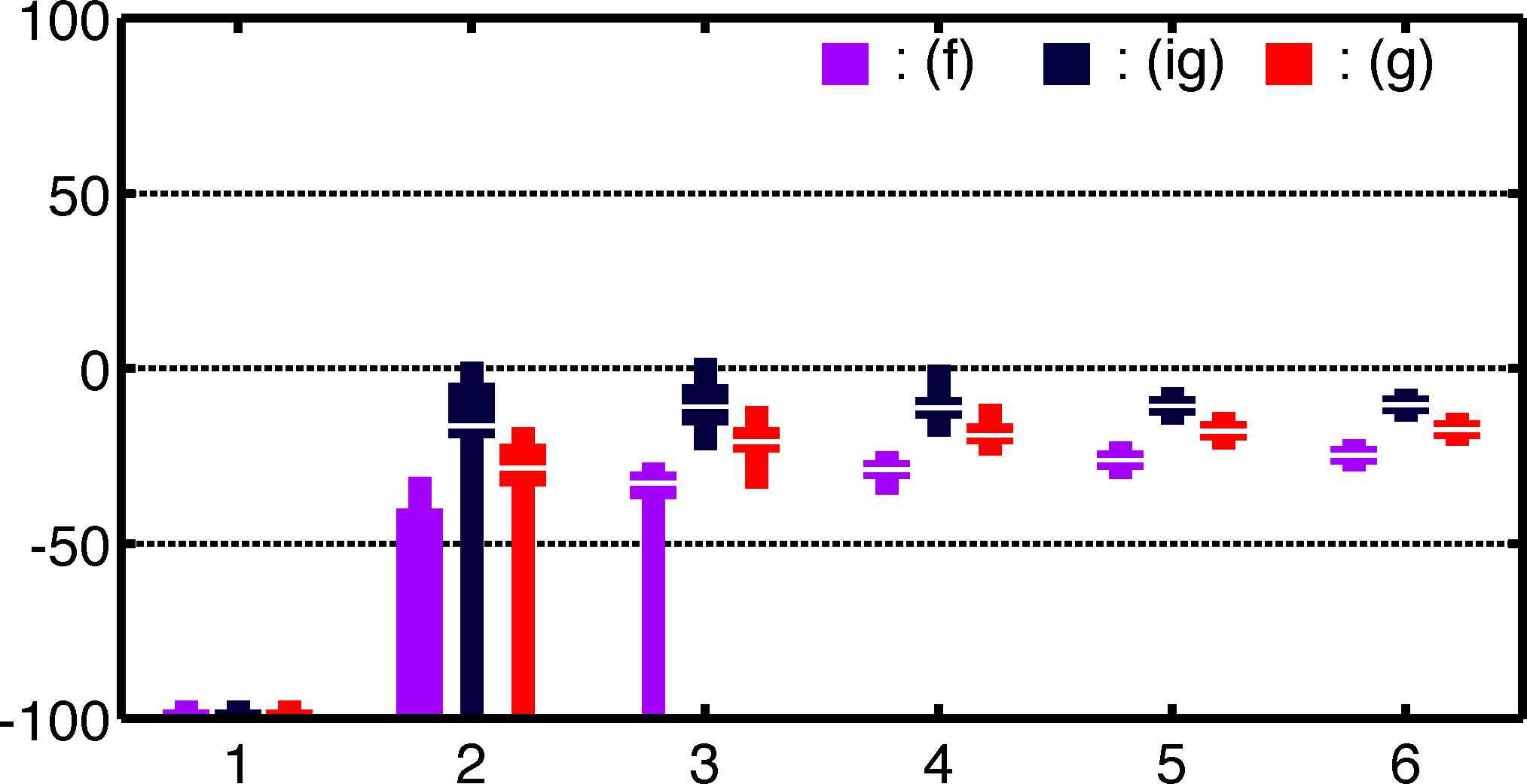} \\ REV, $\theta_0=0.1$\end{center}
\end{minipage} \\ \vskip0.1cm
\begin{minipage}{6cm}\begin{center}
\includegraphics[width=4.2cm,draft=false]{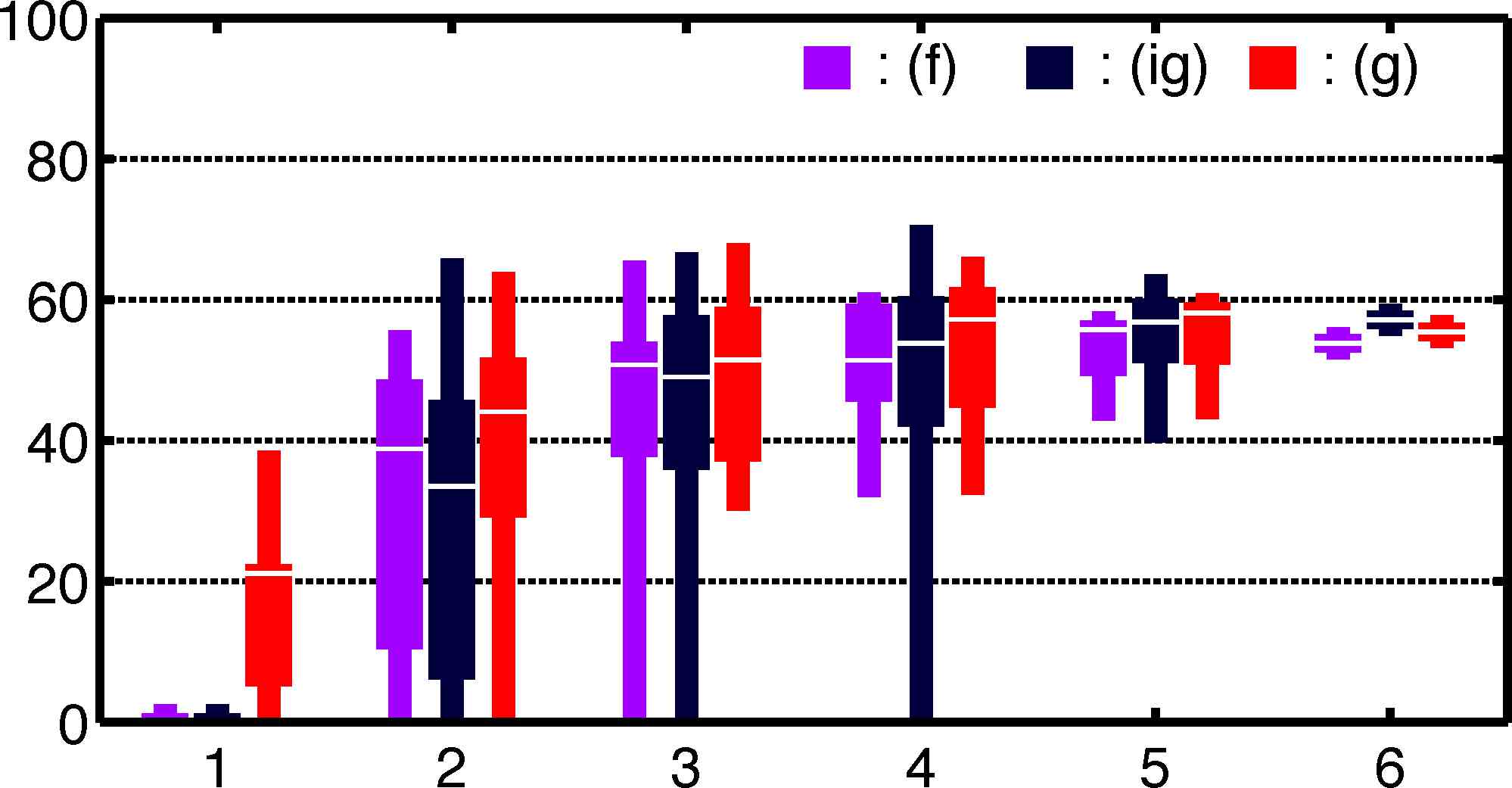} \\ ROV, $\theta_0 = 1$  \end{center}
\end{minipage}
\begin{minipage}{6cm}\begin{center}
\includegraphics[width=4.2cm,draft=false]{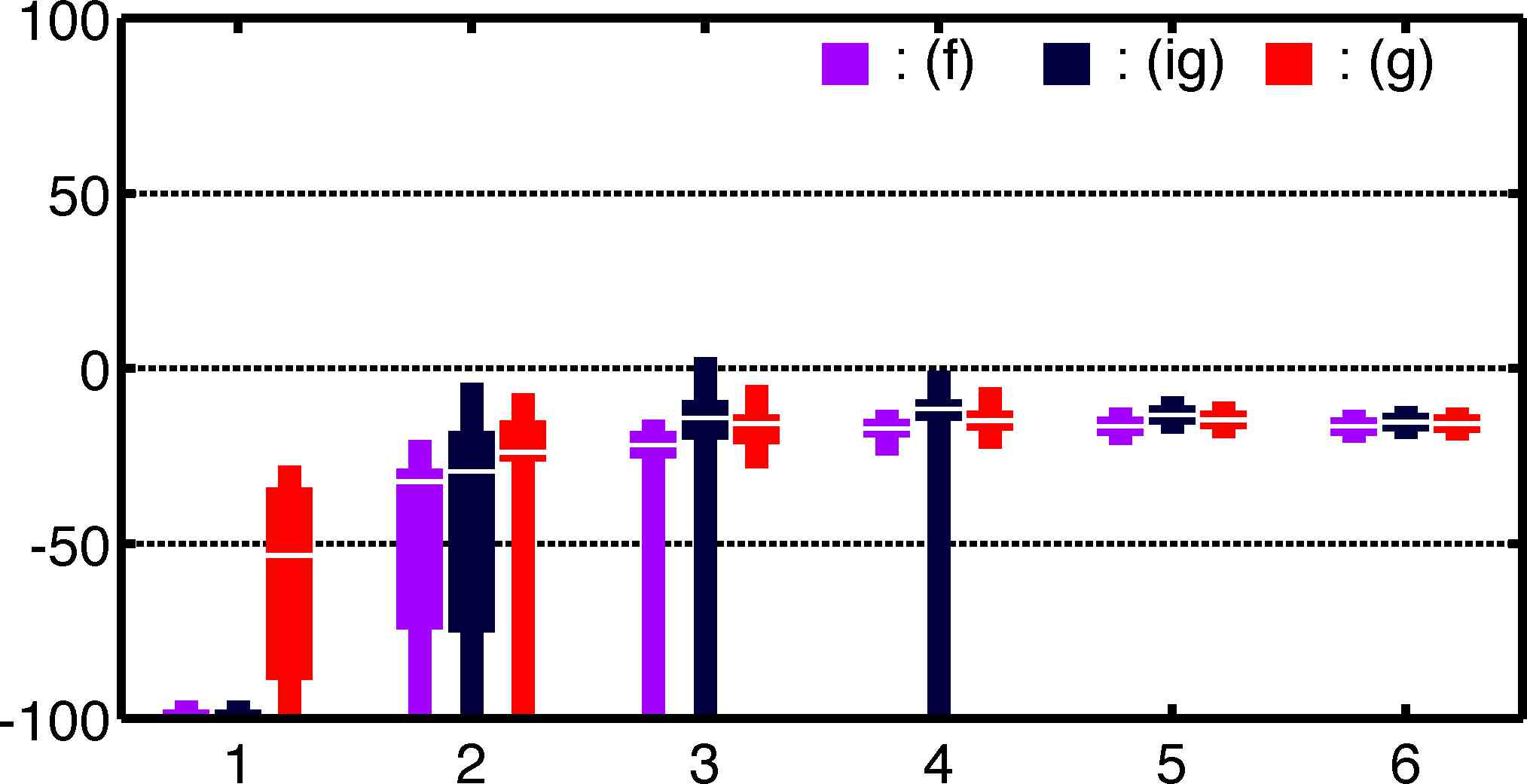} \\ REV, $\theta_0=1$ \end{center}
\end{minipage}\\ \vskip0.1cm
\begin{minipage}{6cm}\begin{center}
\includegraphics[width=4.2cm,draft=false]{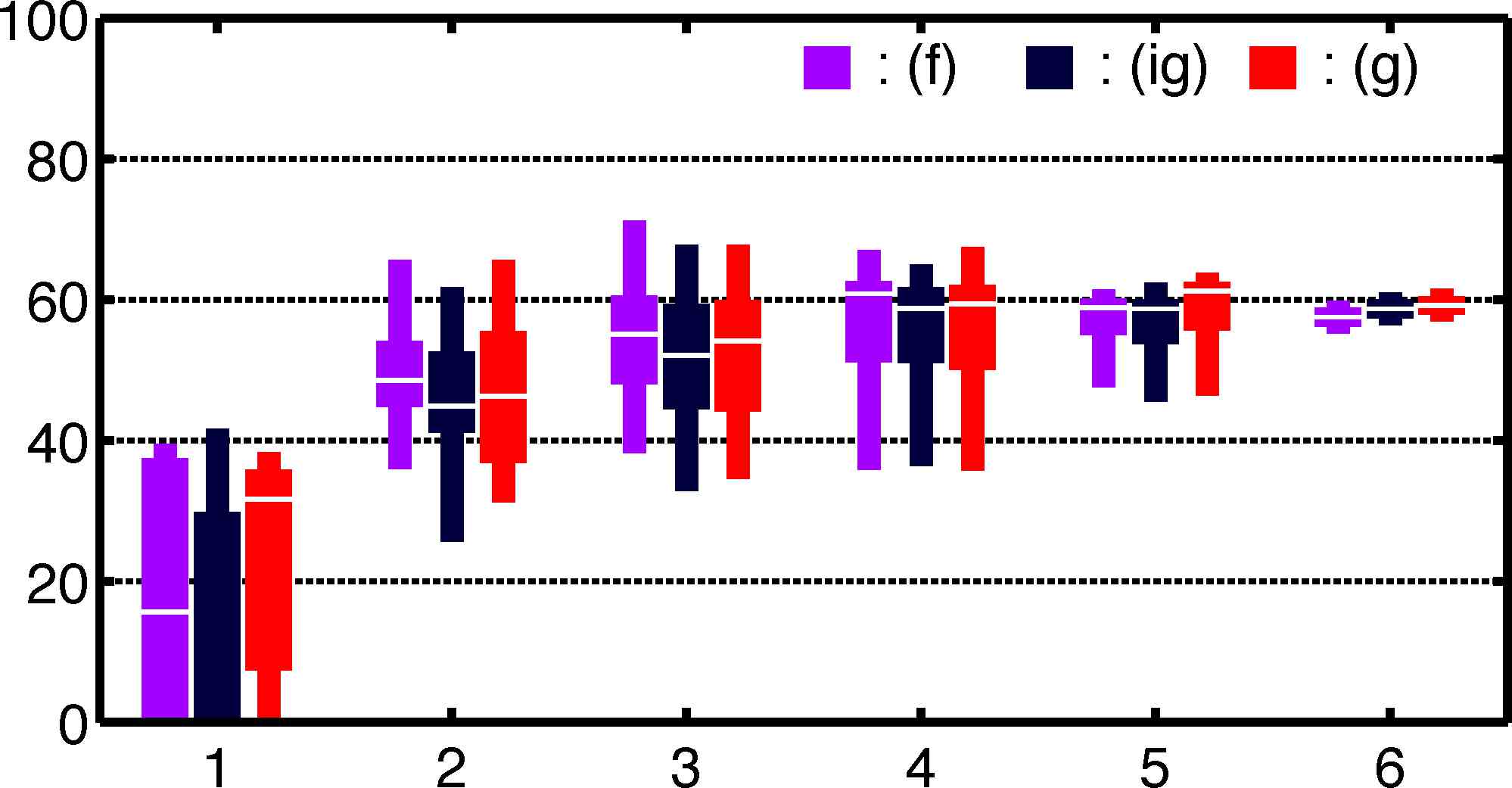}  \\ ROV, $\theta_0 = 10$ \end{center}
\end{minipage}
\begin{minipage}{6cm}\begin{center}
\includegraphics[width=4.2cm,draft=false]{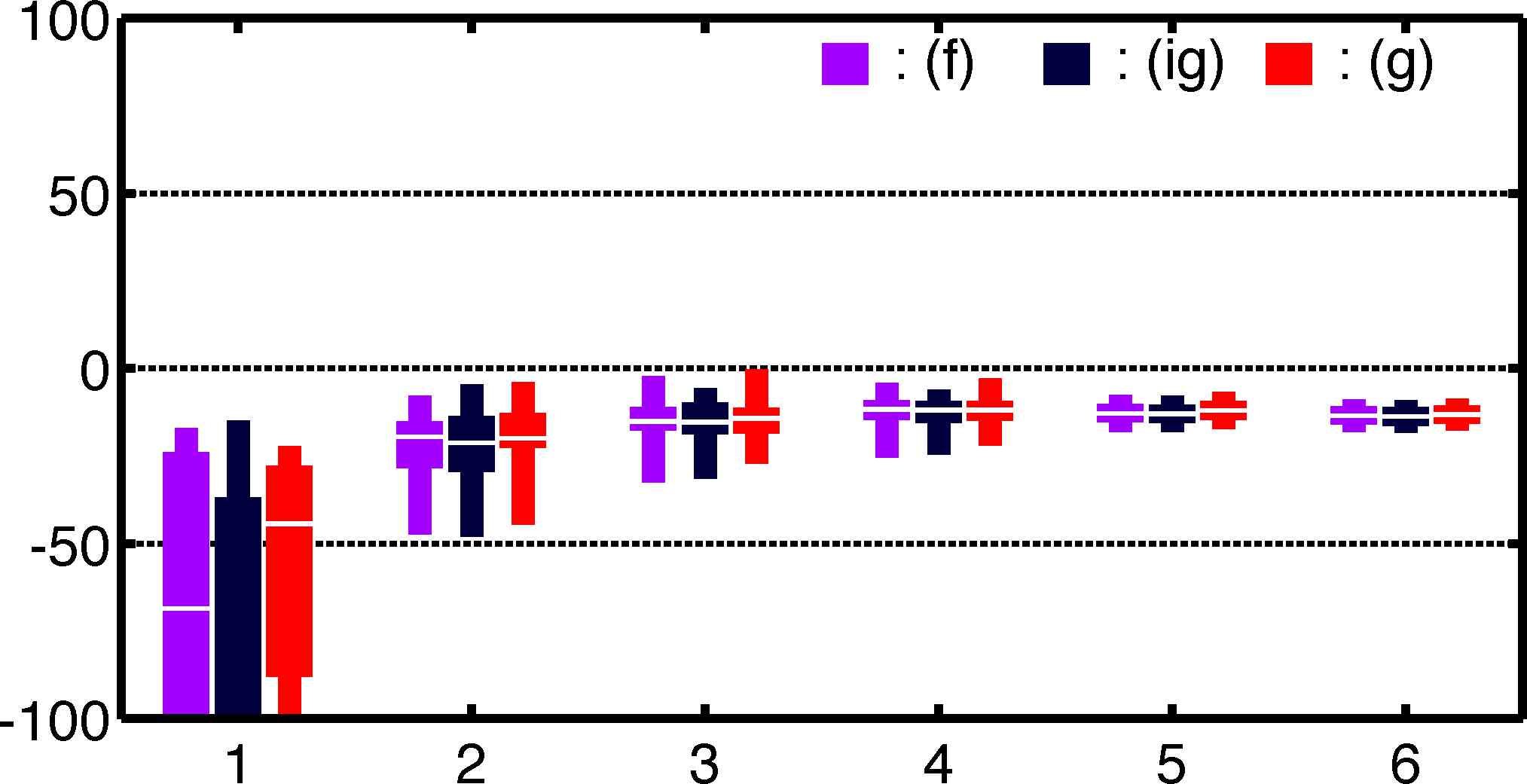} \\ REV, $\theta_0=10$ \end{center}
\end{minipage} \\ \vskip0.1cm
\begin{minipage}{6cm}\begin{center}
\includegraphics[width=4.2cm,draft=false]{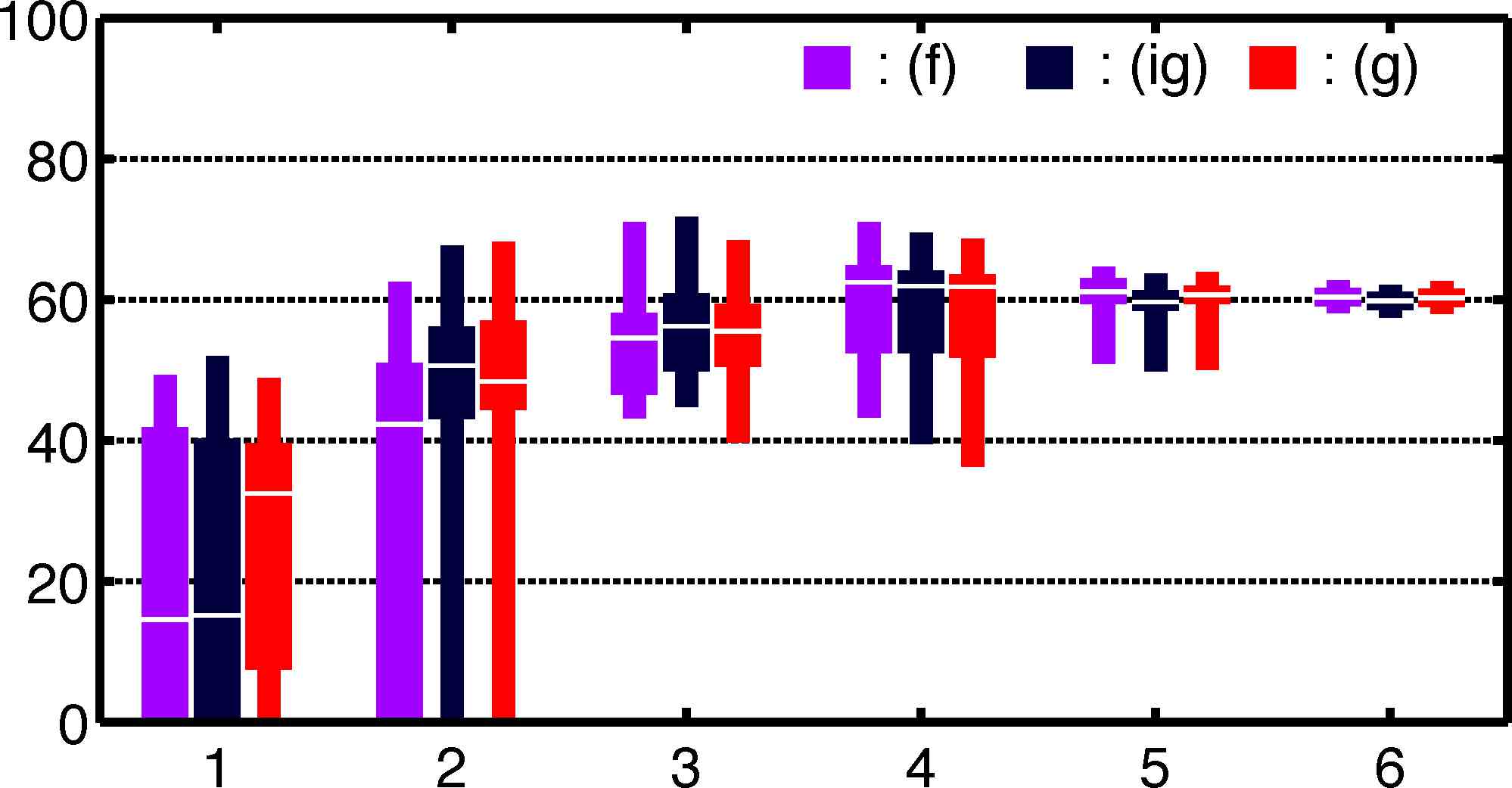} \\ ROV, $\theta_0 = 100$  \end{center}
\end{minipage}
\begin{minipage}{6cm}\begin{center}
\includegraphics[width=4.2cm,draft=false]{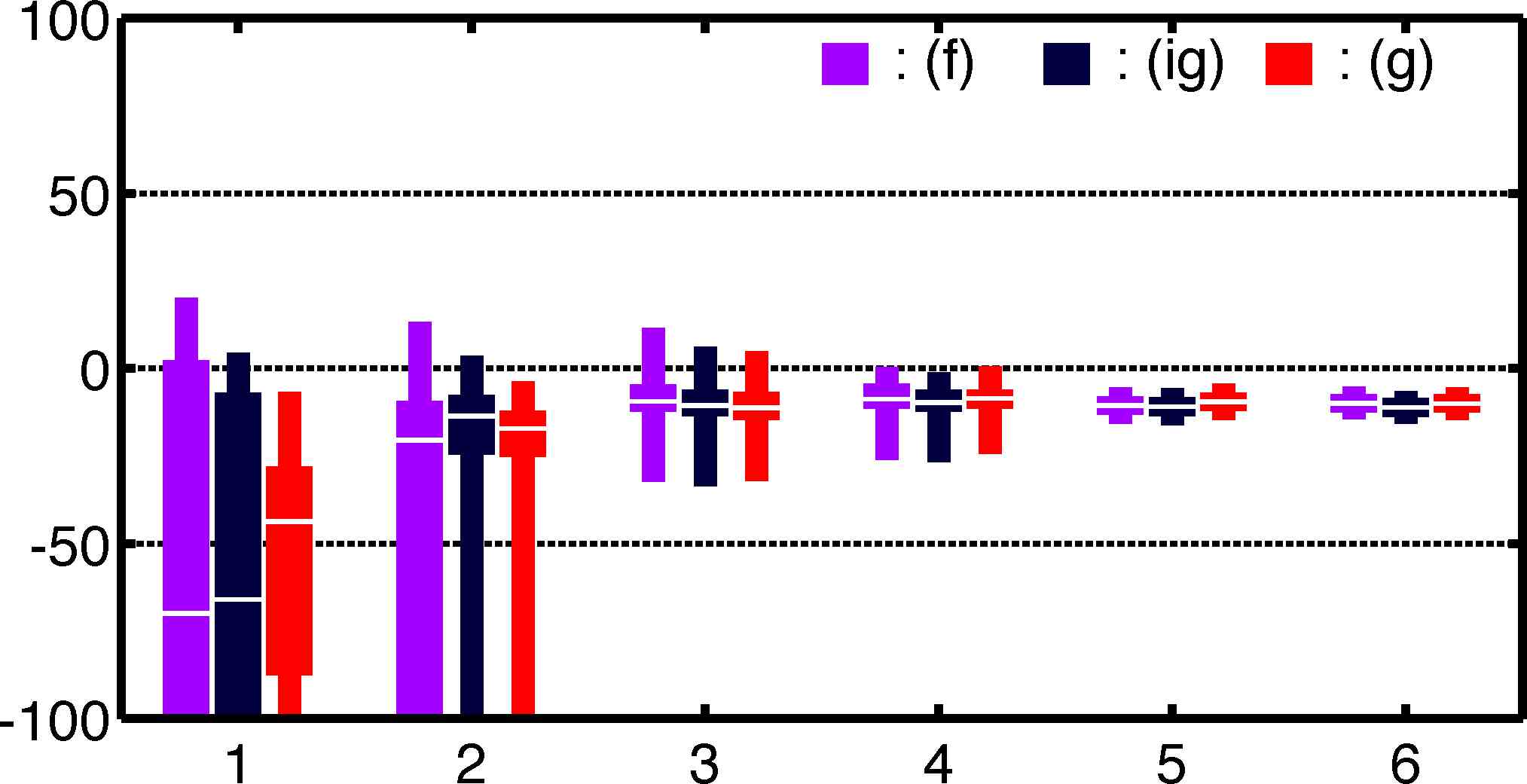} \\ REV, $\theta_0 = 100$ \end{center}
\end{minipage}\\ \vskip0.1cm
\begin{minipage}{6cm}\begin{center}
\includegraphics[width=4.2cm,draft=false]{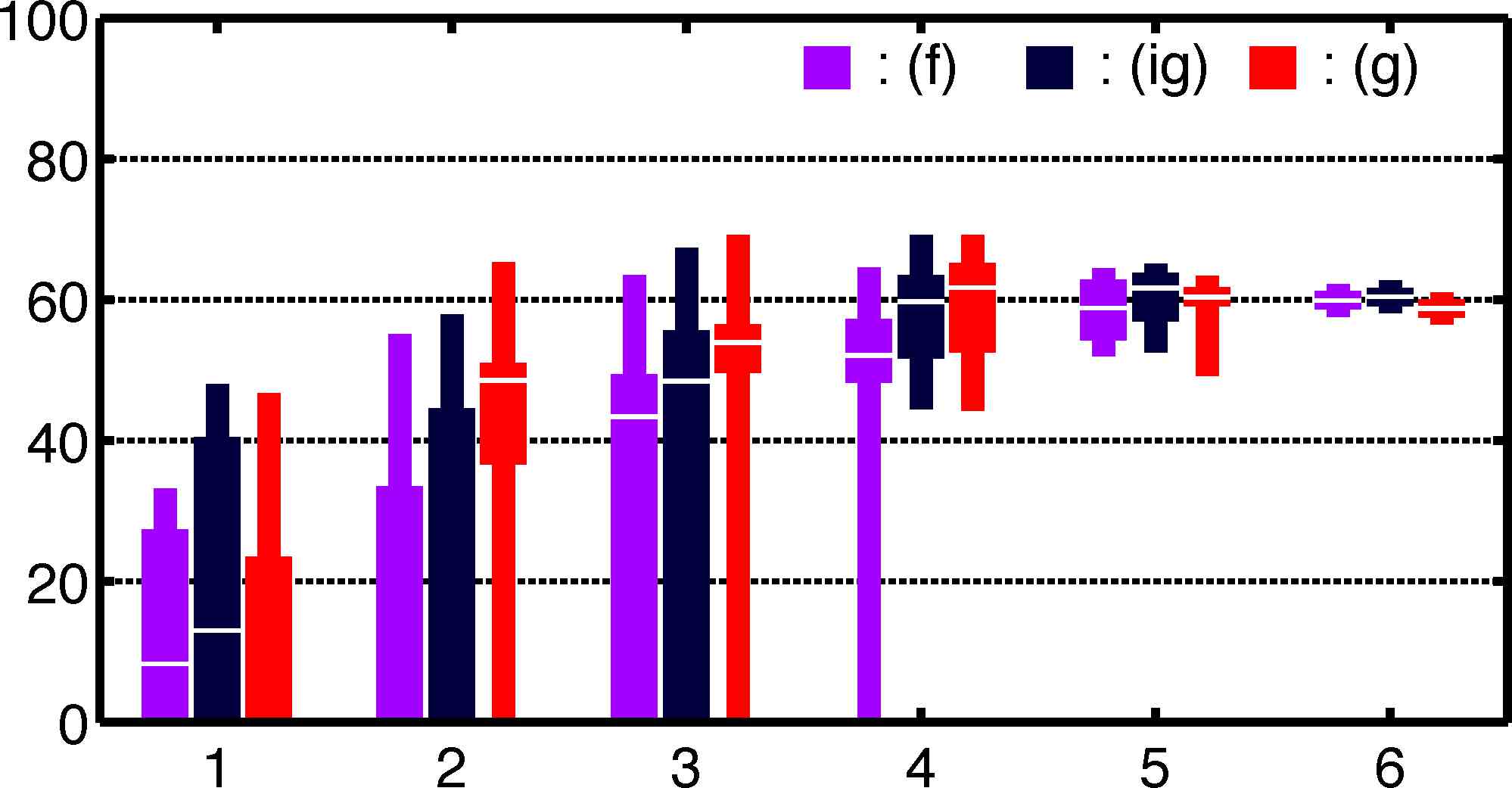} \\ ROV, $\theta_0 = 1000$ \end{center}
\end{minipage}
\begin{minipage}{6cm}\begin{center}
\includegraphics[width=4.2cm,draft=false]{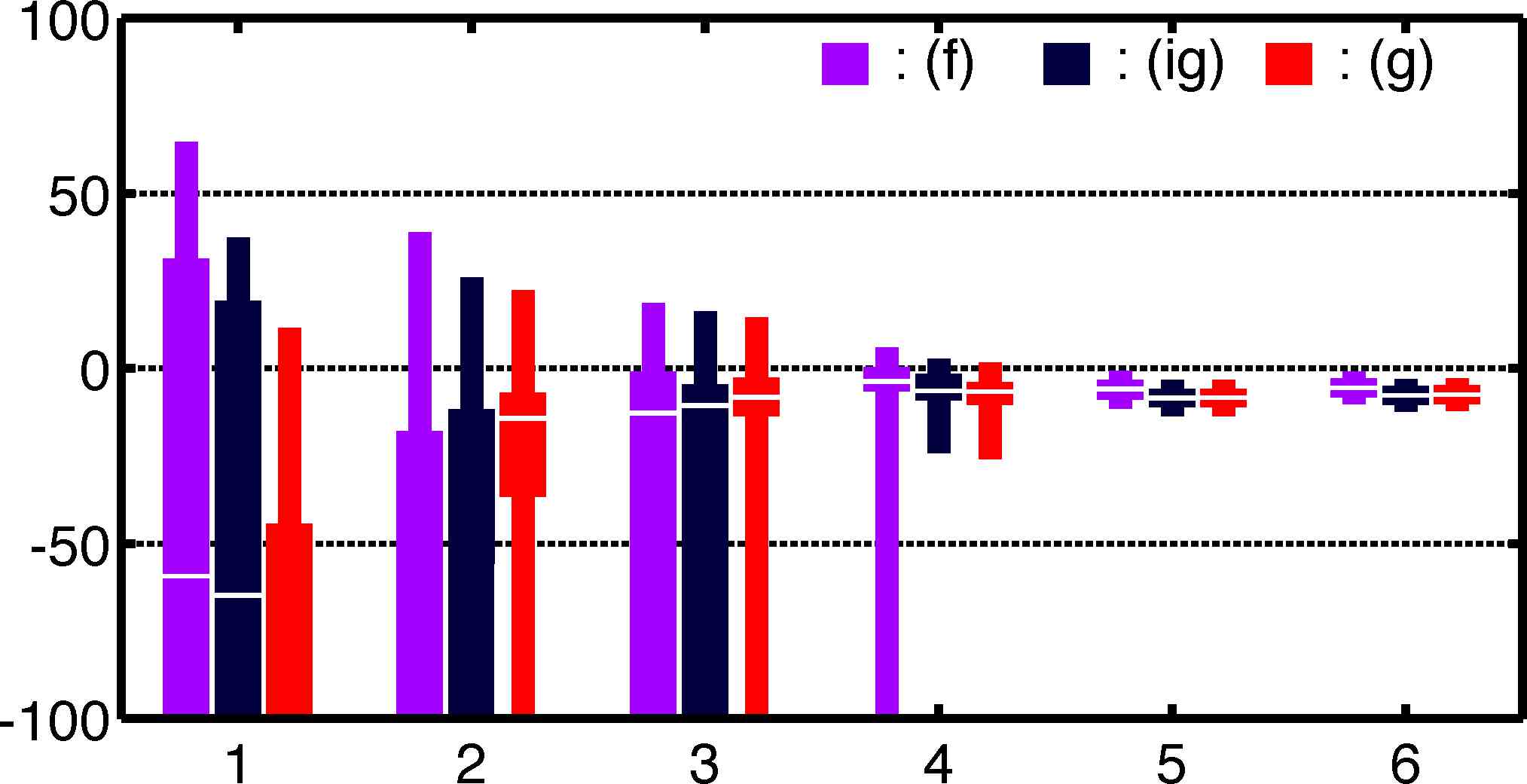} \\ REV, $\theta_0=1000$ \end{center}
\end{minipage}
\end{footnotesize}
\end{center}
 \caption{Box plots of ROV (left) and REV (right) with the rows  from top to bottom corresponding, respectively, to initial prior variances $\theta_0 = 10^k, k=-1,0,1,2,3$. Each subfigure covers source counts 1--6 and the  reconstruction types (g), (ig) and (f). 
} \label{comparison_2}
\end{figure}
\begin{figure}[h!]\begin{center}\begin{footnotesize}
\begin{minipage}{6cm}\begin{center}
\includegraphics[width=4.2cm,draft=false]{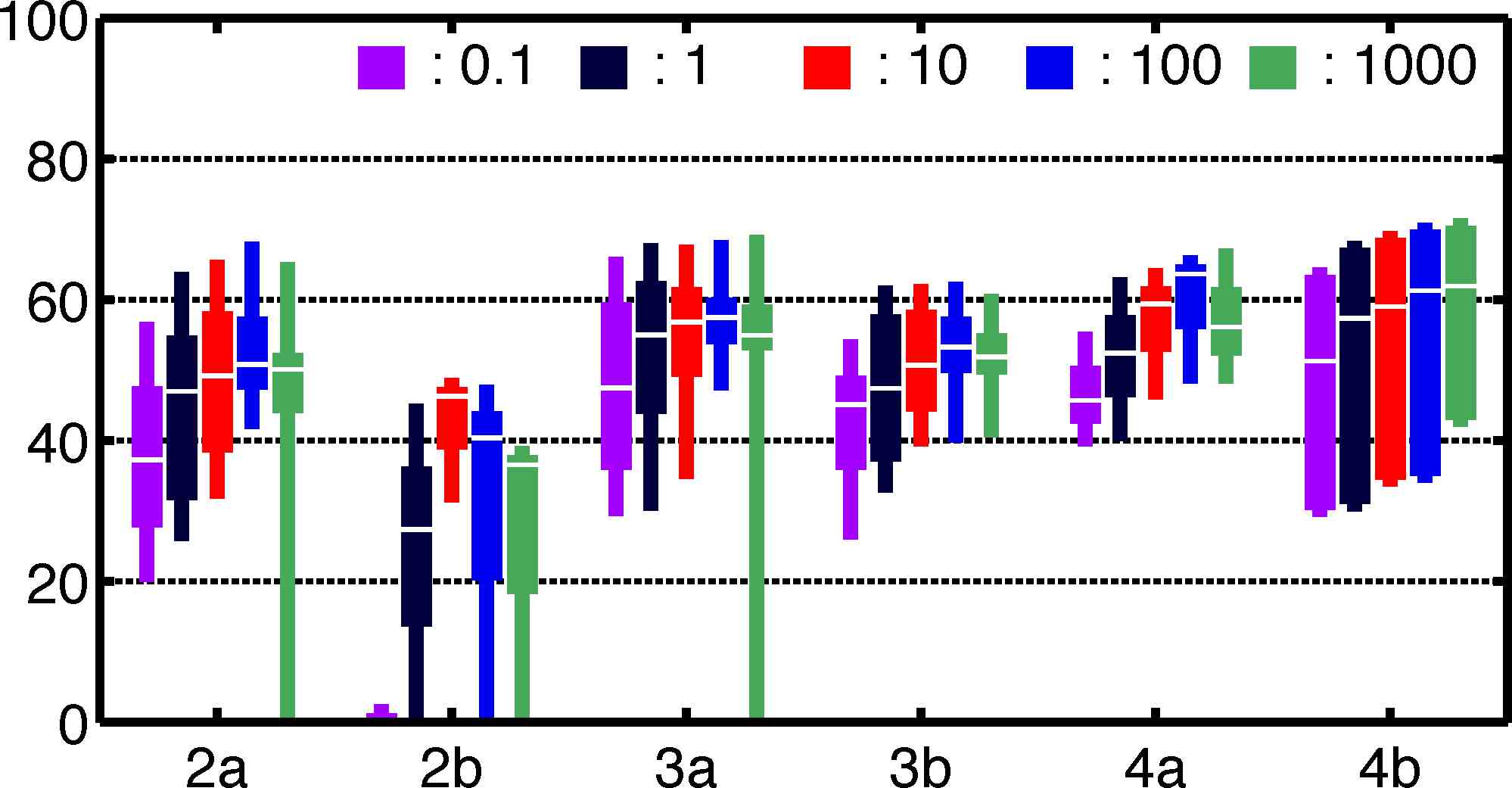} \\ ROV, (g) \end{center}
\end{minipage}
\begin{minipage}{6cm}\begin{center}
\includegraphics[width=4.2cm,draft=false]{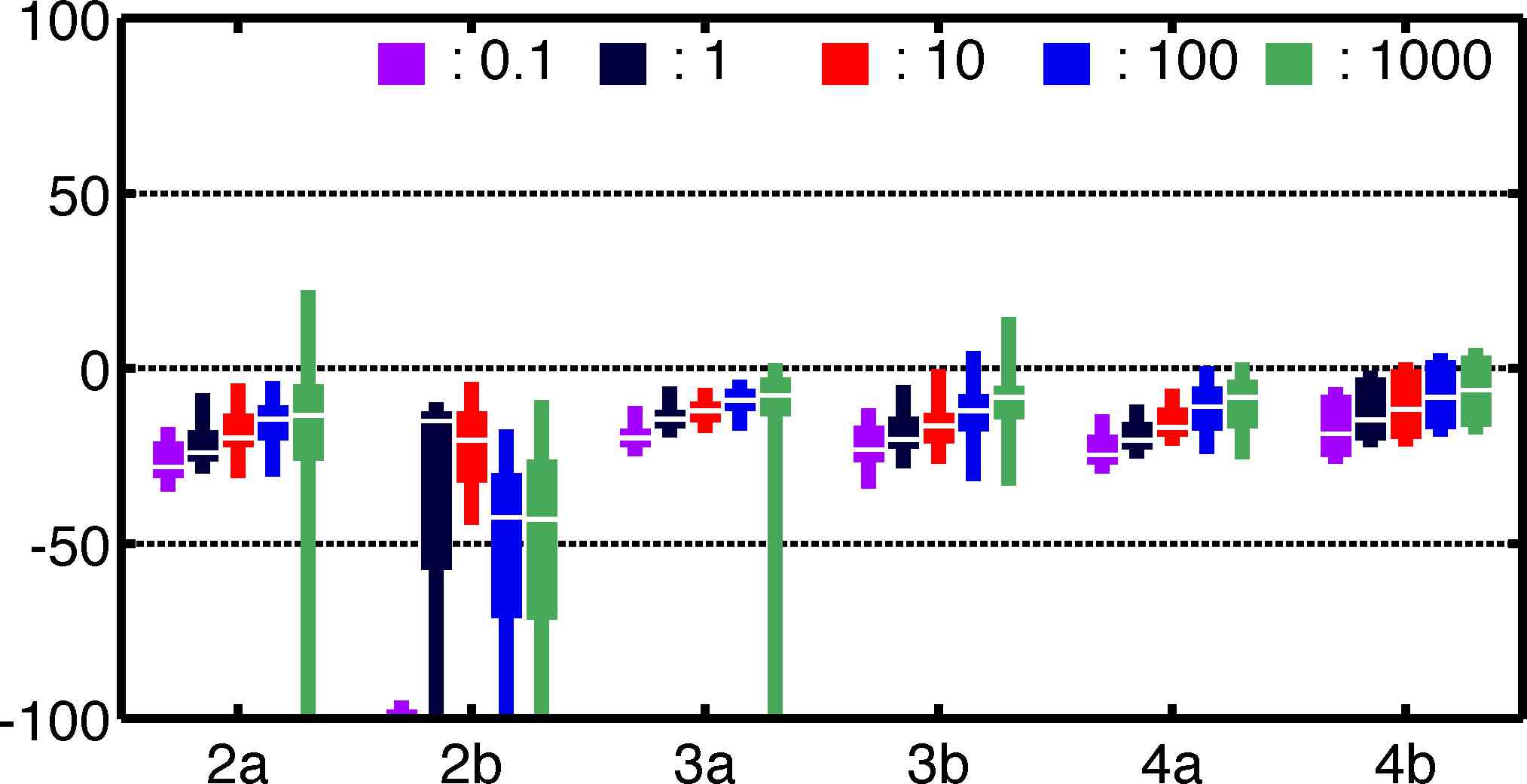} \\ REV, (g) \end{center}
\end{minipage} \\ \vskip0.1cm
\begin{minipage}{6cm}\begin{center}
\includegraphics[width=4.2cm,draft=false]{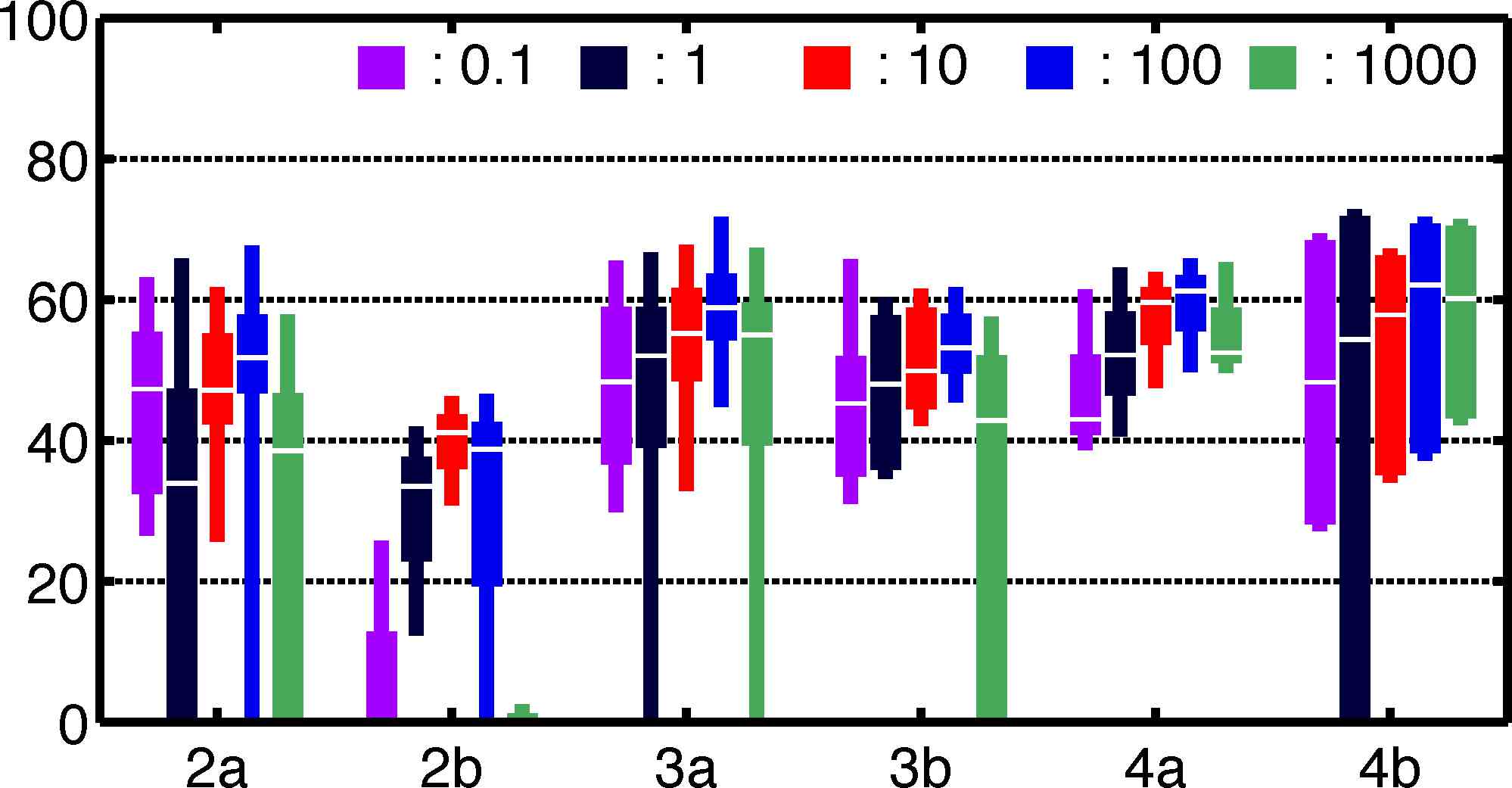} \\ ROV, (ig) \end{center}
\end{minipage}
\begin{minipage}{6cm}\begin{center}
\includegraphics[width=4.2cm,draft=false]{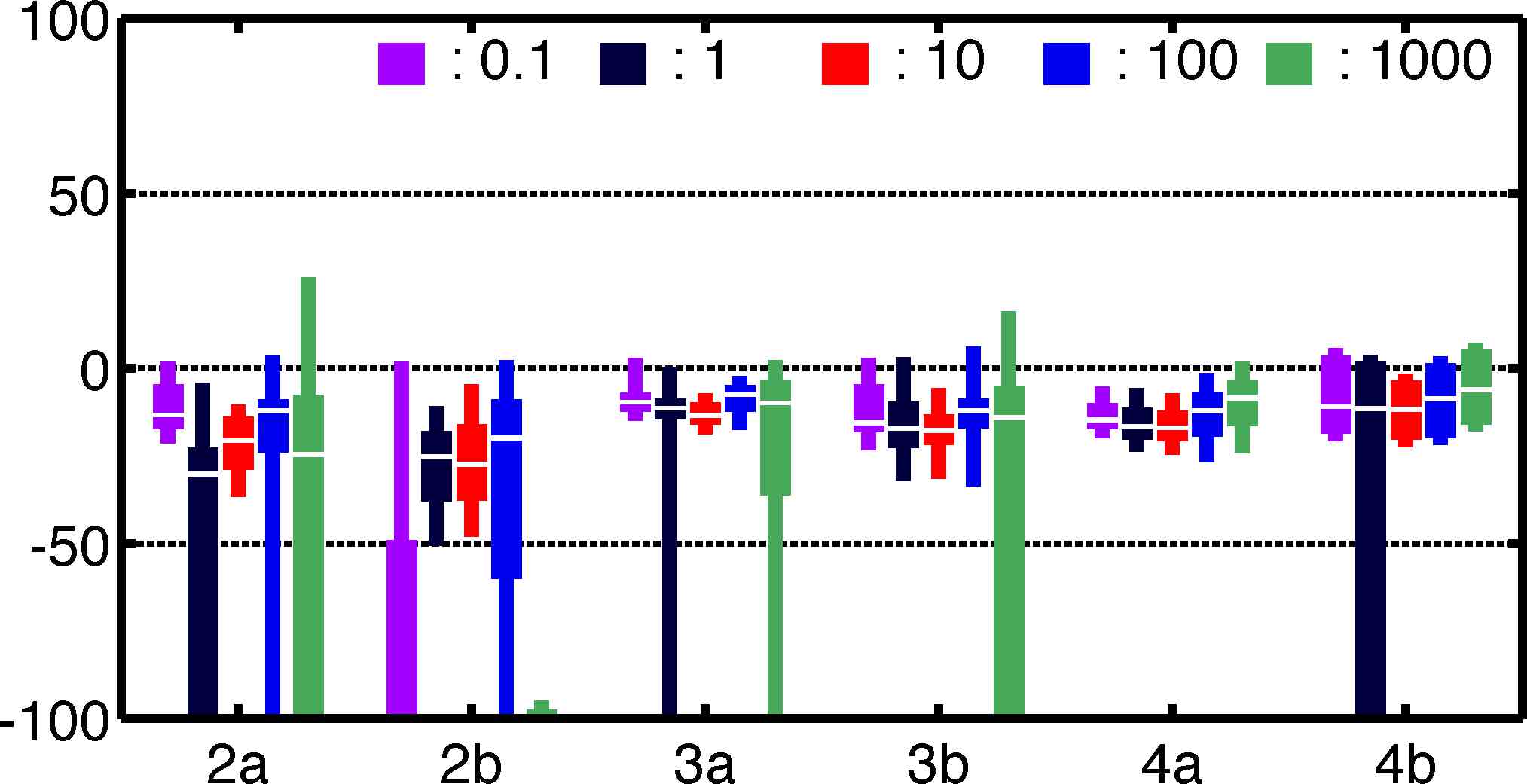} \\ REV, (ig) \end{center}
\end{minipage}\\ \vskip0.1cm
\begin{minipage}{6cm}\begin{center}
\includegraphics[width=4.2cm,draft=false]{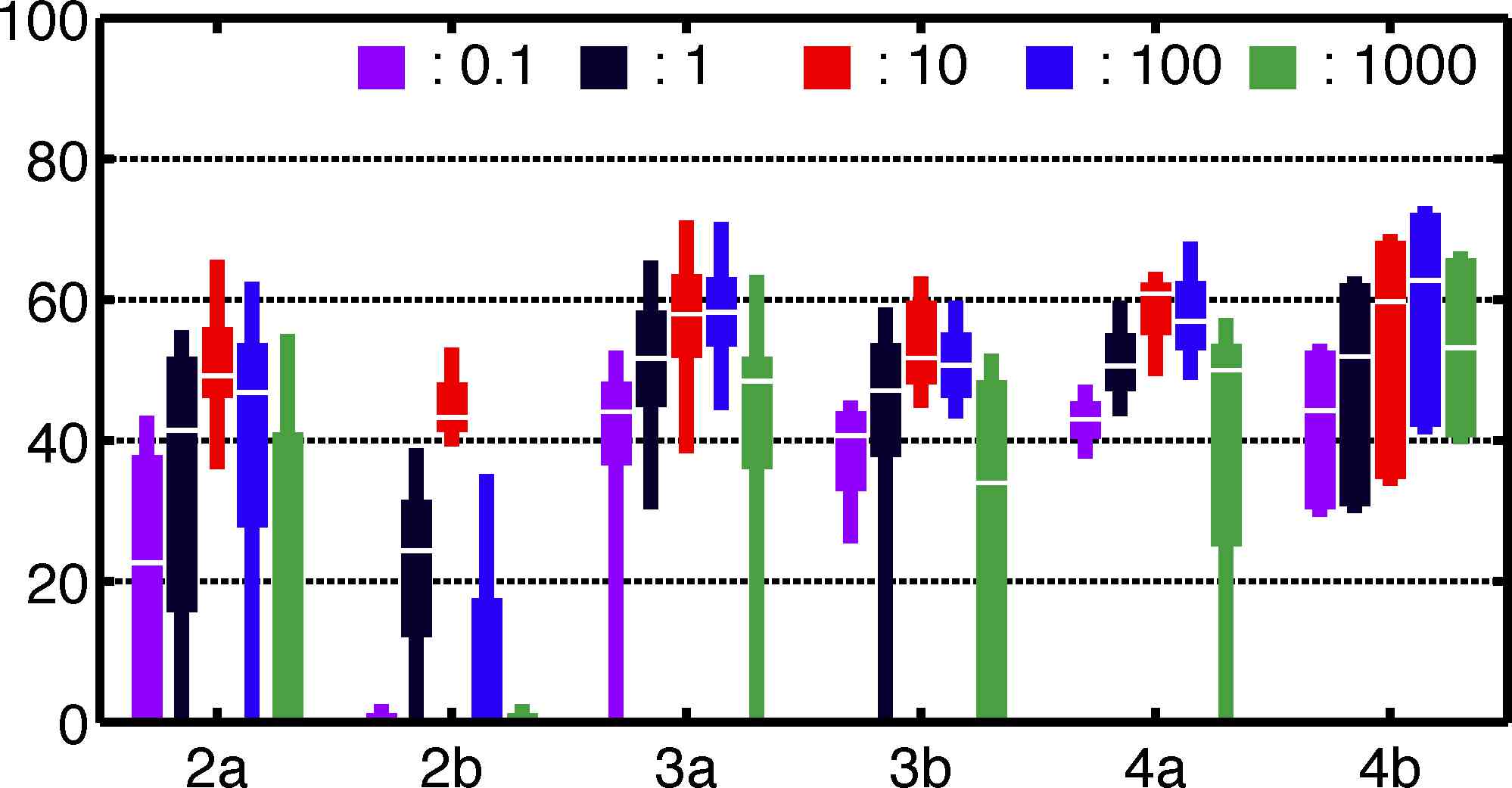} \\ ROV, (f) \end{center}
\end{minipage}
\begin{minipage}{6cm}\begin{center}
\includegraphics[width=4.2cm,draft=false]{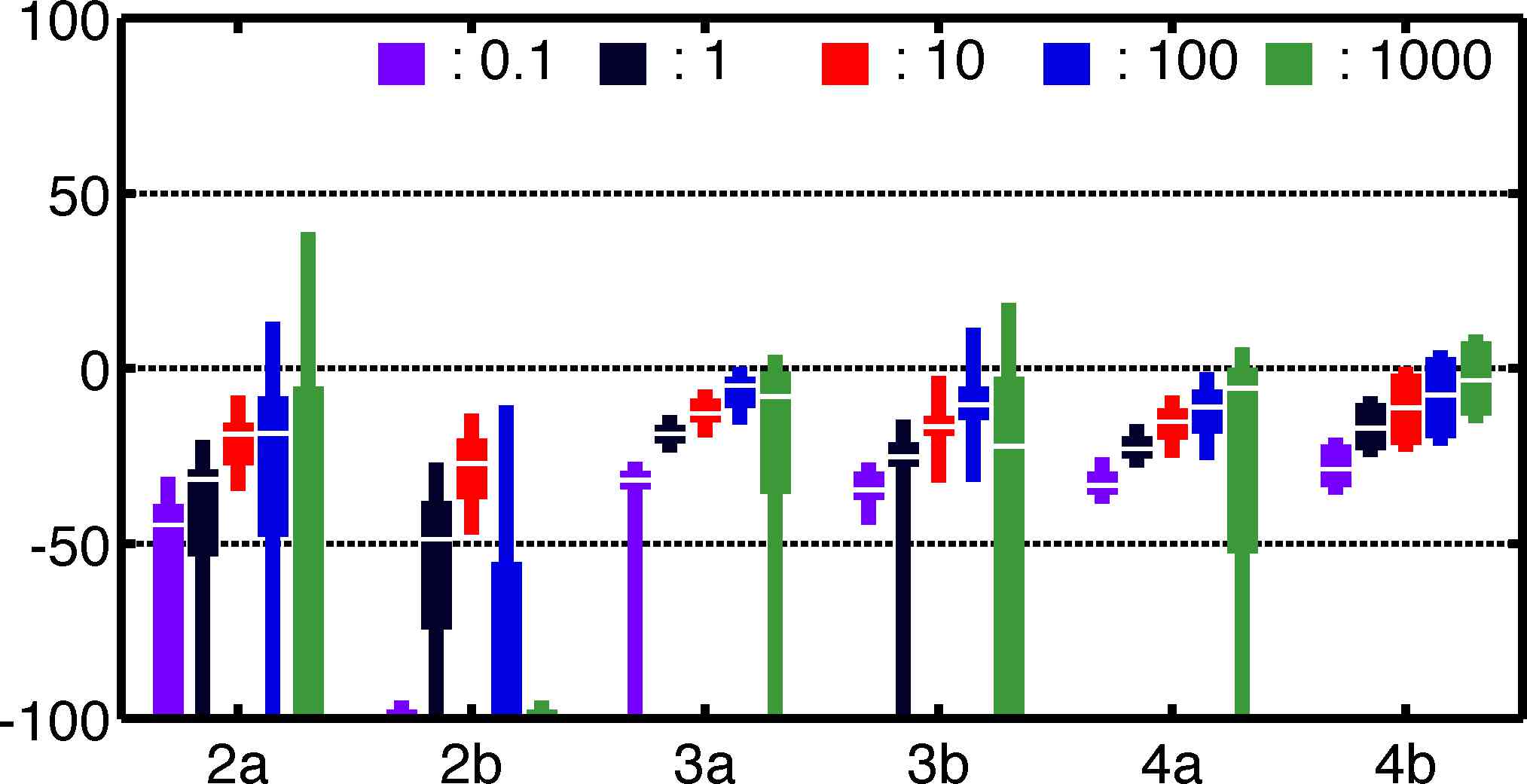} \\ REV, (f) \end{center}
\end{minipage}
\end{footnotesize}
\end{center}
 \caption{Box plots of ROV (left) and REV (right) with the rows  from top to bottom corresponding, respectively, to reconstruction types (g), (ig) and (f). Each subfigure covers source counts 2--4 for a- and b-type configurations and initial prior variances $\theta_0 = 10^k, k=-1,0,2,3$. } \label{comparison_3}
\end{figure}\begin{figure}[h!]\begin{center}\begin{footnotesize}
\begin{minipage}{6cm}\begin{center}
\includegraphics[width=4.2cm,draft=false]{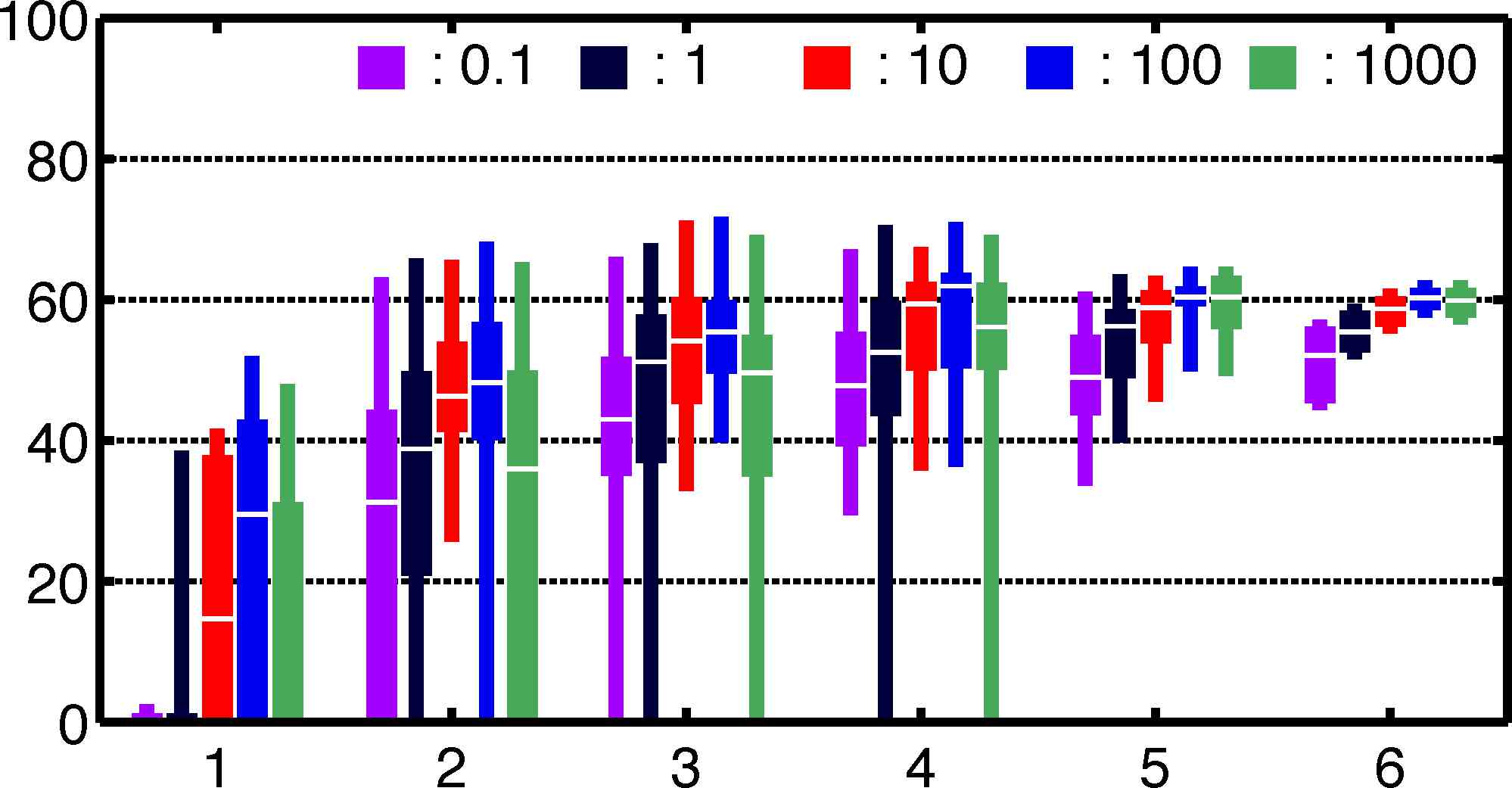} \\ ROV, $\nu=2/5$, $\sigma=1$\end{center}
\end{minipage}
\begin{minipage}{6cm}\begin{center}
\includegraphics[width=4.2cm,draft=false]{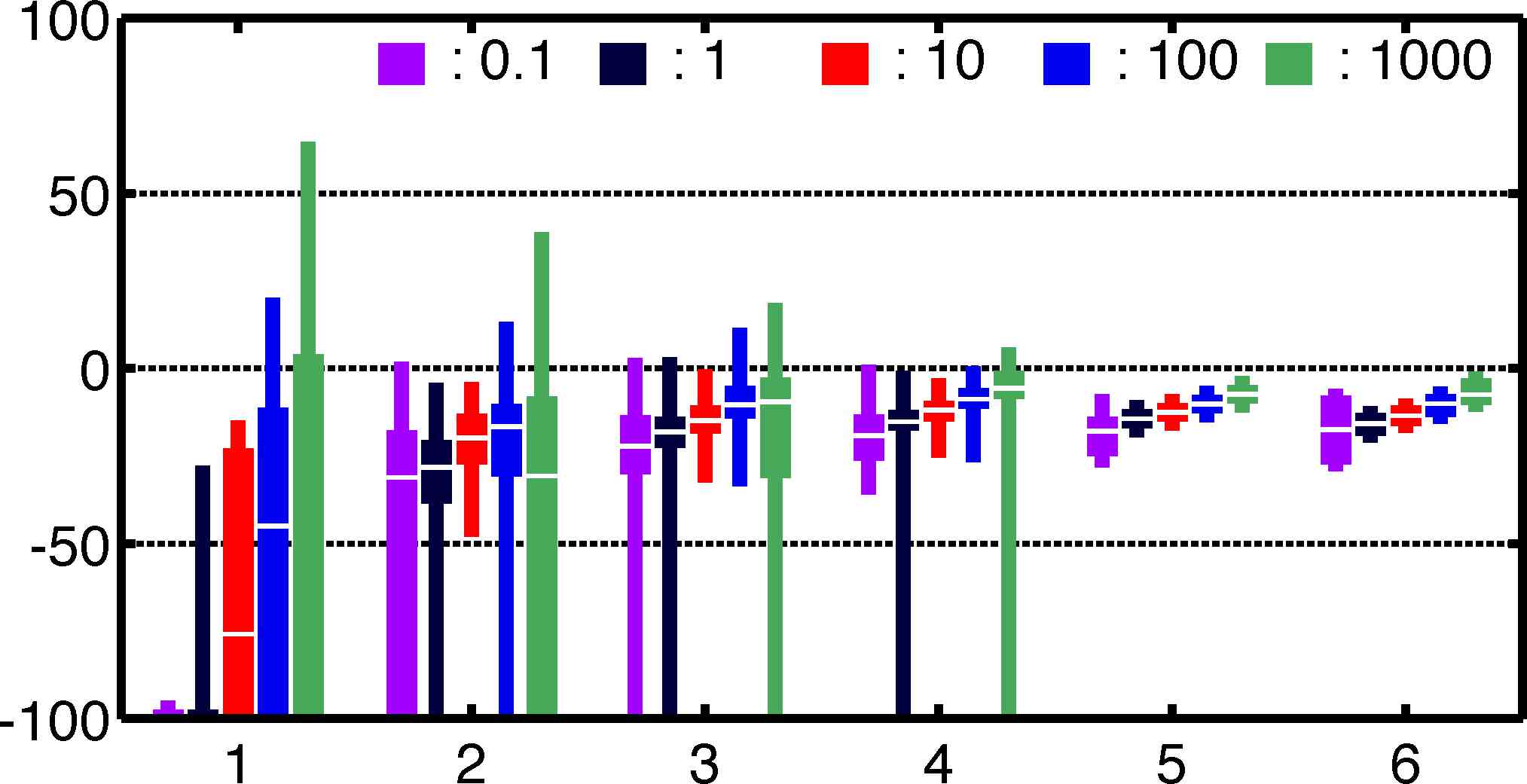} \\ REV, $\nu=2/5$, $\sigma=1$\end{center}
\end{minipage}\\ \vskip0.1cm
\begin{minipage}{6cm}\begin{center}
\includegraphics[width=4.2cm,draft=false]{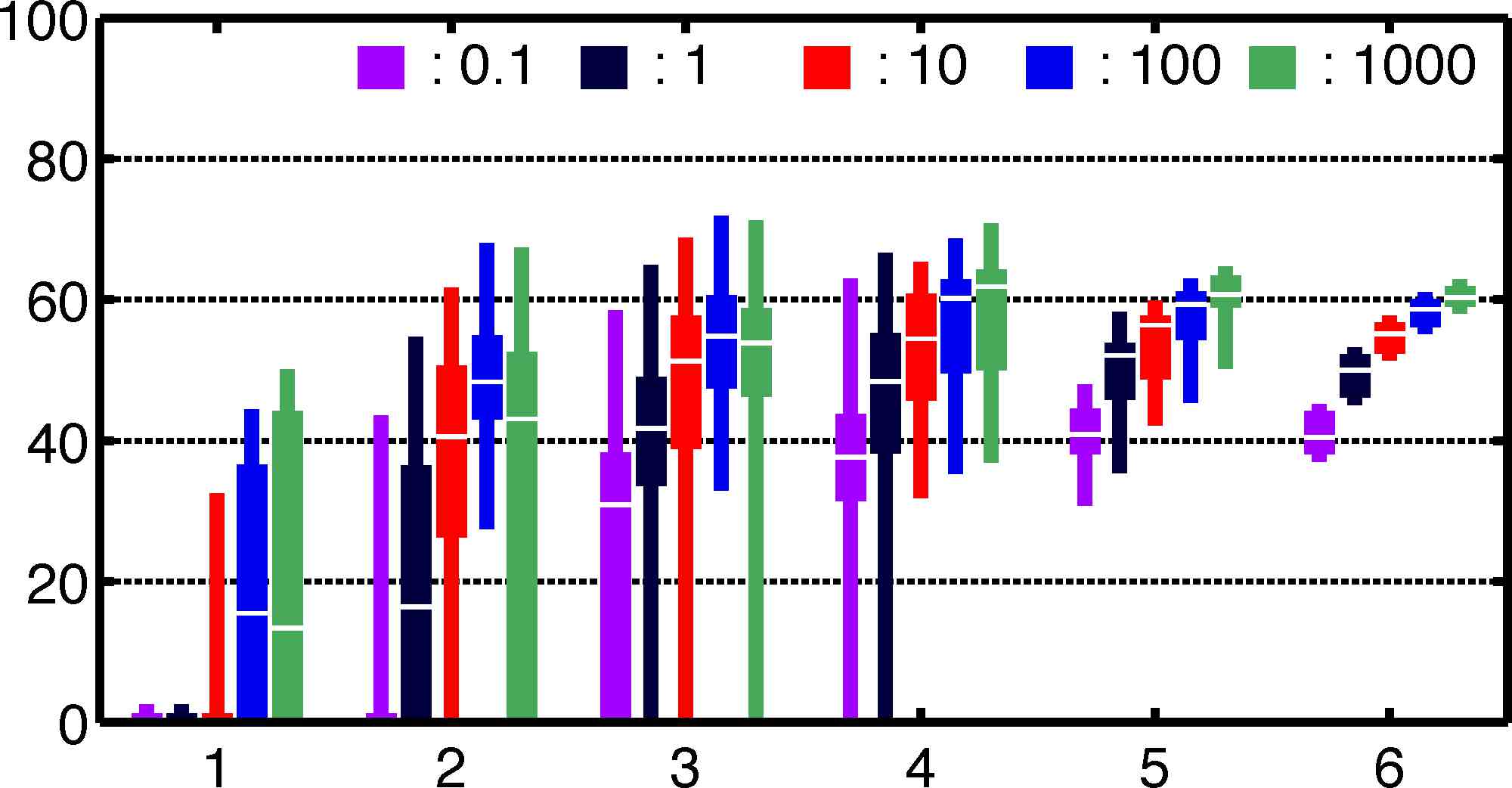} \\ ROV, $\nu=2/5$, $\sigma=2$ \end{center}
\end{minipage} 
\begin{minipage}{6cm}\begin{center}
\includegraphics[width=4.2cm,draft=false]{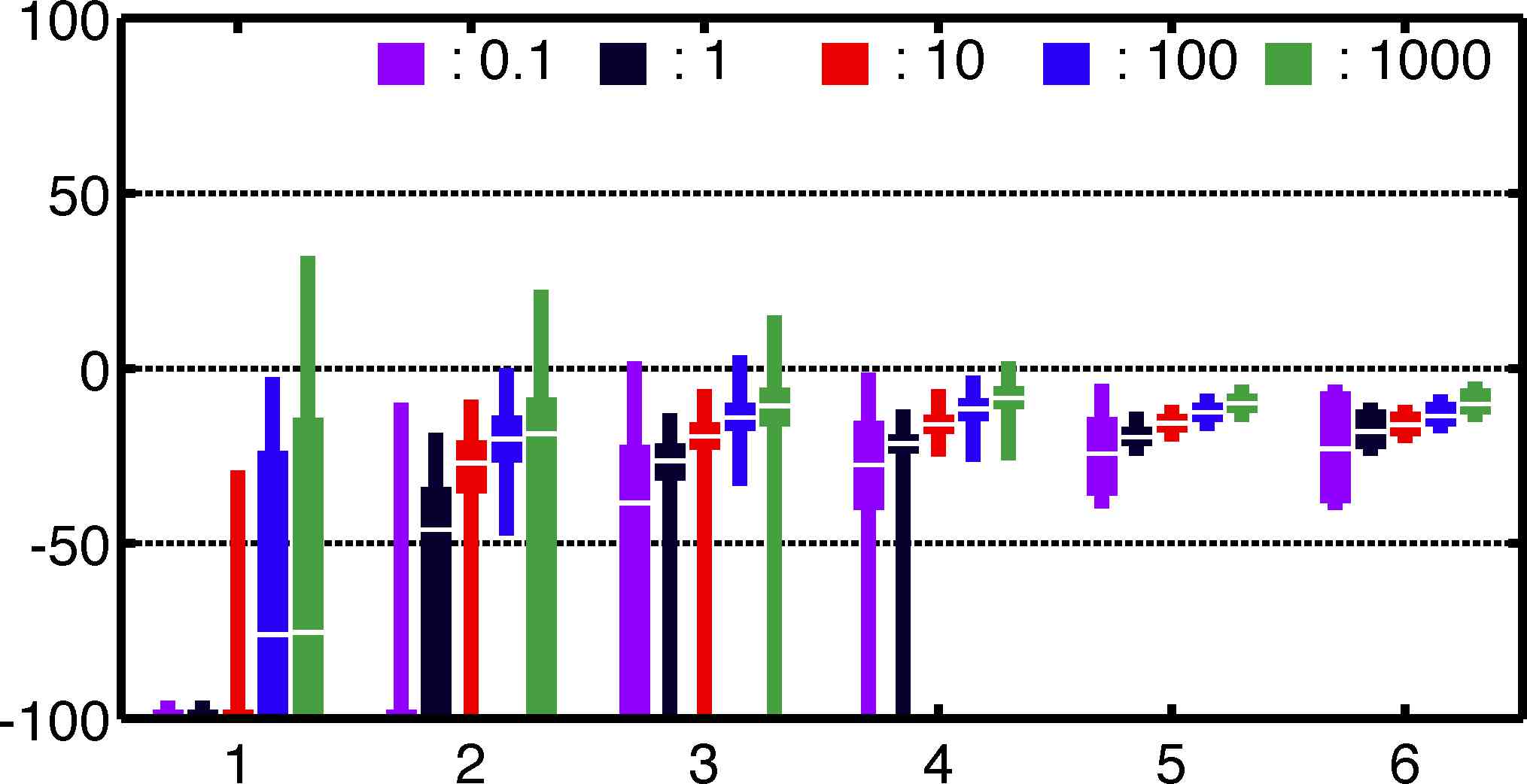} \\ REV, $\nu=2/5$, $\sigma=2$ \end{center}
\end{minipage} \\ \vskip0.1cm
\begin{minipage}{6cm}\begin{center}
\includegraphics[width=4.2cm,draft=false]{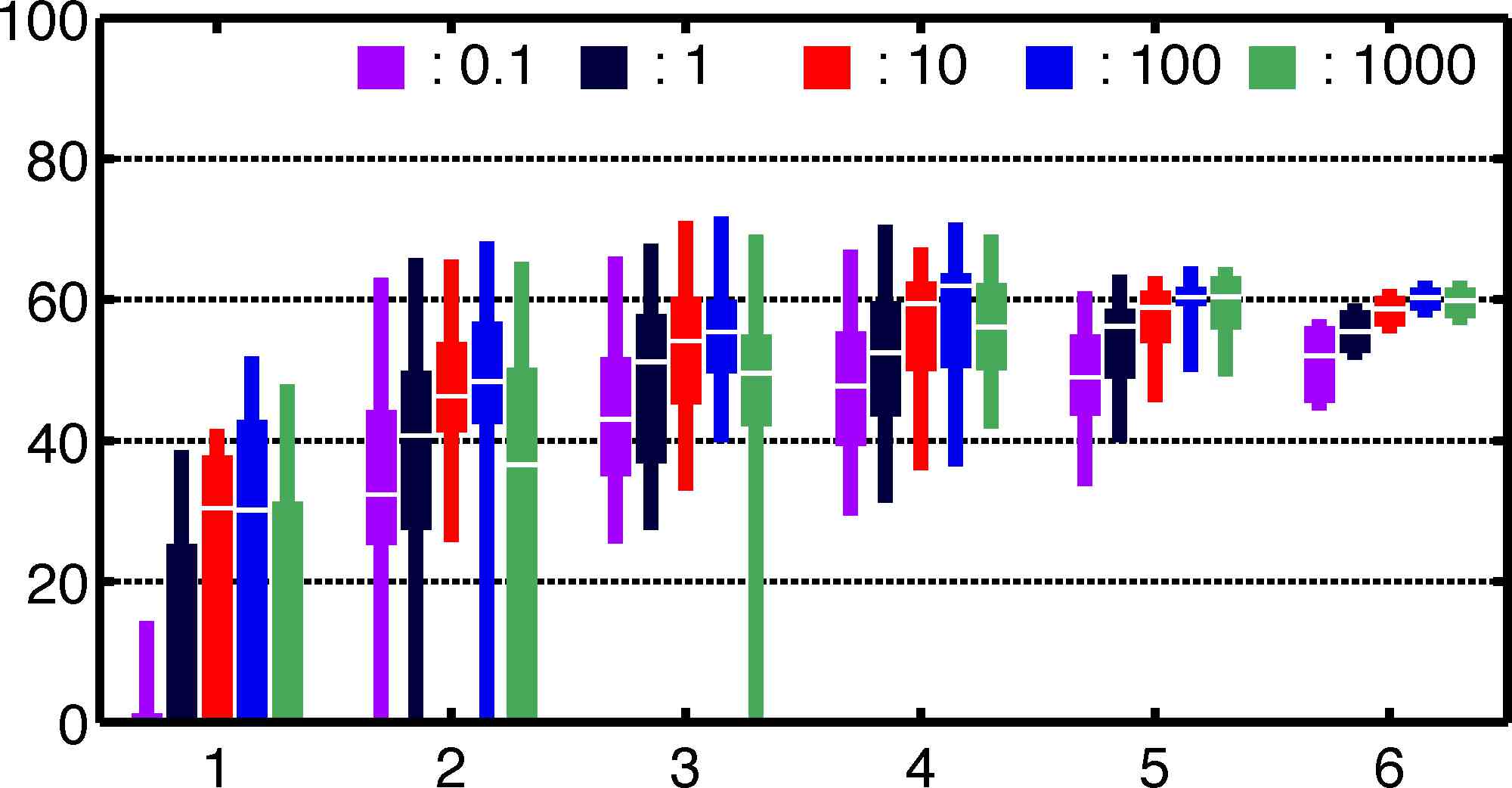} \\ ROV,  $\nu=1/5$, $\sigma=1$ \end{center}
\end{minipage}
\begin{minipage}{6cm}\begin{center}
\includegraphics[width=4.2cm,draft=false]{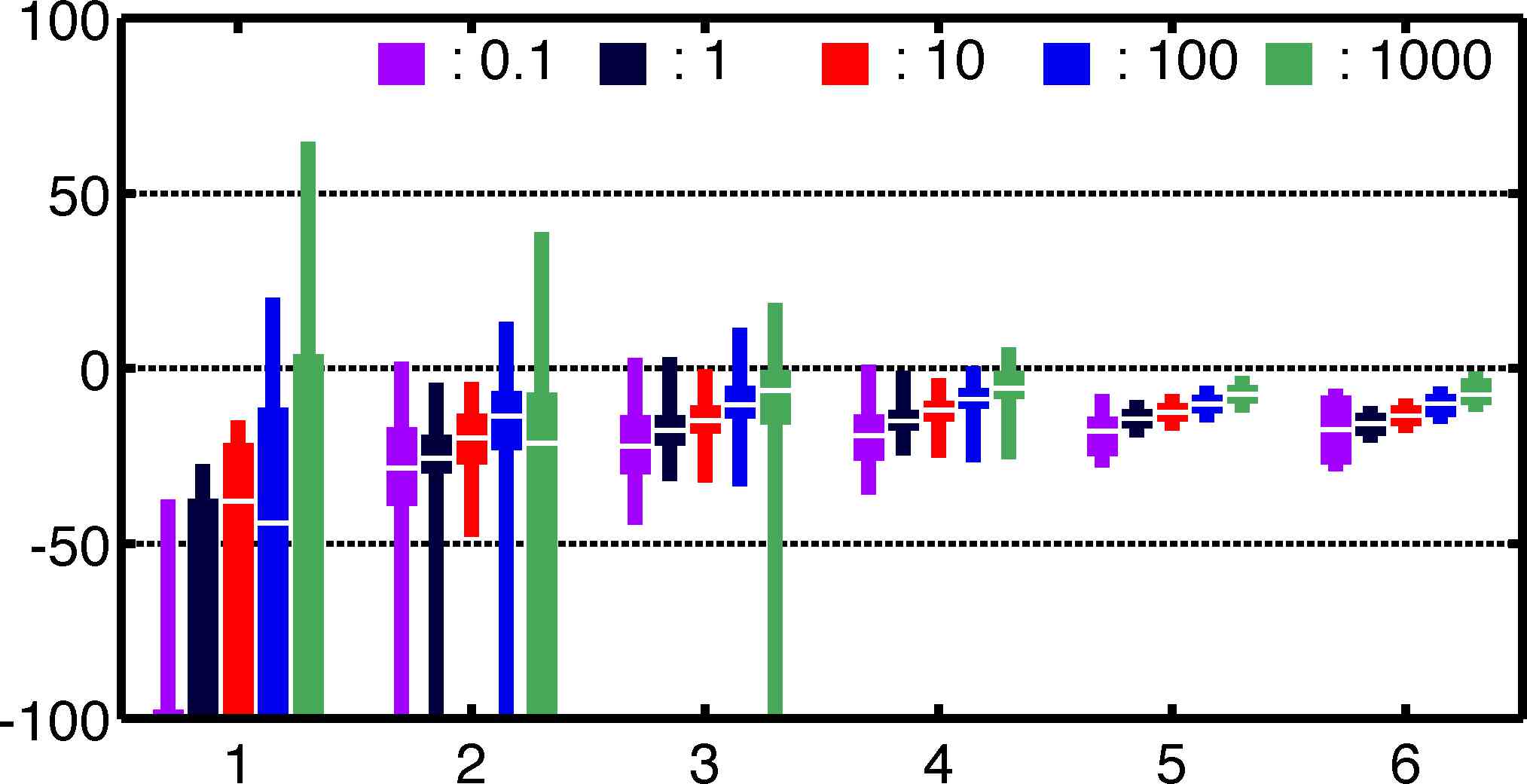} \\ REV,  $\nu=1/5$, $\sigma=1$ \end{center}
\end{minipage} \\ \vskip0.1cm
\begin{minipage}{6cm}\begin{center}
\includegraphics[width=4.2cm,draft=false]{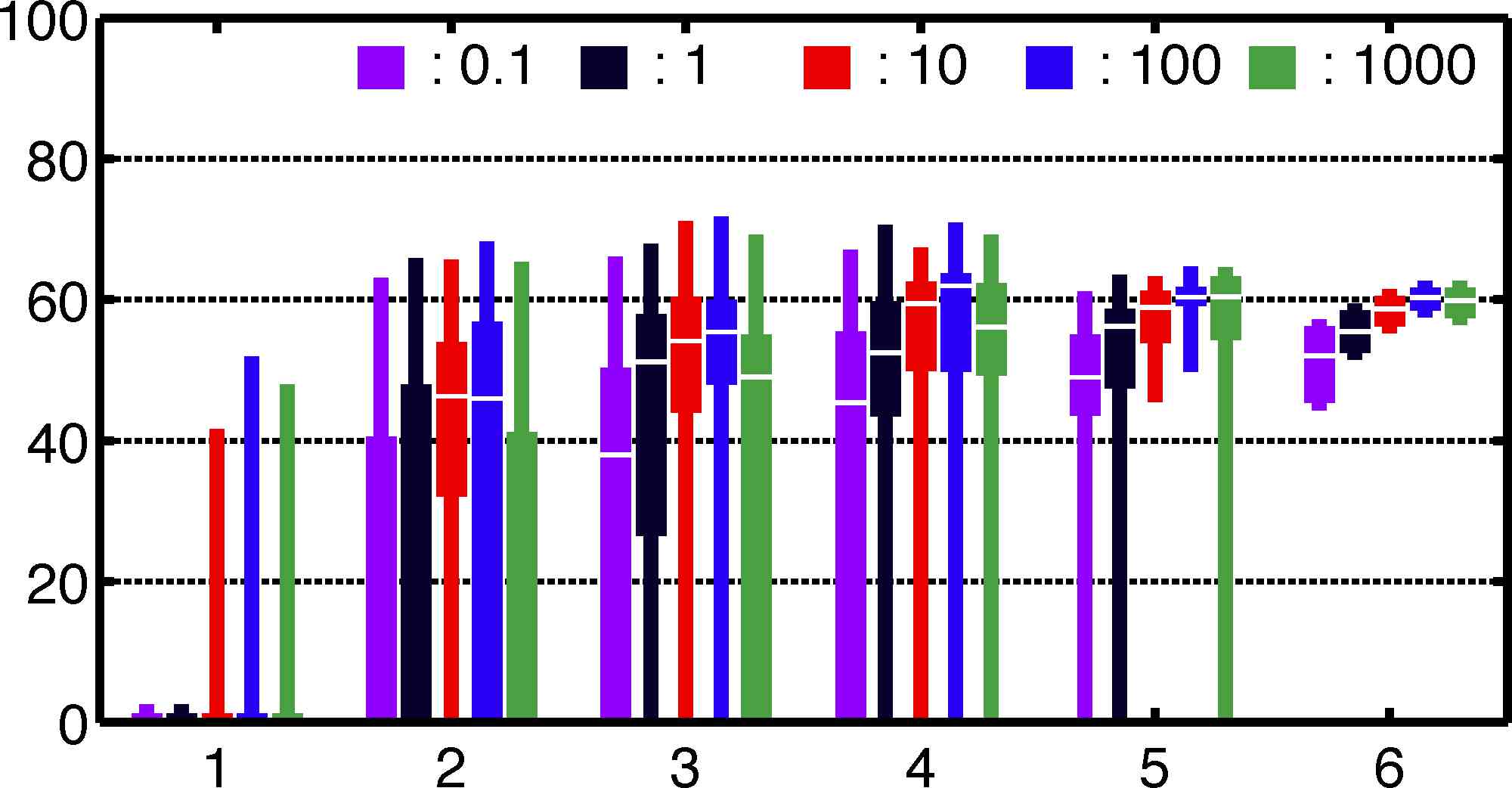} \\ ROV,   $\nu=3/5$, $\sigma=1$ \end{center}
\end{minipage}
\begin{minipage}{6cm}\begin{center}
\includegraphics[width=4.2cm,draft=false]{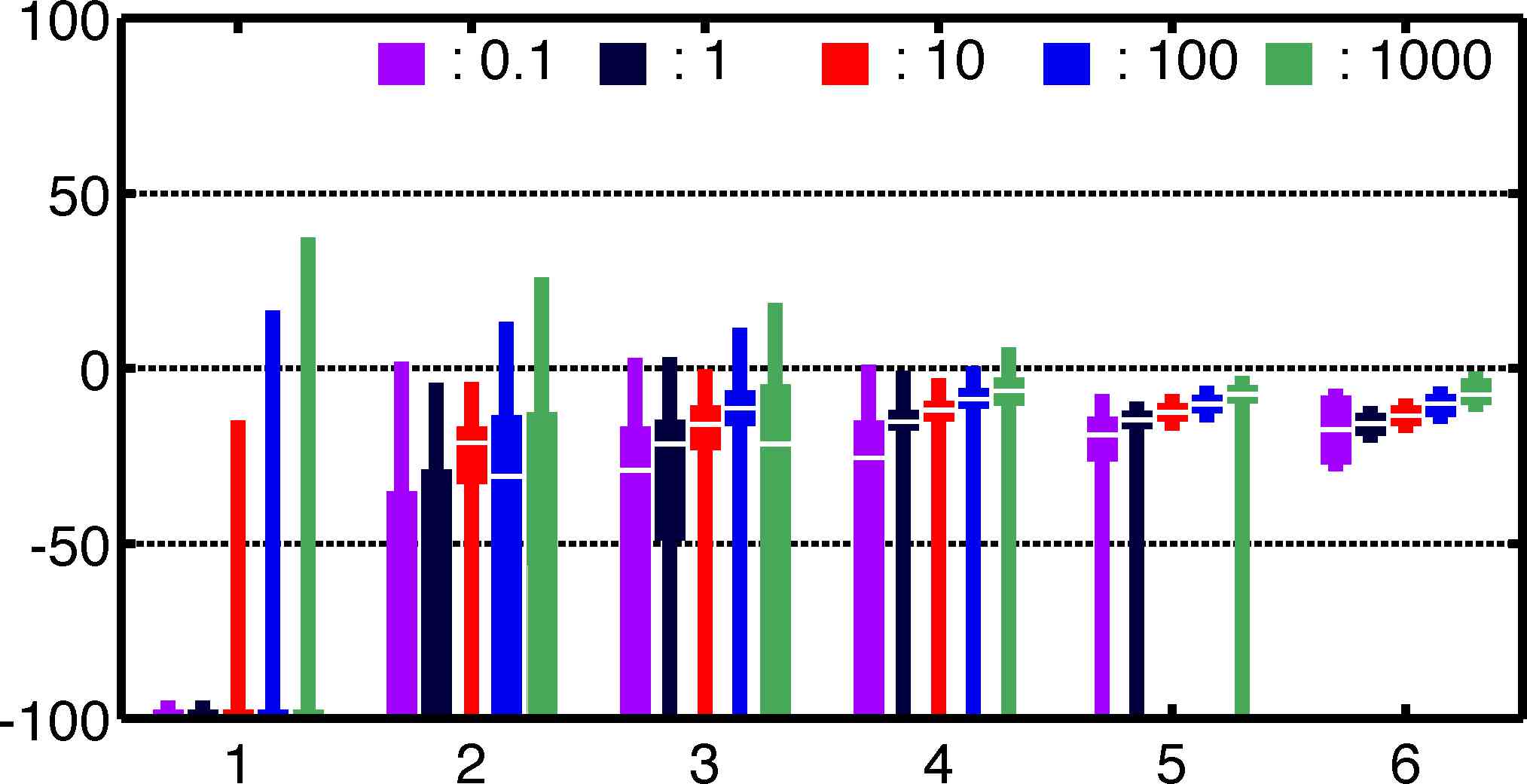} \\ REV,   $\nu=3/5$, $\sigma=1$ \end{center}
\end{minipage}
\end{footnotesize}
\end{center}
 \caption{Box plots of ROV and REV (joint w.r.t.\ reconstruction type) for alternative combinations noise standard deviation $\sigma$ and minimum overlap criterion $\nu$. Each subfigure covers source counts 1--6 and initial prior variances $\theta_0 = 10^k, k=-1,0,2,3$. } \label{alternative_results}
\end{figure}
\begin{figure}[h!]\begin{center}\begin{footnotesize}
\begin{minipage}{3.8cm}\begin{center} \hskip0.3cm
\includegraphics[width=2.414cm,draft=false]{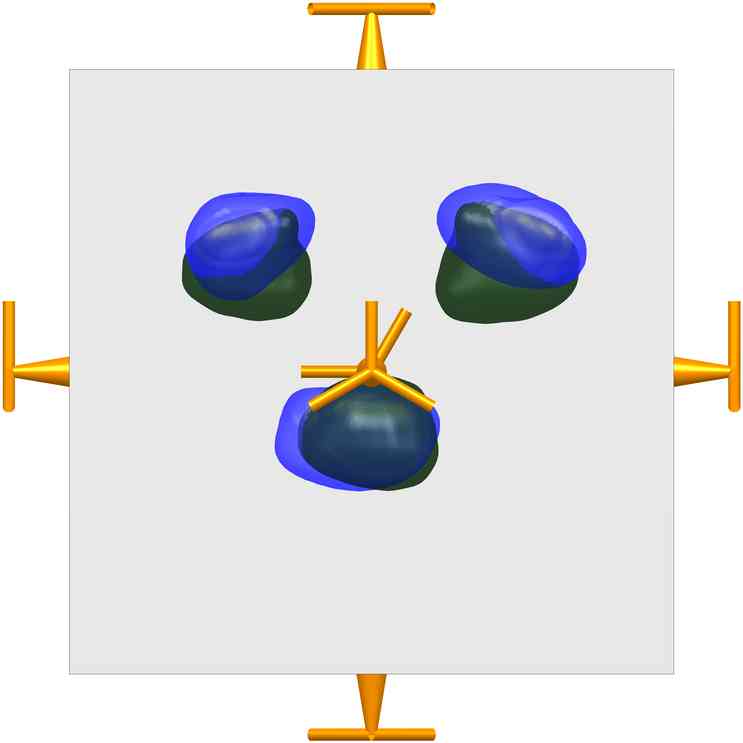} \\ \hskip0.3cm Config.\ {\bf 6}, xy-view \end{center}
\end{minipage}
\begin{minipage}{3.8cm}\begin{center} \hskip0.3cm
\includegraphics[width=2.414cm,draft=false]{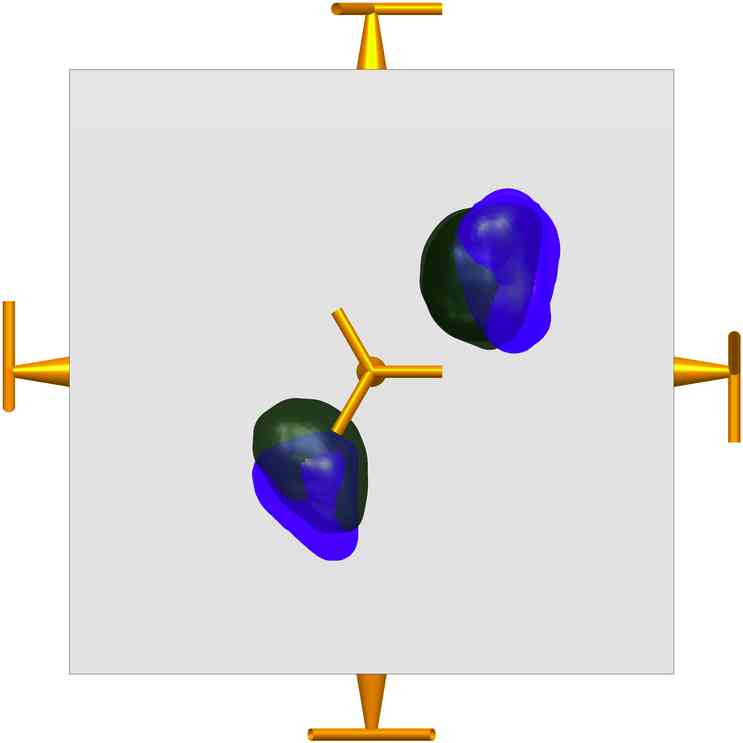} \\ \hskip0.3cm Config.\ {\bf 6}, yz-view \end{center}
\end{minipage}
\begin{minipage}{3.8cm}\begin{center} \hskip0.3cm
\includegraphics[width=2.414cm,draft=false]{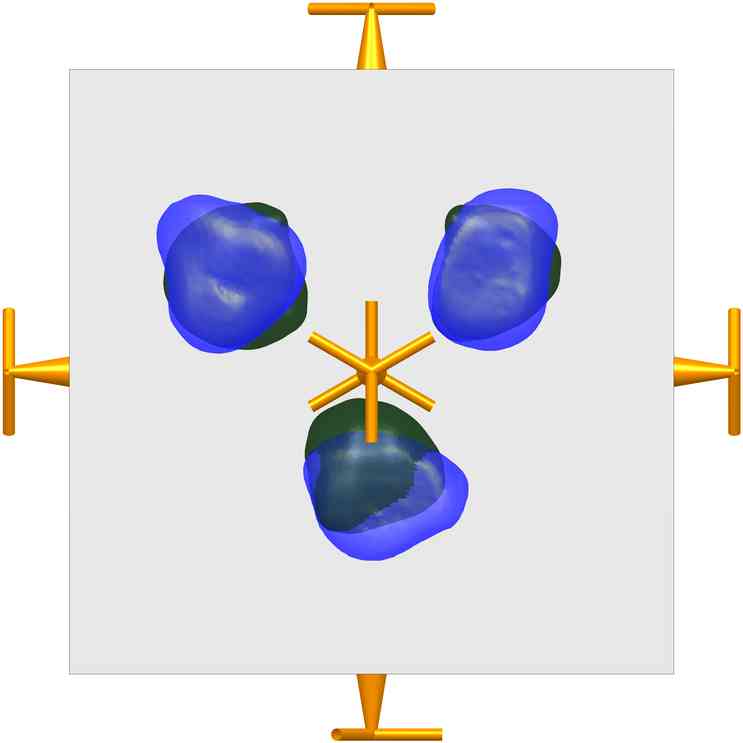} \\ \hskip0.3cm Config.\ {\bf 6}, zx-view \end{center}
\end{minipage}\\ \vskip0.1cm
\begin{minipage}{3.8cm}\begin{center} \hskip0.53cm
\includegraphics[width=2.21cm,draft=false]{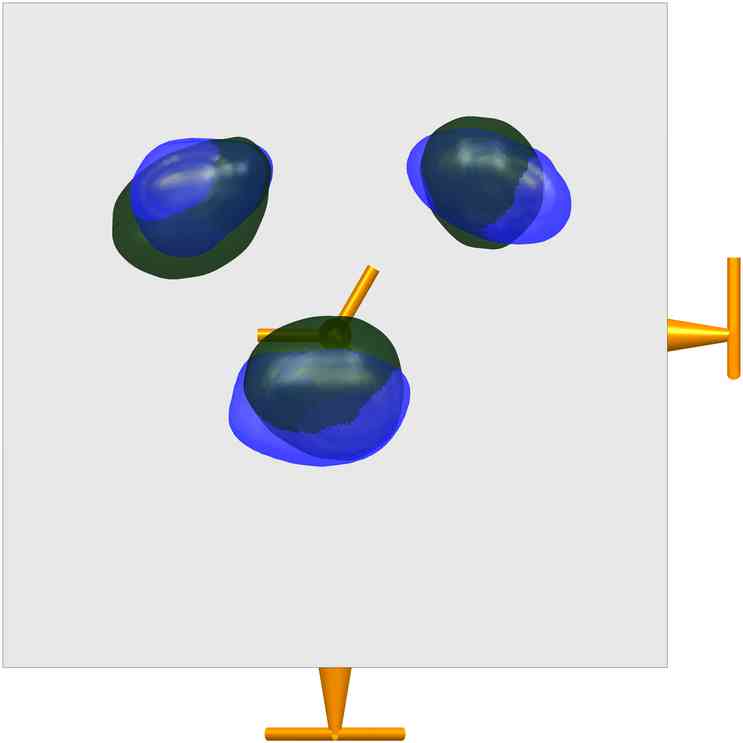} \\  \hskip0.3cm Config.\ {\bf 3a}, xy-view \end{center}
\end{minipage}
\begin{minipage}{3.8cm}\begin{center}
\includegraphics[width=2.21cm,draft=false]{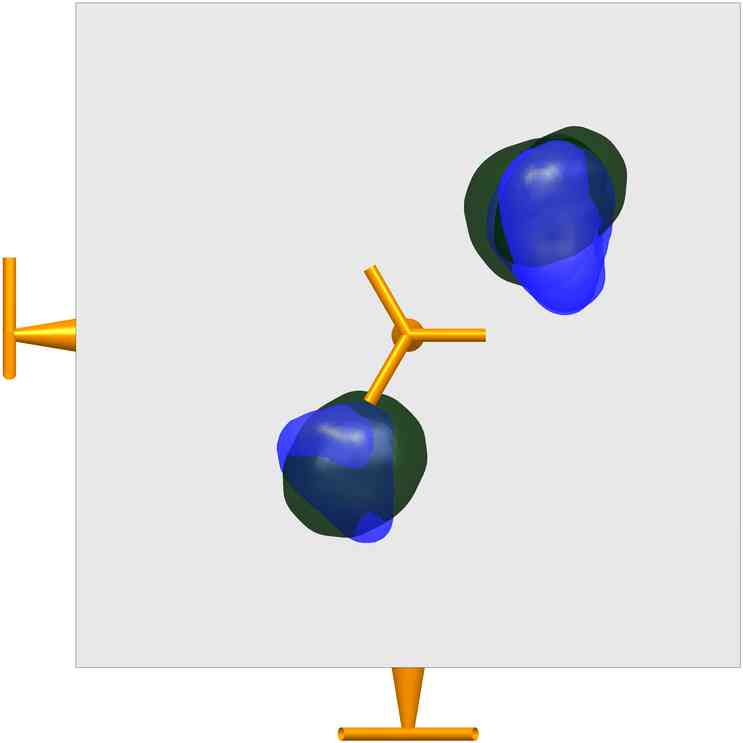} \\  \hskip0.3cm Config.\ {\bf 3a}, yz-view\end{center}
\end{minipage}
\begin{minipage}{3.8cm}\begin{center}
\includegraphics[width=2.21cm,draft=false]{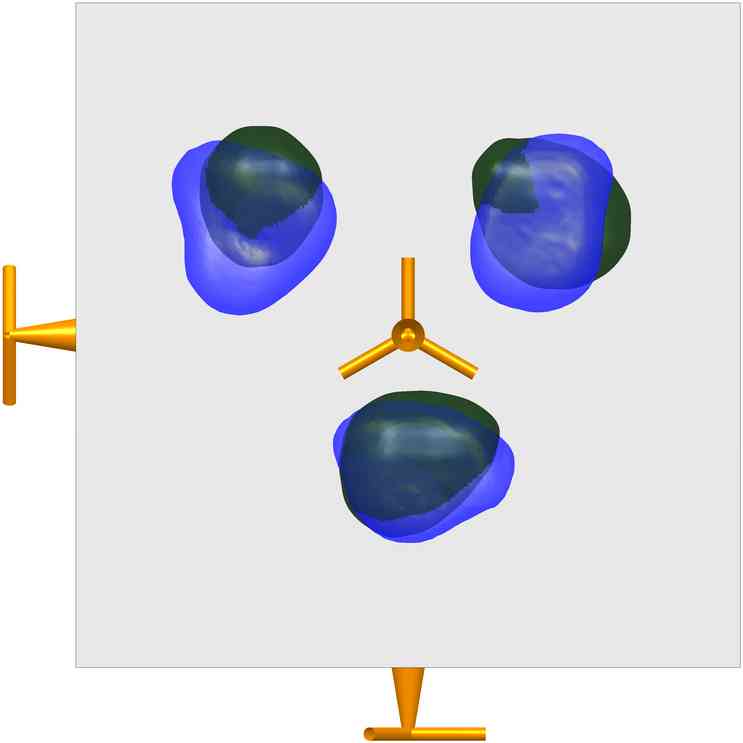} \\ \hskip0.3cm Config.\  {\bf 3a}, zx-view\end{center}
\end{minipage}
\\ \vskip0.1cm
\begin{minipage}{3.8cm}\begin{center} \hskip0.53cm
\includegraphics[width=2.21cm,draft=false]{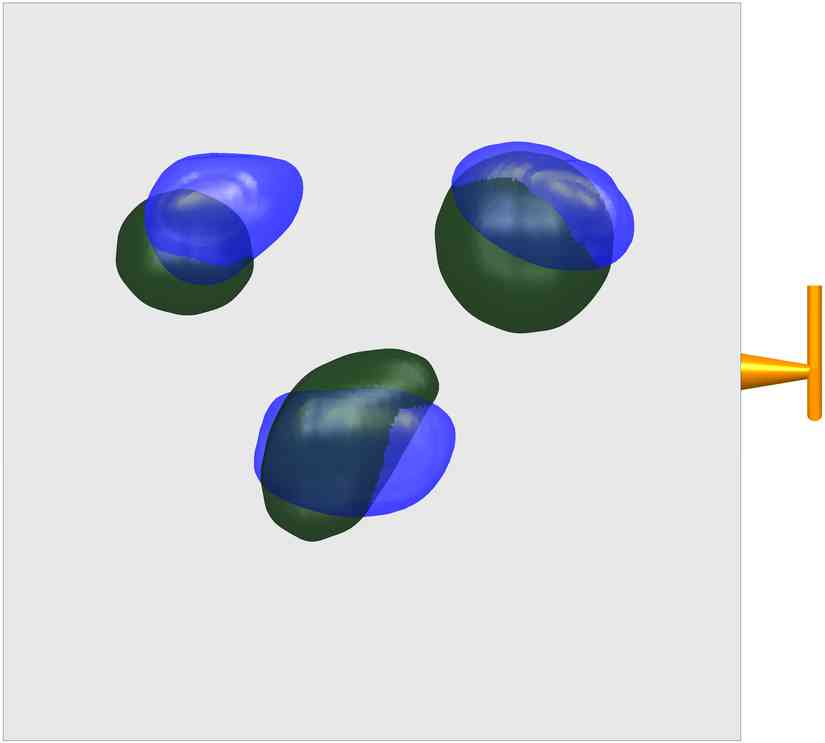} \\  \hskip0.3cm Config.\ {\bf 1}, xy-view \end{center}
\end{minipage}
\begin{minipage}{3.8cm}\begin{center} \hskip0.23cm
\includegraphics[width=2.00cm,draft=false]{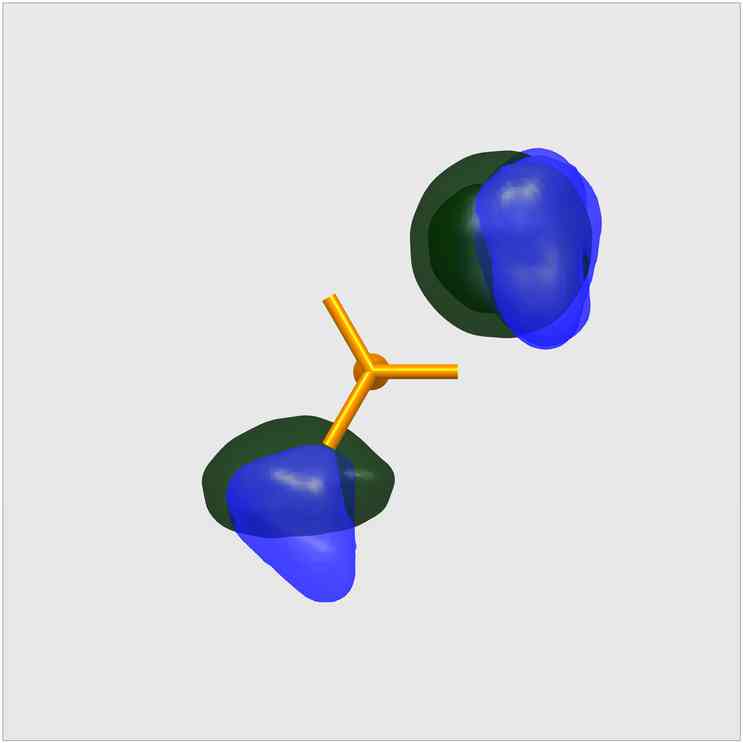} \\  \hskip0.3cm Config.\ {\bf 1}, yz-view \end{center}
\end{minipage}
\begin{minipage}{3.8cm}\begin{center}
\includegraphics[width=2.21cm,draft=false]{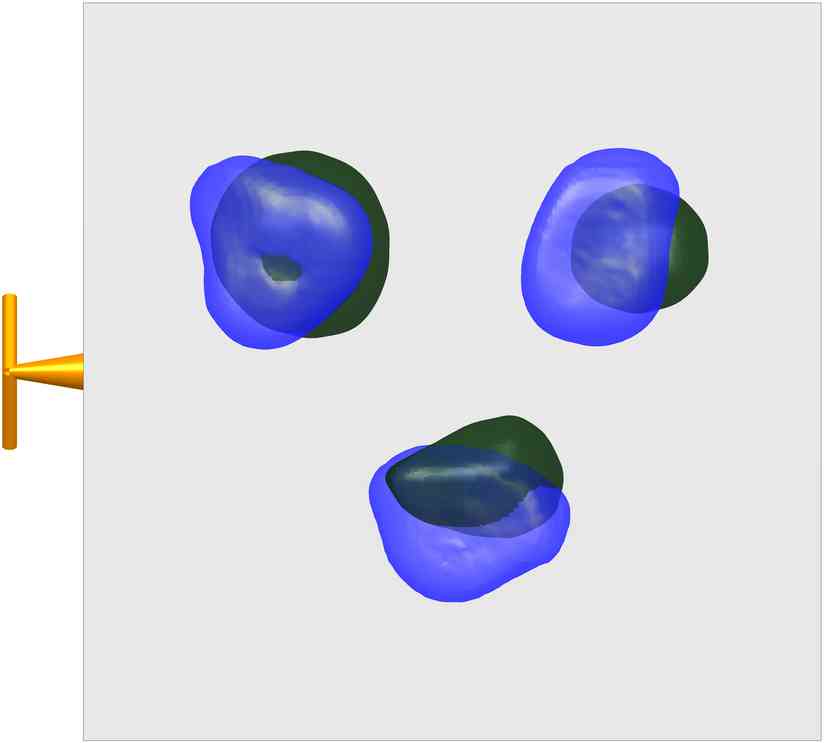} \\ \hskip0.3cm Config.\ {\bf 1}, zx-view \end{center}
\end{minipage}
\end{footnotesize}
\end{center}
 \caption{Visual comparison between the stones $\bigcup_{i=1}^3 \Omega_i$ (light blue, transparent) and the set $\mathcal{R}$ (dark green) covering an equal volume at the extreme part of a (g) type reconstruction obtained with $\theta_0 = 100$. Subfigures cover the source configurations {\bf 1}, {\bf 3a} and {\bf 6}  (from top to bottom, respectively) as well as xy-, yz-, and zx-view (from left to right, respectively).} \label{reconstructions}
\end{figure}

The results regarding ROV and REV have been shown in Figures \ref{comparison_1}--\ref{comparison_3} and in Tables \ref{table_rov} and \ref{table_ranking}. Three examples of the recovered stones with respect to the actual ones have been illustrated in Figure  \ref{reconstructions}. 

Figure \ref{comparison_1} visualizes the sensitivity of the type (g), (ig) and (f) estimates (Section \ref{section_ias_algorithm}) with respect to the choice of  $\theta_0$. Type (g) was found to have been the most robust one based on the amount of successful detections ($\hbox{ROV}>0$ and $\hbox{REV}>-100$). For four and more sources, all detections corresponding to (g) were successful. The case of three sources can also be considered to have been robust as it led to failure only with the extreme value $\theta_0=1000$. Both the accuracy and robustness improved significantly along with  increase in the number of sources from one to three. Further increase led to saturation or slowing down of this tendency. Figure \ref{comparison_2} portrays a mutual comparison of (g), (ig) and (f) for different values of $\theta_0$, showing that the lower the source count, the stronger the mutual differences. Value $\theta_0=10$ yielded the highest number of successful reconstructions and also led to visually the smallest differences between (g), (ig) and (f). Again, values $\theta_0 = 0.1$ and $\theta_0=1000$ resulted in slight under- and over-sensitivity with the lower source numbers 1--3,  indicated by the large errors in the median and very narrow and wide spread of the sample, respectively. This sensitivity was the most pronounced in the case of (f). 
Figure \ref{comparison_3} compares the performances of {\bf a}- and  {\bf b}-type source configurations, suggesting that the former ones led to better results in general with respect to the minimum,  median and spread of the sample. The differences were observed to have been more obvious with respect to ROV than to REV, and the most pronounced between the configurations {\bf 2a} and {\bf 2b}. Table \ref{table_rov} documents the configuration-wise accuracy obtained with the reconstruction type yielding the highest median. Within this context, ROV and REV  corresponded mostly to the types (g) and (ig), respectively. Furthermore, the  relative error of refractive index ($|\hbox{REV}$|) was systematically lower than that of the overlapping volume ($|1-\hbox{ROV}|$). Table \ref{table_ranking} shows the ranking of the source configurations based on Table \ref{table_rov}. The overall ranking regarding both ROV and REV was {\bf 6}, {\bf 5}, {\bf 4a}, {\bf 4b}, {\bf 3a}, {\bf 3b}, {\bf 2a}, {\bf 2b} and {\bf 1} from the highest to lowest, i.e., it was determined primarily by the number of sources and secondarily by the type of the configuration, as intuitively expected. 

Figure \ref{alternative_results} shows ROV and REV (joint w.r.t.\ reconstruction type) for some alternative combinations of the STD of the total noise $\sigma$ and the minimum overlap criterion $\nu$. One can observe that doubling the STD to $\sigma = 2$ $\mu$s decreased the overall quality of the reconstructions, and that the number of accepted  detections was  considerably lower with the tight criterion $\nu = 3/5$  than with the softer ones $\nu = 2/5$ and $\nu = 1/5$. Moreover, a source count below three resulted in a large number of unsuccesfull detections with each tested value of $\nu$.

Finally, Figure \ref{reconstructions} visualizes three (g)-type reconstructions corresponding to the source configurations ${\bf 6}$, ${\bf 3 a}$ and ${\bf 1}$ with $\theta_0 = 100$, comparing the set $\mathcal{R}$ of the recovered stones with that of the original ones $\cup_{i=1}^3\Omega_i$. Based on visual impression, the increase in the number of sources made not only the size and location but also the shape of $\mathcal{R}$  match better with $\cup_{i=1}^3\Omega_i$.

\section{Discussion}

Our results suggest that, in the context of the present tomography approach, the accuracy and robustness of inversion grows and saturates rapidly along with increment in the number of sources. Of the investigated MAP estimate types (g), (ig) and (f), the most suitable ones for reconstructing the anomaly volume and refractive index in terms of the median were (g) and (ig), respectively. Moreover, the former was found to yield the most robust results in general. We propose three or four as the minimum feasible number of sources for robust anomaly detection.  Judging from the saturation of the inversion accuracy, a higher number can be cost-ineffective if the expense of source placement is significant, as in asteroid tomography. The results were observed to be more accurate with respect to the recovery of the refractive index than of the volume, which can be at least partially explained by our simplified assumption of linear signal paths. Ways to improve  the forward simulation include, for example, the finite-difference time-domain (FDTD) method \cite{landmann2010,arnold2011,schneider2012}.

The ranking of the source configurations suggests that a higher number of sources is preferable regardless of positioning and that  ${\bf a}$-type (category I)   configurations are favorable over  ${\bf b}$-types (category II). In practice, source positioning can be, to a large extent, motivated by the geometry and structure of the target; cf.\ the simulations with a realistic (randomized) asteroid geometry in our recent study \cite{pursiainen2013}. 

Thanks to the rapid IAS algorithm, a sample of 315 MAP estimates covering all possible source combinations and $\theta_0=10^k,k=-1,0,1,2,3$ could be produced within a reasonable time. A similar procedure can be used in an authentic situation for pre-inversion data analysis when picking an interval for $\theta_0$. An inversion estimate can then be produced, for instance, through a robust sampling-based (e.g.\ Markov chain Monte Carlo) method \cite{liu2001,calvetti2009,kaipio2004}. Such an inversion procedure can be interesting, for instance, in ultrasonic detection and classification of breast lesions \cite{ruiter2012, ranger2012, duric2007,nowicki2012,mudry2013}. Other potential inversion approaches to be used in the current context of signal sparsity include, for instance, the compressive sensing \cite{gurbuz2009} and transdimensional Bayesian
 methods of geophysics  \cite{sambridge2013,bodin2012,bodin2009}.

The present experiment setup is a simplification of a real-life setting in which the background refractive index will involve more uncertainty. Consequently, the results obtained were somewhat optimistic but served our goal to quantify the effect of the source configuration. In more complicated situations including
 a strongly non-homogeneous background or when the shapes of the anomalies are elongated, planar or heavily non-convex, a higher number of sources or an improved prior model can be necessary. It is also obvious that guaranteeing a given minimum overlap with the actual anomalies can require more sources: e.g., here 50 \% would have  necessitated six sources. Note that our conclusions are conditional at least on the subjective choices of $\sigma = 1$ $\mu$s and $\nu=2/5$, following from our estimate for evident but elusive forward errors and requirement for reasonable correspondence in size between the actual and recovered anomalies.  Additional tests regarding $\sigma$ and $\nu$  supported our conception that the values chosen, indeed, reflect the true forward and inversion accuracy levels, respectively.

Considering asteroid tomography, the current strategy to improve the accuracy of the inversion rather by increasing the number of sources than by extending the forward model is well-motivated, as our recent results suggest that the invertible data are mainly carried by the direct part of the signal \cite{pursiainen2013}: since the back-scattered part is likely to be more noisy due to strong reflections and refractions caused by the overall high permittivity of the asteroid minerals \cite{elshafie2013,virkki2014,bottke2002}, we expect that a forward model including back-scattering \cite{barriot1999} would lead to principally similar results with respect to the number of sources. Forward modelling in asteroid tomography can additionally require propagating the signal from the surface to the orbit which is a missing feature in the current model. Measurements made at the orbit will obviously  include somewhat more noise than data gathered on the surface. For reference, the present ROV results obtained with configuration {\bf 6} are somewhat better than those of  \cite{pursiainen2013} on numerically simulated orbit data.

This study is relevant to the planning of planetary missions in which radio data can be collected akin to CONSERT 
\cite{kofman2007,kofman2004,kofman1998,herique2011,herique2011b,herique2010,herique1999,landmann2010,nielsen2001,barriot1999}. For example, the following setups can be considered: (i) the current one in which one or more fixed or movable signal transmitters or transponders are to be carried by a lander onto asteroid's surface; (ii) an orbiter can transmit a signal pulse and record the back-scattering echo; and, (iii) signals can be transmitted and received between two orbiters located on the opposite sides of the target.  
The results of this study suggest that (i) can provide reliable results with a sparse distribution of sources. Whether (ii) and (iii) can provide similar reliability without a direct surface contact needs to be studied in the future. Another important topic of future research will be the accuracy of direct vs.\ echo measurements. An interesting echo based alternative would be, e.g., to produce conebeam data similar to this study through a highly targeted narrow beam transmitted and received by a movable lander functioning as a scanner. Moreover, the inversion approach of this study can be used up to some extent in alternative tomographic procedures such as in seismic tomography \cite{russell1988}, providing a potential solution for situations in which the target asteroid is impenetrable to a radio signal due to a high concentration of metals, that is characteristic to M-type asteroids \cite{bottke2002}. 

This study also pertains to the three-dimensional tomography of concrete structures 
\cite{rens2000} in which, e.g., air-coupled transducers or transducer arrays can be utilized to collect data on-site over a surface \cite{blum2005,schickert2003,schickert2005}. As indicated by the results, our strategy seems to be able to recover internal defects caused, for example, by casting or grouting processes. Potential source configurations for such a purpose include, for instance, the pillar-compatible ones {\bf 3b} and {\bf 4a}. Moreover, the current measurement strategy is directly applicable to targets made of concrete, motivating a further study on this subject.

As for the future work, the results suggest a further study on the placement of  3--6 sources on the surface of an asteroid. A potential alternative for the hexahedral pattern {\bf 6} will be, among others, a tetrahedral one consisting of four sources. An ongoing study is to utilize our sparse source approach in recovery of  elongated, planar and branched (tree root) shapes through a suitable prior model, such as the total variation  prior \cite{kaipio2004}. Another project in progress aims to investigate the accuracy of the inversion with coarser  materials, such as plaster or concrete, which can include a grained structure, cracks and casting defects hindering the signal and causing considerable model noise. Our general future goal is to provide complementary and more applied knowledge to serve asteroid tomography, ultrasound testing  \cite{schickert2005,bungey2006,raj2002,blitz1996}, tomography of tree roots \cite{attia_al_hagrey2007} as well as medical imaging \cite{nowicki2012,mudry2013}.

\section{Conclusions} 

%The goal of this study was to approximate the minimal number of source positions needed for robust detection of refractive anomalies utilizing real travel time data. The inverse problem was to detect  three onyx stones placed in the central part of a cast synthetic resin cube based on ultrasound travel time measurements. The data were recorded with a standard PUNDIT device measuring the travel time of a 55 kHz signal pulse between two cylindric transducers on opposite faces of the cube. All possible configurations of face centered sources were explored. These were named as {\bf 1}, {\bf 2a}, {\bf 2b}, {\bf 3a}, {\bf 3b}, {\bf 4a}, {\bf 4b}, {\bf 5} and {\bf 6} with the number indicating the source count and the letter marking the type of the configuration. The {\bf a}-types were categorized as of higher interest than the {\bf b}-ones due to more reasonable positioning of the sources. The results were analyzed statistically by covering three different reconstruction (MAP estimate) types (g), (ig) and (f) based on a hierarchical Bayesian inverse model as well as a wide range of prior variances $\theta_0=10^k, k=-1,0,1,2,3$. Iversion accuracy was investigated with respect to ROV (relative overlapping volume) and REV (relative error in value). 

Our results suggest that the accuracy and robustness of the inversion grow and saturate rapidly along with increment in the number of sources. Reconstructions of types (g) and (ig) were found to yield the most accurate median in recovery of the anomaly volume and refractive index,  respectively. Type (g) was observed to result in the most robust inverse results in general. The simultaneous use of four or more source positions resulted in successful (g)-type reconstructions independently of the initial prior variance $\theta_0$ and  of positioning.  Three sources led to almost as good results as four. With a suitably chosen initial value for the prior variance, already two sources yielded an acceptable reconstruction quality. Three or four is proposed as the minimum number of sources for robust anomaly detection. The 
ranking of the source configurations suggests that an increased number of sources is preferable regardless of positioning, and that the ${\bf a}$-type configurations (category I) are favorable over  the ${\bf b}$-types (category II). Future work will address the placing of 3--6 sources on an asteroid surface. A potential alternative to the hexahedral pattern of this study can be, among others, a tetrahedral one consisting of four sources. Even higher numbers of sources can be tested. Our ongoing studies utilize the sparse source approach in the recovery of  various anomaly shapes through a suitable prior model, and investigate the accuracy of inversion with coarser materials such as plaster or concrete.

\ack

This work was supported by the Academy of Finland (Centre of Excellence in Inverse Problems Research and the project "Inverse problems of regular and stochastic surfaces"). SP's work was supported by the Academy of Finland's project number 257288. Thanks to Tapani Honkavaara, Viljami Sairanen, Janne Hirvonen, Arto K\"{o}li\"{o}, Jukka Piironen, Saku Suuriniemi, and Lauri Kettunen for their valuable help.

\section*{References}
\bibliographystyle{iopart-num}
\bibliography{references}

\end{document}